\documentclass[12pt]{article}
\usepackage[english]{babel}
\usepackage{graphicx}
\usepackage{amsmath}
\usepackage{amssymb,color}
\usepackage{bbm}

\def\nn{\nonumber\\}

\def\6#1{{\underline{#1}}}
\def\m6#1{{\underline{#1}\,}}

\newdimen\Tdim
\def\ispan{{\setbox0=\hbox{i}%
\Tdim\ht0\advance\Tdim\dp0\rule[-\dp0]{0pt}{\Tdim}}}
\def\jspan{{\setbox0=\hbox{j}%
\Tdim\ht0\advance\Tdim\dp0\rule[-\dp0]{0pt}{\Tdim}}}
\def\Tspan#1{{\setbox0=\hbox{#1}%
\Tdim\ht0\advance\Tdim\dp0\advance\Tdim.55ex\rule[-\dp0]{0pt}{\Tdim}\box0}}

\def\be{\begin{eqnarray}}
\def\ben{\begin{eqnarray*}}
\def\ee{\end{eqnarray}}
\def\een{\end{eqnarray*}}
\def\Tr{{\rm Tr}}
\newcommand {\tr}{{\rm tr}\,\!}

\def\p{\partial}

\def\D{\mathcal{D}}

\def\=:{=\hspace{-.7em}\raisebox{1.1ex}{.}\hspace{.1em}\raisebox{-0.2ex}{.} }

\newcommand{\NF}{N_{\rm F}}
\newcommand{\NC}{N_{\rm C}}
\newcommand{\hs}[1]{\hspace{#1 mm}}
\newcommand {\1}[1]{\frac{1}{#1}}

\newcommand {\beq}{\begin{eqnarray}}
\newcommand {\eeq}{\end{eqnarray}}
\newcommand {\non}{\nonumber\\}

\makeatletter \@addtoreset{equation}{section}

\renewcommand{\thefootnote}{\fnsymbol{footnote}}
\makeatother

\makeatletter
\newcommand{\thetablename}{Table}
\def\fnum@table{\thetablename\ \thetable}
\makeatother

\begin{document}

\thispagestyle{empty}
\begin{flushright}
DAMTP-2009-27\\
IFUP-TH/2009-07\\
TIT/HEP-595
\end{flushright}
\vspace{3mm}

\begin{center}
{\Large \bf Non-Abelian Vortices in $SO(N)$ and $USp(N)$ \\[10pt]
  Gauge Theories} 
\\[15mm]
{Minoru~{\sc Eto}}$^{1,2}$\footnote{\it e-mail address:
minoru(at)df.unipi.it},
{Toshiaki~{\sc Fujimori}}$^{3}$\footnote{\it e-mail address:
fujimori(at)th.phys.titech.ac.jp},
{Sven Bjarke~{\sc Gudnason}}$^{1,2}$\footnote{\it e-mail address:
gudnason(at)df.unipi.it},
{Kenichi~{\sc Konishi}}$^{1,2}$\footnote{\it e-mail address:
konishi(at)df.unipi.it},  \\[1mm]
{Takayuki~{\sc Nagashima}}$^{3}$\footnote{\it e-mail address:
nagashima.t.aa(at)m.titech.ac.jp},
{Muneto~{\sc Nitta}}$^{4}$\footnote{\it e-mail address:
nitta(at)phys-h.keio.ac.jp},
{Keisuke~{\sc Ohashi}}$^{5}$\footnote{\it e-mail address:
K.Ohashi(at)damtp.cam.ac.uk} and
{Walter~{\sc Vinci}}$^{1,2,5}$\footnote{\it e-mail address:
walter.vinci(at)pi.infn.it}.
\vskip 6 mm

\bigskip\bigskip
{\it
$^1$  
~Department of Physics, University of Pisa,
Largo Pontecorvo, 3, Ed. C, 56127 Pisa, Italy
\\
$^2$  INFN, Sezione di Pisa,
Largo Pontecorvo, 3, Ed. C, 56127 Pisa, Italy \\
$^3$ Department of Physics, Tokyo Institute of
Technology, Tokyo 152-8551, Japan\\
$^4$ Department of Physics, Keio University, Hiyoshi,
Yokohama, Kanagawa 223-8521, Japan\\
$^5$ Department of Applied Mathematics and Theoretical Physics, \\
University of Cambridge, CB3 0WA, UK\\
}

\bigskip

\bigskip

{\bf Abstract}\\[5mm]
{\parbox{13cm}{\hspace{5mm}
Non-Abelian BPS vortices in $SO(N)\times U(1)$ 
and $USp(2N)\times U(1)$ gauge theories are constructed in  maximally color-flavor locked vacua. We study in detail their moduli and transformation properties under the exact symmetry of the system.  Our results generalize non-trivially those found earlier 
in supersymmetric $U(N)$ gauge theories.  The structure of the moduli spaces 
turns out in fact to be considerably richer here   than what was found in the $U(N)$ theories. 
 We find that vortices are generally of the semi-local type,  with power-like tails of  profile functions.
}}
\end{center}
\newpage
\pagenumbering{roman}
\tableofcontents
\newpage
\pagenumbering{arabic}
\setcounter{page}{1}
\setcounter{footnote}{0}
\renewcommand{\thefootnote}{\arabic{footnote}}
\section{Introduction}

Solitons play an important role in a wide range of physics, 
from {condensed-matter and fluid dynamics} to cosmology 
and particle physics {\cite{Abrikosov}-
\cite{Vilenkin}}. 
A quiet revolution in our understanding of soliton vortices 
has been taking place in the last few years, 
in the context of supersymmetric gauge theories, 
triggered by the discovery of  {\it vortices 
of non-Abelian type} \cite{HT,ABEKY}. 
The latter represent continuous families of vortex solutions 
carrying various moduli corresponding to the internal orientations 
(related to the exact flavor symmetry of the system) 
as well as other zero-modes. 
It is possible that such non-Abelian vortices 
are a key to unravel the mystery of confinement 
in Quantum Chromodynamics (QCD).  

Motivated by this, together with physical interests, 
many related questions have been investigated 
systematically by several groups
{\cite{Isozumi:2004vg}-
\cite{Collie:2008mx}}. 
The moduli-matrix formalism was introduced in
Refs.~{\cite{Isozumi:2004vg}-
\cite{Isozumi:2004jc}} in order to  
exhaust all possible moduli. 
The moduli and transformation properties of the moduli 
in the cases of  {composite} vortices (higher-winding vortices) 
\cite{HashiTong, ASY,Eto:2006cx,Duality} 
and semi-local vortices \cite{SYSemi,SemiL}, have been studied.  
A new type of (Seiberg-like) duality was found among pairs of models 
having related vortex moduli, sharing a common sigma-model-lump limit
\cite{SemiL}. 
Systems having vortex solutions carrying more than one non-Abelian
modulus factor have been {studied} recently \cite{DKO}. {Furthermore, vortices were found to provide us with a deep physical intuition about some well-known correspondence between theories in the four and two dimensions \cite{Dorey,HT2,Shifman:2004dr}. So far, however, 
most studies} have been limited to 
the case of  $U(N)$ gauge theories, 
with a few but notable exceptions \cite{FK,FGK,Eto:2008qw}. 

In a brief note, some of the present authors have presented 
a general prescription for constructing the 
Bogomol'nyi-Prasad-Sommerfield (BPS) vortices 
in color-flavor locked vacua of a more general class of theories, 
with a gauge group of the form, $G = G^{\prime}\times U(1)$, 
where $G^{\prime}$ is any semi-simple group \cite{Short}. 
Some explicit expressions for the moduli matrix construction 
of the minimum-winding vortex in $SO(N)$, $USp(2N)$ models were given
there.  

It is the purpose of this paper to discuss 
the properties of the non-Abelian BPS vortices 
in $SO(N)$ and $USp(2N)$ gauge theories in more detail. 
The moduli space in each case is carefully analyzed, 
both for the fundamental (or minimal) vortices and 
for the winding-number two vortices. 
The study of the non-minimal vortices and 
their transformation properties is particularly important 
from the point of view that the latter has a simple group-theoretic
nature, in terms of a dual group. 
When the model is embedded in a larger, underlying gauge group, 
spontaneously broken to the model under study, 
the vortex transformation properties endow the monopoles 
appearing at the extremes of the vortices with non-Abelian moduli. 
The latter can be interpreted as gauge modulations 
in the dual,  confinement phase \cite{Duality}. 
       
    The paper is organized as follows. 
In Section~\ref{Basics} the model is presented 
and the vortex Bogomol'nyi equations are put in a 
simple form by the introduction of the moduli matrix. 
The basic characterization of vortices 
in $SO(N)$ and $USp(2N)$ theories which follows 
from this general construction is discussed. 
    Section~\ref{sec:local vortex} is dedicated to the study of vortex solutions of 
the Abrikosov-Nielsen-Olesen (ANO) \cite{Abrikosov,Nielsen:1973cs} type
(sometimes called local vortices), 
their moduli space and its structure. {We make use of concrete examples (the lowest-rank gauge groups) for the sake of clarity}. 
    {The analysis is then extended in Section~\ref{sec:semilocal} 
to a larger set of BPS-saturated vortex solutions which includes  the so-called semi-local vortices \cite{Achucarro}}. 
The structure of the moduli space including these points 
becomes {much richer}. 
Again, {we discuss in some detail a few concrete cases} 
with the lowest-rank gauge groups. 
{An index theorem for the dimension of the moduli space} for vortices with a general gauge group $U(1) \times G'$ 
is discussed in Appendix \ref{sec:index_th}.

Two issues of considerable interest 
seem to emerge from our study, 
which are only briefly discussed here. 
One is the question of the Goddard-Nuyts-Olive-Weinberg(GNOW)
quantization/duality of the non-Abelian vortices, 
which is deeply related to the original problem 
of understanding non-Abelian {\it monopoles} \cite{GNO}. 
{Another is the appearance of ``fractional vortices'', which seems to be very common
 when one studies vortices in models other than the $U(N)$ gauge theories. } 
Although the results of the present paper provide us 
with a concrete starting point and important ingredients for the analysis of these questions, 
in order to keep the length of the paper  to a reasonable size and for the ease of reading, 
we reserve a more thorough discussion of these two issues 
for separate, forthcoming {papers}.

\section{Vortex equations and basics \label{Basics}}

\subsection{The moduli matrix and BPS equations}

In this section we study vortex solutions 
in four-dimensional gauge theories with an 
$SO(N)\times U(1)$ or $USp(N)\times U(1)$ gauge group\footnote{ 
  The case of local vortices with the gauge groups $SO(N)\times U(1)$
  has first been considered in Ref.~\cite{FGK}.}, 
and with $N_{\rm F}$ scalars in the fundamental
representation.
Sometimes the gauge group will be indicated in a more general way, 
as $G = G' \times U(1)$ with $G'$ being any simple group; 
the prescription for writing down the BPS vortex solutions 
in all these cases has in fact been given in Ref.~\cite{Short}. 
However, below we shall concentrate on the gauge groups $SO(N)\times
U(1)$ and $USp(N)\times U(1)$. 
An integer $M$ will be used to indicate the gauge group, 
such that $N = 2M$ or $N = 2M+1$, 
for even $SO(N)$ and $USp(N)$ or odd $SO(N)$, respectively.

The Lagrangian density reads
\beq
\mathcal{L} = \Tr_c\left[
-\frac{1}{2e^2}F_{\mu\nu}F^{\mu\nu}
-\frac{1}{2g^2}\hat{F}_{\mu\nu}\hat{F}^{\mu\nu}
+\D_\mu H\left(\D^\mu H\right)^\dag
-\frac{e^2}{4}\left|X^0t^0 - 2\xi t^0\right|^2
-\frac{g^2}{4}\left|X^at^a\right|^2 \right]\ , \label{Lagrang}
\eeq
with the field strength, gauge fields and covariant derivative denoted as 
\begin{align}
F_{\mu\nu} &= F_{\mu\nu}^0t^0 \ , &
F_{\mu\nu}^0 &= \partial_\mu A_\nu^0 - \partial_\nu A_\mu^0 \ , \qquad
\hat{F}_{\mu\nu} = \partial_\mu A_\nu - \partial_\nu A_\mu +
i\left[A_\mu,A_\nu\right],   \nonumber \\
A_\mu &= A_\mu^at^a \ , &
\D_\mu &= \partial_\mu + iA_\mu^0t^0 + iA_\mu^at^a \,.
\end{align}
$A_\mu^0$ is the gauge field associated with $U(1)$
and $A_\mu^a$ are the gauge fields of $G'$. 
The matter scalar fields are written 
as an $N \times N_{\rm F}$ complex 
color (vertical) -- flavor (horizontal) mixed matrix $H$. 
It can be expanded as
\beq
X = HH^\dag = X^0t^0 + X^at^a + X^\alpha t^\alpha \ , \qquad
X^0 = 2\,\Tr_c\left(HH^\dag t^0\right) \ , \qquad
X^a = 2\,\Tr_c\left(HH^\dag t^a\right) \ , 
\eeq
where the traces with subscript $c$ are over the color indices. 
$e$ and $g$ are the $U(1)$ and $G'$ coupling constants, respectively,
while $\xi$ is a {real} constant. 
$t^0$ and $t^a$ stand for the 
$U(1)$ and $G'$ generators, respectively, 
and finally,  
$t^\alpha \in \mathfrak{g}'_{\perp}$, where
$\mathfrak{g}'_{\perp}$ is the orthogonal complement 
of the Lie algebra $\mathfrak{g}'$ in $\mathfrak{su}(N)$. 
We normalize the generators according to 
\beq 
t^0 = \frac{{\bf 1}_{N}}{\sqrt{2N}} \ , \qquad
\Tr(t^a t^b) = \frac{1}{2} \delta^{ab} \ . 
\eeq

We have chosen in Eq.~(\ref{Lagrang}) a particular, 
critical quartic scalar coupling 
equal to the (square of the) gauge coupling constants, i.e.~the BPS limit. 
Indeed such a Lagrangian can be regarded as 
the truncated bosonic sector of 
an ${\cal N}=2$ supersymmetric gauge theory.\footnote{{
  The full supersymmetric bosonic sector contains an additional set of squarks in the anti-fundamental representation of the gauge group, and an adjoint scalar field. We can consistently forget about them, as they are trivial on the BPS vortices. Although we shall not make explicit use of any of the consequences of
  ${\cal N}=2$ supersymmetry (the missing sector is truly relevant at the quantum level), this way of regarding our system is useful for providing a convenient choice of the potential and its stability against radiative
  corrections. }}
The constant $\xi$ would in this case be the Fayet-Iliopoulos parameter. 
In order to keep the system in the Higgs phase, we take $\xi>0$. 
The model has a gauge symmetry acting from the left on $H$ 
and a flavor symmetry acting from the right.
First we  note that this theory has a continuous Higgs vacuum
which was discussed in detail in Ref.~\cite{Eto:2008qw}. In this paper,
we choose to work in a particular point of the vacuum manifold:
\beq
\langle H \rangle = \frac{v}{\sqrt{N}}{\bf 1}_{N} \ ,
\qquad \xi = \frac{v^2}{\sqrt{2N}}\ , \label{eq:vacuum} 
\eeq
namely, in the maximally ``color-flavor-locked'' Higgs phase of the theory.  We have set  $N_{\rm F}=N$  which  is the minimal number of flavors allowing  such a vacuum \footnote{Notice that this not the minimal choice for the existence of a vacuum which supports BPS vortices. In fact, such a minimal number is $N_{\rm F}=1$ in the $SO$ case and $N_{\rm F}=2$ in the $USp$ case. However, in this case there is a residual Coulomb phase.  The vortices  actually reduce to those appearing in theories with a lower-rank gauge group.}.  The existence of a continuous vacuum degeneracy  implies 
the emergence of vortices of semi-local type as we shall see 
shortly. 

Performing the Bogomol'nyi completion, the energy (tension) reads
\begin{align} T &= \int d^2x \,\Tr_c\left[
\frac{1}{e^2}\left|F_{12}-\frac{e^2}{2}\left(X^0t^0 -
2\xi t^0\right)\right|^2
+\frac{1}{g^2}\left|\hat{F}_{12}-\frac{g^2}{2}\,X^at^a\right|^2
+4\left|\bar{\mathcal{D}}H\right|^2-2\xi F_{12}t^0 \right]
\nonumber\\
&\ge -\xi \int d^2x\, F^0_{12} \ ,
\end{align}
where $\bar{\mathcal{D}} \equiv
\frac{\mathcal{D}_1+i\mathcal{D}_2}{2}$ 
is used along with the standard complex coordinates $z = x^1+ix^2$ 
and all fields are taken to be independent of $x^3$. 
When the inequality is saturated (BPS condition), the tension is simply
\beq T = 2\sqrt{2N}\pi\xi\nu = 2\pi v^2 \nu\ , \qquad
\nu = -\frac{1}{2\pi\sqrt{2N}}\int d^2x\,F_{12}^0
\ ,  \label{eq:tension} \eeq
where  $\nu$ is the $U(1)$ winding number of the vortex.
This leads immediately to the BPS equations for the vortex
\begin{align}
\bar{\D}H &= \bar{\partial}H + i\bar{A}H = 0 \ , \label{BPS1} \\
F_{12}^0 &= e^2\left[\, \Tr_c\left(HH^\dag t^0\right) - \xi \, \right] \ ,
\label{BPS2} \\
F_{12}^a &= g^2\, \Tr_c\left(HH^\dag t^a\right) \ . \label{BPS3}
\end{align}
 The matter BPS equation (\ref{BPS1}) can be solved  
\cite{Isozumi:2004vg,Eto:2005yh,Eto:2006pg} by the Ansatz
\beq 
H = S^{-1}(z,\bar{z})H_0(z) \ , \qquad 
 \bar{A} = -iS^{-1}(z,\bar{z})\bar{\partial}S(z,\bar{z}) \ , 
\label{eq:H_A}
\eeq
where $S$ belongs to the complexification of the gauge
group,  $S\in \mathbb{C}^{*}\times {G'}^\mathbb{C}$. 
$H_0(z)$, holomorphic in $z$, is called the \emph{moduli matrix}
\cite{Isozumi:2004jc}, which contains all moduli parameters of the
vortices as will be seen below.

A gauge invariant object can be constructed as $\Omega = SS^\dag$. It
will, however, prove convenient to split this into the $U(1)$ part and
the $G'$ part, such that $S = s\, S'$ and analogously 
$\Omega = \omega\, \Omega'$, $\omega  = |s|^2$, $\Omega' =S'{S'}^\dag$.   
In terms of $\omega$ the tension (\ref{eq:tension}) can be rewritten
as  
\begin{align}
T = 2\pi v^2\nu = 2v^2\int d^2x\  \partial\bar{\partial}\log\omega  \ , \qquad
\nu = \frac{1}{\pi}\int d^2x\, \partial\bar{\partial}\log\omega \ ,
\end{align}
and  $\nu$ determines  the asymptotic behavior of the Abelian field 
as
\beq 
\omega = ss^\dag \sim \left|z\right|^{2\nu}, \hs{10} \mbox{for}\ 
  \left|z\right|\to\infty\ . 
\eeq 

The minimal vortex solutions can be written down \cite{Short} by
making use of the holomorphic invariants for the gauge group $G'$ made
of $H$, which we denote $I_{G'}^i(H)$. If the $U(1)$ charge of the
$i$-th invariant is denoted by $n_i$, $I_{G'}^i(H) $ satisfies     
\beq
I_{G'}^i(H) = I_{G'}^i\left(s^{-1}{S'}^{-1}H_0\right) = 
s^{-n_i}I_{G'}^i(H_0(z)) \ , 
\eeq
while the boundary condition is
\beq
I_{G'}^i(H)\bigg|_{|z|\to\infty}=I_{\rm vev}^i  \, e^{i\nu n_i\theta}
\ ,  
\eeq
where $\nu \, n_i$ is the number of {\it zeros} of $I_{G'}^i$. 
This leads then to the following asymptotic behavior
\beq 
I_{G'}^i(H_0) = s^{n_i}I_{G'}^i(H) 
   \,\,\,{\stackrel {|z|\to\infty} {\longrightarrow}} \,\,\,
 I_{\rm   vev}^i z^{\nu n_i} \,. 
\eeq
It implies that $I_{G'}^i(H_0(z))$,  being holomorphic in  $z$,  are
actually polynomials. Therefore $\nu \, n_i$ must be positive integers
for all $i$:
\beq 
\nu \, n_i \in \mathbb{Z}_{+} \qquad \to \qquad \nu = \frac{k}{n_0}
\ , \qquad k\in\mathbb{Z}_{+} \ , \label{defnu} 
\eeq
with 
\beq 
n_0 \equiv {\rm gcd}\left\{n_i\,|I_{\rm vev}^i\neq 0\right\} \ ,
\eeq
where ``gcd'' stands for the greatest common divisor. The $U(1)$
gauge transformation $e^{2\pi i /n_0}$  leaves $I_{G'}^i(H)$ invariant
and thus the true gauge group is 
\beq 
G = \left[U(1)\times G'\right]/\mathbb{Z}_{n_0} \ , 
\eeq
where $\mathbb{Z}_{n_0}$ is the center of the group $G'$.  
The minimal winding in $U(1)$ found here,  $\frac{1}{n_0}$,
corresponds to the minimal element of  
$\pi_{1}(G) = {\mathbb Z}$, as it represents a minimal loop in the
group manifold $G$.  
As a result we find the following non-trivial constraints for $H_0$ 
\beq 
I_{G'}^i(H_0) = I_{\rm vev}^i \, z^{\frac{k n_i}{n_0}} +
\mathcal{O}\left(z^{\frac{k n_i}{n_0} -1}\right) \ . 
\eeq


Let us now obtain the explicit constraints for the gauge
groups $SO$ and $USp$.  The invariants are
\beq 
(I_{SO,USp})^r_{\phantom{r}s} = (H^{\rm T}J H)^r_{\phantom{r}s} \
, \qquad 1\leq r\leq s \leq N \ , \label{eq:meson}
\eeq
which finally yields what we call the weak constraint for the moduli
matrix,
\beq
H_0^{\rm T}(z)J H_0(z) = z^{\frac{2k}{n_0}} J + 
\mathcal{O}\left(z^{\frac{2k}{n_0}-1}\right)\ .
\label{eq:cd_SO,USp}
\eeq
Here $J$ is the invariant tensor of  $G'$: 
\beq 
J_{\rm even} = \begin{pmatrix}
  0&{\bf 1}_{M}\\\epsilon \,{\bf 1}_{M}&0
\end{pmatrix} \ , \qquad
J_{\rm odd} = \begin{pmatrix}
  0&{\bf 1}_M&0\\{\bf 1}_M&0&0\\0&0&1 \end{pmatrix} \ ,
\label{invarianttensor}
\eeq
where in the first matrix  $\epsilon = +1$ for $SO(2M)$ 
and  $\epsilon = -1$ for $USp(2M)$\footnote{{The symbol $\epsilon$ will appear many times below. It will always take one of the two values, depending on the choice of the gauge group}}, 
while the second matrix is for the $SO(2M+1)$ theory. 
The integer $n_0$ for each group is listed in Table \ref{tab:n_0}.
\begin{table}[t]
\begin{center}
\begin{tabular}{c|ccc}
 & $SO(2M)$ & $USp(2M)$ & $SO(2M+1)$ \\
\hline
$n_0$ & 2 & 2 & 1\\
\end{tabular}
\caption{{\small $n_0$ for $SO(N)$ and $USp(2M)$}}
\label{tab:n_0}
\end{center}
\end{table}
Vortices represented by Eq.~(\ref{eq:cd_SO,USp}) include also
semi-local vortices. 

In terms of $\Omega$ the BPS-equations (for the gauge-fields)
(\ref{BPS2}) and (\ref{BPS3}) can be expressed as 
\begin{align}
\partial\bar{\partial}\log\omega &=
\frac{m_e^2}{4} \left(1- \frac{1}{N\omega} 
\Tr_c\left(\Omega_0{\Omega'}^{-1}\right)\right) \ , \label{master1} \\
\bar{\partial}\left(\Omega'\partial{\Omega'}^{-1}\right) &=
\frac{m_g^2}{8\,\omega} \left(\Omega_0{\Omega'}^{-1} -
J^\dag(\Omega_0{\Omega'}^{-1})^{\rm T}J\right)\ ,  \label{master2}
\end{align}
where $\Omega_0 \equiv \frac{N}{v^2}H_0 H_0^\dag$ and 
$m_e=\frac{ev}{\sqrt{N}}$, $m_g=\frac{gv}{\sqrt{N}}$ are 
masses around the vacuum (\ref{eq:vacuum}).
The equations (\ref{master1}) and (\ref{master2}) are called 
\emph{master equations} for the gauge group $G'=SO(N)$ and $USp(2M)$
with the respective invariant tensor $J$. Both sides of these
equations transform covariantly under the following transformation: 
\begin{align}
S(z,\bar{z}) \rightarrow V_e(z) V'(z) S(z,\bar{z}) \ , \quad 
H_0(z) \rightarrow V_e(z) V'(z) H_0(z) \ ,  \quad 
V_e(z) \in {\mathbb C}^*\ ,\  V'(z) \in {G'}^{\mathbb C} \ .
\label{eq:V-transf}
\end{align}
This transformation does not change the original fields $H$ and $A$
(see equation (\ref{eq:H_A})). Therefore, the solutions to the
equations (\ref{master1}) and (\ref{master2}) are equivalent if they
are related by the transformation (\ref{eq:V-transf}). We denote this
the \emph{V-equivalence relation}. 
The master equations (\ref{master1}) and (\ref{master2}) 
should be solved such that the solution approaches the vacuum
configuration at the boundary $|z|\rightarrow\infty$. Therefore,   
one must enforce the following asymptotic behavior 
on \footnote{
For vortices satisfying the strong condition (\ref{eq:strong_cond}), 
$\Omega_\infty$ reduces to $\Omega_0$ and 
the next to leading terms of $\log \Omega$ 
are ${\cal O}\left(e^{-m_{e,g}|z|}\right)$ as will be explained later.}
$\Omega=\omega \Omega'$, 
\begin{eqnarray}
 \log \Omega = \log \Omega_\infty 
 +{\cal O}\left( \frac{1}{m_{e,g} z},\,\frac{1}{m_{e,g} \bar z}\right).
\end{eqnarray}
Here the leading contribution 
$\Omega_\infty=\omega_\infty \Omega_\infty'$  is given 
as the unique solution to the $D$-term conditions $X^0=X^a=0$ with a
given $H_0(z)$. 
They are obtained by the K\"ahler quotient method and are found for
the gauge groups $G'=SO(N),USp(N)$ in Ref.~\cite{Eto:2008qw} to be: 
\begin{eqnarray}
 \Omega_\infty' =
H_0(z)\frac{{\bf 1}_{N}}{\sqrt{I^\dagger_{G'} I_{G'}}}H_0(z)^\dagger,\quad
\omega_\infty=\frac{1}{v^2}{\rm Tr}\left[\sqrt{I^\dagger_{G'} I_{G'}}\right],
\label{eq:lumpsolution} 
\end{eqnarray}
where the $G'$-invariant $I_{G'}=I_{G'}(H_0)=H_0^{\rm T}(z)J H_0(z)$.  
With this boundary condition,
the master equations are expected to have a
unique (and smooth) solution with a given $H_0(z)$. 
Namely, we expect that vortex configurations are completely
characterized by $H_0(z)$. The validity of this expectation will be
discussed in Sec \ref{sec:dim-moduli}.  

\subsection{GNOW quantization for non-Abelian vortices}\label{sec:special_mm}

Our task is to find all possible moduli matrices which satisfy the
weak condition (\ref{eq:cd_SO,USp}).
In general this is not easy.  But certain special solutions can
be found readily, and each such solution is characterized by a 
{\it weight vector of the dual group}, and 
 are labelled by a set of integers $\nu_a$
$\left(a=1,\cdots,{\rm rank}(G') \right)$
\beq
H_0 (z) = z^{\nu {\bf 1}_N + \nu_a {\mathcal H}_a}
\in U(1)^{\mathbb C} \times {G'}^{\mathbb C}\ ,
\eeq
where $\nu=k/n_0$ is the $U(1)$ winding number and ${\mathcal H}_a$
are the generators of the Cartan subalgebra of $\mathfrak{g}'$. 
These special solutions satisfy the strong condition
(\ref{eq:strong_cond}), given below, with $z_i=0$. 
$H_0$ must be holomorphic in $z$ and {\it single-valued}, which gives
the constraints for a set of integers $\nu_a$
\beq
\left( \nu {\bf 1}_N + \nu_a {\mathcal H}_a \right)_{ll} 
\in {\mathbb Z}_{\ge 0} \quad \forall \,l\ .
\label{eq:quant1}
\eeq
Suppose that we now consider scalar fields in an $r$-representation of
$G'$. The constraint is equivalent to
\beq
\nu + \nu_a \mu_a^{(i)}\in {\mathbb Z}_{\ge 0} \quad \forall \,i\ ,
\label{eq:quant2}
\eeq
where 
$\vec \mu^{(i)}=\mu_a^{(i)}$ $\left(i=1,2,\cdots,{\rm dim}(r)\right)$ 
are the weight vectors for the $r$-representation of $G'$.
Subtracting pairs of adjacent weight vectors, one arrives at the
quantization condition 
\beq
\vec \nu \cdot \vec \alpha \in {\mathbb Z}\ ,  \label{GNOWQ}
\eeq
for every root vector  $\alpha$  of $G'$.

Eq.~(\ref{GNOWQ}) is formally identical to the well-known
Goddard-Nuyts-Olive-Weinberg (GNOW) quantization condition \cite{GNO}
for the monopoles, and to the vortex flux quantization rule found in
Ref.~\cite{KS}. There is however a crucial difference here, as compared to
the case of \cite{GNO} or \cite{KS}. 
Because of an exact flavor (color-flavor diagonal $G_{\rm C+F}$)
symmetry  present here, which is broken by individual vortex
solutions, our vortices possess continuous moduli. {As will be seen later, at least in the local case these moduli are normalizable, and there are no conceptual problems in their quantization. On the contrary,
vortices in Ref. \cite{KS} do not have any continuous modulus, while in the case of ``non-Abelian monopoles''  \cite{GNO}  these interpolating modes suffer from the well-known problems of non-normalizability. Another way the latter difficulty manifests itself is that the na\"{i}ve ``unbroken'' 
group cannot be defined globally due to a topological obstruction \cite{CDyons}} in the  monopole backgrounds.  

The solution of the quantization condition (\ref{GNOWQ}) is that 
\beq
 \tilde{\vec\mu} \equiv \vec\nu/2 \ ,
\eeq
is any of the {\it weight vectors} of the dual group of $G^{\prime}$. 
The dual group, denoted as $\tilde G'$, is defined by the dual root 
vectors \cite{GNO} 
\beq
\vec \alpha^* = \frac{\vec\alpha}{\vec\alpha \cdot \vec\alpha}\ .
\eeq
We show examples of  dual pairs of groups $G'$, $\tilde G'$  in Table
\ref{table:dual_group}. 
\begin{table}[tb]
\begin{center}
\begin{tabular}{c|c}
$G'$ & $\tilde G'$  \\
\hline
\hline 
$SU(N)$  &  $SU(N)/{\mathbb Z}_{N}$ \\
 $U(N)$  &  $ U(N)$ \\
$SO(2M)$ & $SO(2M)$ \\
$USp(2M) $ & $SO(2M+1)$ \\
$SO(2M+1)$ & $USp(2M)$ 
\end{tabular}
\caption{Some pairs of dual groups \label{table:dual_group}}
\end{center}
\end{table}
Note that (\ref{eq:quant2}) is stronger than (\ref{GNOWQ}), it has to be
zero or a positive integer. This positive quantization condition allows for only a few weight vectors. 
For concreteness, let us consider scalar fields in the fundamental
representation, and choose a basis where the Cartan generators of 
$G'=SO(2M),SO(2M+1),USp(2M)$ are given by 
\beq
{\cal H}_a = {\rm diag}
\Big(\underbrace{0,\cdots,0}_{a-1},\frac{1}{2},
\underbrace{0,\cdots,0}_{M-1},-\frac{1}{2},0,\cdots,0\Big)\ ,
\eeq
with $a=1,\cdots,M$. In this basis, special solutions $H_0$ have the
form\footnote{
  The integers $k_{a}^{\pm}$ and $k$ here coincide with  $n_{a}^{\pm}$
  and $n^{(0)}$, respectively, of Ref.~\cite{FGK}.} 
for $G'=SO(2M)$ and $USp(2M)$
\beq
H_0^{\left( \tilde{\mu}_1,\cdots, \tilde{\mu}_M \right)} = {\rm diag}\left(
z^{k_1^+},\cdots,z^{k_M^+},z^{k_1^-},\cdots,z^{k_M^-}
\right)\ ,
\label{eq:special_2M}
\eeq
while for $SO(2M+1)$
\beq 
H_0^{\left( \tilde{\mu}_1,\cdots, \tilde{\mu}_M \right)} = {\rm diag}
\left(
z^{k_1^+},\cdots,z^{k_M^+},z^{k_1^-},\cdots,z^{k_M^-},z^k
\right)\ ,
\label{eq:sp_points_so_odd}
\eeq
where $k_a^{\pm}=\nu \pm \tilde{\mu}_a$. 

For example, in the cases of $G'=SO(4),USp(4)$ with a $\nu=1/2$
vortex, there are four special solutions with 
$\vec{\tilde{\mu}} = (\frac{1}{2},\frac{1}{2}),(\frac{1}{2},-\frac{1}{2}),
(-\frac{1}{2},\frac{1}{2}),(-\frac{1}{2},-\frac{1}{2})$
\beq
H_0^{(\frac{1}{2},\frac{1}{2})} &=& 
{\rm diag}(z,z,1,1) = z^{\frac{1}{2}{\bf 1}_4 + 1\cdot{\cal H}_1 + 1\cdot{\cal H}_2},\\
H_0^{(\frac{1}{2},-\frac{1}{2})} &=& 
{\rm diag}(z,1,1,z) = z^{\frac{1}{2}{\bf 1}_4 + 1\cdot{\cal H}_1 - 1\cdot{\cal H}_2},\\
H_0^{(-\frac{1}{2},\frac{1}{2})} &=& 
{\rm diag}(1,z,z,1) = z^{\frac{1}{2}{\bf 1}_4 - 1\cdot{\cal H}_1 + 1\cdot{\cal H}_2},\\
H_0^{(-\frac{1}{2},-\frac{1}{2})} &=& 
{\rm diag}(1,1,z,z) = z^{\frac{1}{2}{\bf 1}_4 - 1\cdot{\cal H}_1 - 1\cdot{\cal H}_2}.
\eeq
These four vectors are the same as the weight vectors of two Weyl spinor
representations 
${\bf 2} \oplus {\bf 2}'$
of $\tilde G'=SO(4)$ for $G'=SO(4)$,
and the same as those of the Dirac spinor representation
${\bf 4}$ of $\tilde G'=Spin(5)$ for $G'=USp(4)$. 

\begin{figure}[htb]
\begin{center}
\includegraphics[height=2cm]{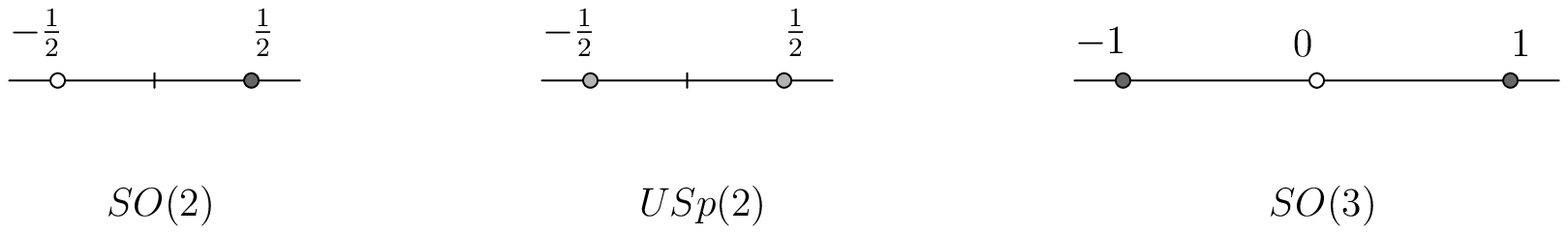}\\\ \\
\includegraphics[height=5.3cm]{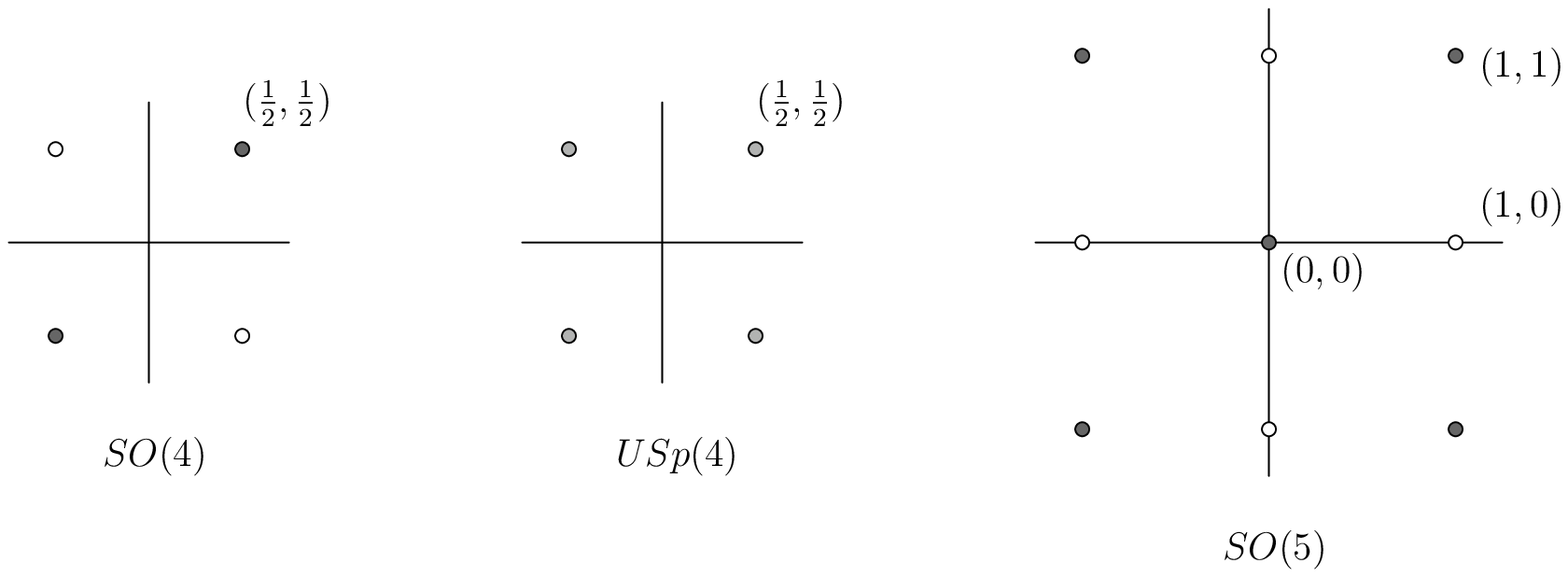}\\\ \\
\includegraphics[height=4cm]{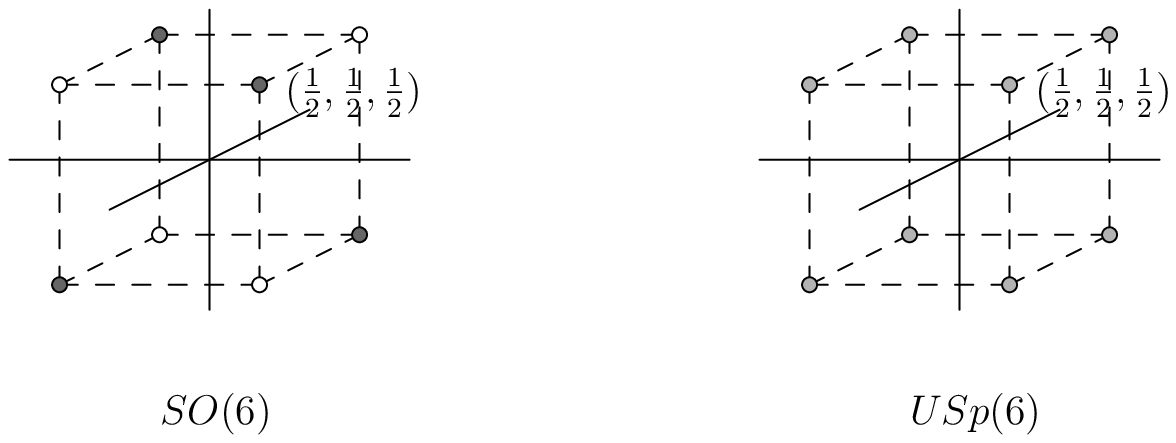}
\caption{The special points for the $k=1$ vortex.}
\label{fig:patches_k1}
\end{center}
\end{figure}

The second example is $G'=SO(5)$ with $\nu=1$. We have nine special
points which are described by
$\vec{\tilde\mu} = (0,0)$ and $(1,0),(0,1),(-1,0),(0,-1)$ and
$(1,1),(1,-1),(-1,1),(-1,-1)$ and thus correspond to
\beq
H_0^{(0,0)} &=& 
{\rm diag}(z,z,z,z,z) = z^{1 \cdot {\bf 1}_5 + 0\cdot{\cal H}_1 + 0\cdot{\cal H}_2},\\
H_0^{(1,0)} &=& 
{\rm diag}(z^2,z,1,z,z) = z^{1 \cdot {\bf 1}_5 + 2\cdot{\cal H}_1 + 0\cdot{\cal H}_2},\\
H_0^{(0,1)} &=& 
{\rm diag}(z,z^2,z,1,z) = z^{1 \cdot {\bf 1}_5 + 0\cdot{\cal H}_1 + 2\cdot{\cal H}_2},\\
H_0^{(-1,0)} &=& 
{\rm diag}(1,z,z^2,z,z) = z^{1 \cdot {\bf 1}_5 - 2\cdot{\cal H}_1 + 0\cdot{\cal H}_2},\\
H_0^{(0,-1)} &=& 
{\rm diag}(z,1,z,z^2,z) = z^{1 \cdot {\bf 1}_5 + 0\cdot{\cal H}_1 - 2\cdot{\cal H}_2},\\
H_0^{(1,1)} &=& 
{\rm diag}(z^2,z^2,1,1,z) = z^{1 \cdot {\bf 1}_5 + 2\cdot{\cal H}_1 + 2\cdot{\cal H}_2},\\
H_0^{(1,-1)} &=& 
{\rm diag}(z^2,1,1,z^2,z) = z^{1 \cdot {\bf 1}_5 + 2\cdot{\cal H}_1 - 2\cdot{\cal H}_2},\\
H_0^{(-1,1)} &=& 
{\rm diag}(1,z^2,z^2,1,z) = z^{1 \cdot {\bf 1}_5 - 2\cdot{\cal H}_1 + 2\cdot{\cal H}_2},\\
H_0^{(-1,-1)} &=& 
{\rm diag}(1,1,z^2,z^2,z) = z^{1 \cdot {\bf 1}_5 - 2\cdot{\cal H}_1 - 2\cdot{\cal H}_2}.
\eeq
These nine vectors are the same as the weight vectors of
the vector representation ${\bf 4}$ and the antisymmetric
representation ${\bf 5}$ of the dual group $\tilde
G'=USp(4)$. 
The weight vectors corresponding to the $k=1$ vortex 
in various gauge groups are {given in Fig.~\ref{fig:patches_k1}}.

\subsection{${\mathbb Z}_2$ parity}\label{sec:Z_2}

As discussed in Ref.~\cite{FGK}, the vortices in $G'=SO(N)$ theory
are characterized by the first homotopy group
\beq
\pi_1\left(\frac{SO(N)\times U(1)}{{\mathbb Z}_{n_{0}}}\right) = 
{\mathbb Z} \times {\mathbb Z}_2\ , \qquad  
n_{0}=1\ \  (N \,\, {\rm odd})\ , \quad 
n_{0}=2\ \  (N \,\, {\rm even})\ , \label{soparity}
\eeq
while those of $G'=USp(2M)$ theory correspond to non-trivial elements of 
\beq
\pi_1\left(\frac{USp(2M) \times U(1)}{{\mathbb Z}_{2}}\right) =
{\mathbb Z} \ . 
\eeq
The vortices in $G'=SO(N)$ carry a ${\mathbb Z}_2$ charge in addition
to the usual additive vortex charges. The ${\mathbb Z}_2$ charge can
be seen from the dual weight vector $\vec{\tilde\mu}$. 
As a simple example, let us consider the case of $SO(4)$. The dual
weight vectors are listed in Table~\ref{table:weight_so4_so5}. 
Let us compare two states: namely
$(\tilde\mu_1,\tilde\mu_2)=(1/2,1/2)$ and  
$(\tilde\mu_1,\tilde\mu_2)=(1/2,-1/2)$.
The difference between them is
$\delta(\tilde\mu_1,\tilde\mu_2)=(0,1)$: thus one of them can be
obtained from the other by a $2\pi$ rotation in the $(24)$-plane in
$SO(4)$.  As a path from unity to a $2\pi$ rotation is a
non-contractible loop, they have different ${\mathbb Z}_2$ charges.

On the other hand, the difference between
$(\tilde\mu_1,\tilde\mu_2)=(1/2,1/2)$ and 
$(\tilde\mu_1,\tilde\mu_2)=(-1/2,-1/2)$ is
$\delta(\tilde\mu_1,\tilde\mu_2)=(1,1)$, hence this is homotopic to
the trivial element of ${\mathbb Z}_2$. Therefore, the vortices can be
classified by the ${\mathbb Z}_2$-parity, $Q_{{\mathbb Z}_2}=\pm 1$. 
In Figs.~\ref{fig:patches_k1} and \ref{fig:patches_k2}, the dark
points correspond to vortices with $Q_{{\mathbb Z}_2}=+1$ while the
empty circles correspond to those with $Q_{{\mathbb Z}_2}=-1$. 
\begin{table}[ht]
\begin{center}
\begin{tabular}{ccc}
\begin{tabular}{rr|r}
$\tilde\mu_1$ & $\tilde\mu_2$ & $Q_{{\mathbb Z}_2}$ \\
\hline
$\frac{1}{2}$ & $\frac{1}{2}$ & $+1$ \\
$\frac{1}{2}$ & $-\frac{1}{2}$ & $-1$ \\
$- \frac{1}{2}$ & $\frac{1}{2}$ & $-1$ \\
$- \frac{1}{2}$ & $- \frac{1}{2}$ & $+1$ \\
\end{tabular}
&&
\begin{tabular}{rr|r}
$\tilde\mu_1$ & $\tilde\mu_2$ & $Q_{{\mathbb Z}_2}$ \\
\hline
$0$ & $0\phantom{\}}$ & $+1$\\
$\pm \{ 1$ & $0\}$ & $-1$\\
$\pm \{ 1$ & $\pm 1\}$ & $+1$
\end{tabular}
\end{tabular}
\caption{$k=1$ $SO(4)$ vortices (left), $k=1$ $SO(5)$ or  $k=2$ $SO(4)$ (right).}
\label{table:weight_so4_so5}
\end{center}
\end{table}

\begin{table}[ht]
\begin{center}
\begin{tabular}{ccc}
\begin{tabular}{rcr|r}
$\tilde\mu_1$ & $\tilde\mu_2$ & $\tilde\mu_3\phantom{\}}$ & $Q_{{\mathbb Z}_2}$ \\
\hline
$\pm \{\frac{1}{2}$ & $\frac{1}{2}$ & $\frac{1}{2}\}$ & $\pm 1$\\
$\pm \{ \frac{1}{2}$ & $\frac{1}{2}$ & $-\frac{1}{2}\}$ & $\mp 1$
\end{tabular}
&&
\begin{tabular}{lrr|r}
$\phantom{\pm\{}\tilde\mu_1$ & $\tilde\mu_2$ & $\tilde\mu_3\phantom{\}}$ & $Q_{{\mathbb Z}_2}$ \\
\hline
$\phantom{\pm\{}$0 & 0 & 0\phantom{\}} & $-1$\\
$\pm \{ 1$ & $0$ & $0 \}$ & $+1$\\
$\pm \{ 1$ & $1$ & $0 \}$ & $-1$\\
$\pm \{ 1$ & $-1$ & $0 \}$ & $-1$\\
$\pm \{ 1$ & $1$ & $1 \}$ & $+1$\\
$\pm \{ -1$ & $1$ & $1 \}$ & $+1$
\end{tabular}
\end{tabular}
\caption{$k=1$ $SO(6)$ cases (left),  $k=1$ $SO(7)$ or $k=2$ $SO(6)$ (right).}
\end{center}
\end{table}
{The $\mathbb Z_{2}$ parity of each special point is defined, in general, as follows}:
\beq
Q_{\mathbb Z_{2}}(k_{i}^{+},k_{i}^{-})=(+)^{\sum_{i} k_i^{+}}\times (-)^{\sum_{i} k_i^{-}}=(-)^{\sum_{i} k_i^{-}},
\eeq
{or equivalently in terms of the weight vectors:}
\beq
Q_{\mathbb Z_{2}}(H_{0}^{(\tilde \mu_{i},\dots,\tilde \mu_{M})})=(-)^{\nu M-\sum_{i} \tilde \mu_i}.
\eeq

\subsection{Local versus semi-local vortices}
\label{sec:local and semilocal}

One is often interested in knowing which of the moduli parameters
describe the so-called local (or  the ANO-) vortices 
\cite{Abrikosov,Nielsen:1973cs}),  which have the profile functions with exponential tails. 
For example, the thoroughly studied $U(N)$ non-Abelian vortices are of
the local type when the model has a unique vacuum: this is indeed the
case when the number of flavors is the minimal one, i.e.~just
sufficient for the color-flavor locked vacuum ($\NF = N$ Higgs fields
in the ${\bf N}$ representation of $SU(N)$).   
For $\NF$ greater than $N$, the vacuum moduli space contains
continuous moduli 
$Gr_{\NF,N} \simeq SU(\NF)/[SU(\NF-N)\times SU(N)\times U(1)]$ 
and, as a consequence, the generic non-Abelian vortex solution is of
the ``semi-local'' type \cite{Achucarro,Hindmarsh:1991jq}, with
power-like tails.\footnote{``Local vortex'' and ``semi-local vortex''
  are clearly misnomers, but as they seem to have stuck among the
  experts in the field, we shall use them in this paper.}   
A characteristic feature of the semi-local vortices is their size
moduli, which are non-normalizable \cite{SYSemi,SemiL}. A lesson from
the $U(N)$ non-Abelian vortices is that the semi-local vortices become
local (ANO-like) vortices, when all the size moduli are set to zero.

Our model with $G'=SO(N)$ or $USp(2M)$, even with our choice $N_{\rm F}=N$,  that is the minimum number of flavors 
that allows for a color-flavor locked vacua,   possesses  always a non-trivial vacuum moduli space. 
In fact, in the class of theories considered here, its dimension is given by the following general formula
\beq
{\rm dim}_{{\mathbb C}}\left[{\cal M}_{\rm vac}\right] = 
N N_{\rm F} - {\rm dim}_{\mathbb C} 
\left[U(1)^{\mathbb C} \times G'{}^{\mathbb C}\right] > 0\ .
\eeq
This strongly suggests that even for $\NF = N$, generic configurations
are of the ``semi-local'' type. 
The K\"ahler metric and its potential on the vacuum moduli space 
have been obtained in Ref.~\cite{Eto:2008qw}.

The distinction between local and semi-local vortices can be made
by using the moduli matrix. In order to see this, the asymptotic
behavior of the configurations must be clarified.  
First note that the vacuum moduli spaces of our models are K\"ahler
manifolds ${\cal M}_{\rm vac}$ and our gauge theories reduce to 
non-linear sigma models whose target space is ${\cal M}_{\rm vac}$,  
when the gauge couplings are sent to infinity. In this limit, vortices generally reduce to the so-called sigma model
lumps \cite{Polyakov:1975yp} (sometimes also called two-dimensional 
Skyrmions or sigma model instantons) characterized by 
\[ \pi_2 ({\cal M}_{\rm vac})\;, \]
i.e.~a wrapping around a 2-cycle inside ${\cal M}_{\rm vac}$. 
By rescaling sizes, taking the strong coupling limit can be
interpreted as picking up the asymptotic behavior, and thus,
even for a finite gauge coupling, asymptotic configurations of
semi-local vortices are well-approximated by lumps
\cite{Hindmarsh:1991jq}.

Consider the lump solutions of the non-linear sigma model on 
${\cal M}_{\rm vac}$.  Let us take holomorphic $G$-invariants
$\{I_G^I\}$ as inhomogeneous coordinates of ${\cal M}_{\rm vac}$ and denote
its K\"ahler potential by $K=K(I_G,I_G^*)$.
A lump solution is then given by a holomorphic map
\begin{eqnarray}
 z\in {\mathbb C} \quad \rightarrow \quad I_G^I =f^I (z) 
\in {\cal M}_{\rm vac}\ . 
\label{rationalmap}
\end{eqnarray}
with single-valued functions $\{f^I(z)\}$.
For finite-energy solutions, the boundary $|z|=\infty $ is mapped to a
single point $I_G^I=v^I \in {\cal M}_{\rm vac}$. So the maps $\{f^I(z)\}$ are
asymptotically of the form
\begin{eqnarray}
 f^I(z)=v^I + \frac{u^I}z+{\cal O}\left(z^{-2}\right),\quad
 u^I \in {\mathbb C} \ .
 \label{rationalmaptail}
\end{eqnarray}
The corresponding energy density $\cal E$ has a power behavior 
\begin{eqnarray}
\mathcal{E} = 2K_{J\bar{J}}(I_G, {\bar{I}_G})\, \partial I_G^J(z) \, 
\bar{\partial} {\bar{I}_G^{\bar{J}}}(\bar z) = 
\frac{2}{|z|^4} \, K_{J\bar{J}}(v,\bar v) \, u^J  \bar{u}^{\bar{J}}
+{\cal O}\left(|z|^{-5}\right)\ ,
\end{eqnarray} 
where we assume that $\{I_G^I\}$ is a local coordinate system in
the vicinity of the point $I_G^I=v^I$ and the manifold is smooth at
that point.
As mentioned above, this asymptotic behavior is valid for that of 
the vortices as well.
Since $\{I_G^I\}\simeq\{I^i_{G'}\}/U(1)^\mathbb C$ 
in the case $G=G'\times U(1)$, the holomorphic maps and the moduli
matrix are related by 
\begin{align}
\{f^I(z)\} \simeq \{I_{G'}(z)\}/\sim \ ,
\end{align}
where ``$\sim$'' is defined as the equivalence relation
\begin{align}
I_{G'}^i(z) \sim P(z) I_{G'}^i(z) \ , \qquad {\rm with}\ P(z)\in
\mathbb{C}[z] \ . 
\end{align}
Hence, the asymptotic tail of the configurations is generically
power-like, i.e.~the generic vortices are of the semi-local type.

Although this is in general the case, it might happen that  all the holomorphic functions  
$\{ I_{G'}^i(H_0(z))\}$ have common zeros and that  the quotient above is
ill-defined.  In such a case, from the point of view of $f^I(z)$, we completely lose
the information about the common zeros accompanied by some vorticity. 
Namely, the signature of the corresponding vortices vanishes from
their polynomial tails and $\pi_2({\cal M}_{\rm
  vac})$ becomes trivial\footnote{{The price of the loss of vorticity
    in the map (\ref{rationalmap}) is the appearance of small lump
    singularities, which manifest themselves as spikes (delta
    functions) in the energy density.}}.  
Specifically, it can happen that all the holomorphic invariants are
proportional to a polynomial $P(z)$:   
\beq
f^I(z)={\rm const.}\quad \Leftarrow \quad  
I_{G'}^i(H_0(z))= P(z)^{\frac{n_i}{n_0}}\quad
{\rm for~ all~} i\ , \label{eq:decoupling}
\eeq
or possibly that there exists only one such 
holomorphic invariant.  
In the case of the $U(N)$, with $\NF=N$ i.e.~the model considered
earlier, ${\cal M}_{\rm vac}$ is just a single point.
Even in the $SO$ and $USp$ cases, we do not consider any non-trivial
element of the second homotopy group of ${\cal M}_{\rm vac}$  
but we fix a point of ${\cal M}_{\rm vac}$ at $|z| \to \infty$. 
Therefore, one must return to the master equations to examine the
asymptotic behavior.
The moduli matrix satisfying Eq.~(\ref{eq:decoupling}) could  be 
transformed to a trivial one such that $\Omega_0={\bf 1}_N$ in 
Eqs.~(\ref{master1}) and (\ref{master2}), by using an {extended 
$V$-transformation  allowing for
negative powers of $z$, with a singular determinant  ${\rm det}\,(V(z))=P(z)^{-1}$}.
After this operation the master equation would take the form of a
Liouville-type equation with point-like sources;\footnote{
  In the well-known Abelian case $G=U(1)$, this transformed master
  equation is nothing but Taubes' equation. 
  This transformation for non-Abelian cases means that 
  all information about orientational moduli are also localized at the
  zeros, in other words, the moduli matrix can be reconstructed from
  the data at the zeros in the case of local vortices
  \cite{Baptista:2008ex}.
  For semi-local vortices, this is clearly not the case. 
} 
hence the asymptotic tail is indeed exponential.
In other words, the conditions (\ref{eq:decoupling}) mean that the
(static) vortex is decoupled from any massless mode in the Higgs
vacuum and hence the dominant contribution to its configuration comes 
from massive modes in the bulk. The corresponding vortices are purely of local type. 
Conversely, {we can clearly identify a local vortex and its position 
by looking at common zeros, although a composite state of a 
semi-local vortex and a local vortex also has a polynomial
tail.}
The above observations can briefly be summarized as follows. The
asymptotic behavior of a vortex is classified by the lightest modes in
the bulk coupled to its configuration.
To summarize,  a vortex is necessarily of
  the local type, when the vacuum moduli space is just a point
  (i.e.~a unique vacuum). Semi-local vortices are present only if  the
  vacuum moduli space is non-trivial (i.e.~having continuous moduli).  

{Once we have clarified the origin of the  of polynomial tails, it is easier to identify the non-normalizable modes and the results in Ref.~\cite{Eto:2008qw} for lumps can be readily applied to vortices. 
 {\it  Semi-local vortices always have 
non-normalizable moduli, which live on the tangent bundle of the moduli space of vacua}\footnote{$v^I$ are nothing but
  vacuum moduli and all of the $u^I$'s are not always independent and
  consist of overall semi-local moduli like an overall size modulus.  {The interpretation as a tangent bundle can be derived from Eq. (\ref{rationalmaptail})}}} 
\begin{eqnarray}
 (v^I,u^I)\in T{\cal M}_{\rm vac}\ .
\end{eqnarray}

In our case, $G'=SO(N), USp(N)$, 
with the common $U(1)$ charge of the scalar fields $H$,  
all the $G^{\mathbb C}$ invariants $I_G^I (H)$ can be written using
the meson $I_{SO,USp}$ in Eq.~(\ref{eq:meson}).  
For instance, since ${\rm Tr}[I_{SO,USp}]\not=0$ in the chosen vacuum,
we can construct 
\begin{align}
 I_G^{(r,s)} (H) \equiv \frac{(I_{SO,USp}(H))^{r}{}_s }{{\rm Tr}[I_{SO,USp}(H)]} 
= \frac{\left(H^{\rm T}JH\right)^{r}{}_s}{{\rm Tr}[H^{\rm T}JH]}\ ,\qquad
1\le r\le s\le N \ . 
\label{eq:G'inv/G'inv}
\end{align}
The condition for (winding $k$) local vortices  is thus: 
\beq 
I_{SO,USp}(H_0) = H_0^{\rm T}(z)J H_0(z) = \left(\prod_{i=1}^{k}(z -
z_i)^{\frac{2}{n_0}}\right)J \ .
\label{eq:strong_cond}
\eeq
This will be called the strong condition, in contrast to the weak
condition (\ref{eq:cd_SO,USp}) which characterizes a more general
class of solutions including semi-local vortices.

In the next section we will discuss moduli spaces defined by requiring
the strong condition.
One can regard this condition being physically required by modifying
our model in such a way that the continuous directions of the vacuum
are indeed being lifted. 
For instance, it is not difficult to add an appropriate superpotential 
$\delta W$ to our model, introducing a chiral multiplet $A$ which is a
traceless $N$-by-$N$ matrix taking value in the $\mathfrak{usp}$ 
($\mathfrak{so}$) algebra in the $SO$ case ($USp$ case), 
viz.~$A^{\rm T}J = J A$, and having a $U(1)$ charge $-2$: 
\begin{eqnarray}
\delta W \propto {\rm Tr}[A H^{\rm T}JH J] \ ,
\end{eqnarray}
however such a term would nevertheless reduce the amount of
supersymmetry.  
As we will see in some cases, the strong condition can give rise to
singularities in the moduli space, which will be inherited into the 
target space of an effective action for the local vortices.

\section{Local vortices and their orientational moduli}
\label{sec:local vortex}

In this section we study local non-Abelian vortices in detail 
leaving the analyses of semi-local vortices for the next section. 
The {\it local} non-Abelian vortices carry non-Abelian charges under
the color-flavor symmetry group.  
The corresponding moduli parameters are referred to as the internal
orientations (or orientational modes) of the vortices. 
{Let us consider a single local vortex}. The
strong condition is 
\beq 
H_0^{\rm T}(z)J H_0(z) = (z-z_{0})^{\frac{2}{n_0}} J\ .
\label{eq:strong_cond_k=1}
\eeq
{The parameter $z_{0}$ represents the vortex center and 
is a part of the vortex moduli.  Fixing  $z_{0}=0$, the solutions to the above condition still possesses the orientational modes.}
In fact, once a moduli matrix satisfying Eq.~(\ref{eq:strong_cond_k=1}) is 
been found, other solutions are readily obtained by acting on it  with the
color-flavor symmetry transformations $G'_{\rm C+F}$:
\beq
H_0'(z) \equiv H_0(z) U \ ,\quad U \in G'_{\rm C+F}\ .
\eeq
However, $H_0(z)$ is defined only {\it modulo} $V$-equivalence,
therefore if there exists a $V$-transformation such that 
\beq
V(z) H_0'(z) = H_0(z)\ ,\quad V(z) \in G'{}^{\mathbb C}\ ,
\eeq
then $H_0'(z)$ and $H_0(z)$ should be regarded as physically the same
configuration. Therefore, in order to identify the orientational
moduli, one needs to identify the  flavor rotations which cannot
be undone by any $V$-transformation. 
In the case of $k=1$ local vortices with $G' = SO(2M), USp(2M)$,
this  discussion  is sufficient to describe the moduli spaces
completely. In the $SO(2M+1)$ case, and for higher-winding vortices, however, this is not the case. It is there that the moduli matrix formalism shows its power.

\subsection{The single ($k=1$) local vortex for $G' = SO(2M), USp(2M)$}


The strong condition (\ref{eq:strong_cond_k=1}) with $n_0=2$ is
satisfied by all special moduli matrices given in
Eq.~(\ref{eq:special_2M}). 
For simplicity, let us start with the moduli matrix described by the
dual weight vector 
$\vec{\tilde \mu} = (\frac{1}{2},\frac{1}{2},\cdots,\frac{1}{2})$, i.e. 
\beq
H_0^{(\frac{1}{2},\frac{1}{2},\cdots,\frac{1}{2})}(z) =
{\rm diag}(~\underbrace{z,\cdots,z}_M,~\underbrace{1,\cdots,1}_M~)\ .
\eeq
The color-flavor rotation $G'_{\rm C+F}$ generates other moduli
matrices in a $G'_{\rm C+F}/U(M)$-orbit. 
It is obvious that the action of the $U(M)$ subgroup of
$G'=SO(2M),USp(2M)$  
\beq
U_0 = 
\left(
\begin{array}{cc}
u^{\rm T} & \\
& u^{-1}
\end{array}
\right)\in G'_{\rm C+F}\ ,\quad
u \in U(M) \ ,
\eeq
can be undone by a $V$-transformation (\ref{eq:V-transf}) due to the
fact that 
$H_0^{(\frac{1}{2},\cdots,\frac{1}{2})}U_0 = U_0H_0^{(\frac{1}{2},\cdots,\frac{1}{2})} 
\simeq H_0^{(\frac{1}{2},\cdots,\frac{1}{2})}$.
Therefore, we find the orientational moduli as parametrizing the following spaces \cite{Short}
\beq
{\cal M}_{\rm ori} = \frac{G'_{\rm C+F}}{U(M)_{\rm C+F}} = 
\frac{SO(2M)}{U(M)} \;\quad {\rm or}\; \quad \frac{USp(2M)}{U(M)} \ ,
\label{eq:orientation_space}
\eeq
both of which are Hermitian symmetric spaces 
\cite{Hergason,Higashijima:2001vk}. 
The real dimension of the moduli spaces is $M(2M\mp1)-M^2 + 2 =
M(M \mp 1) + 2$. Where the two corresponds to the
position of the vortex.

In order to see explicitly $G'_{\rm C+F}/U(M)$, let us take the
following element of $G'$ 
\beq
U = 
\begin{pmatrix}
{\bf 1}_M & -b^\dagger_{A,S} \\ 
&{\bf 1}_M
\end{pmatrix}
\begin{pmatrix}
\sqrt{{\bf 1}_M+b^\dagger_{A,S} b_{A,S}} & \\ 
&\left(\sqrt{{\bf 1}_M+b_{A,S}\,b^\dagger_{A,S}}\right)^{-1}
\end{pmatrix}
\begin{pmatrix}
{\bf 1}_M &  \\ 
b_{A,S} & {\bf 1}_M
\end{pmatrix} \ ,
\eeq
where $b_S$ ($b_{A}$)  is an arbitrary $M$-by-$M$ symmetric
(antisymmetric)\footnote{Similar symbols will be used below to
  indicate a symmetric or antisymmetric constant matrix.} matrix 
for the $SO(2M)$ ($USp(2M)$)  case. 
The first two matrices in $U$ can be
eliminated by $V$-transformations, such that the action of $U$ brings 
the moduli matrix $H_0^{(\frac{1}{2},\cdots,\frac{1}{2})}$ onto the
following form 
\beq
H_0^{(\frac{1}{2},\cdots,\frac{1}{2})}(z) U&\stackrel{V}{\to}& 
H_0^{(\frac{1}{2},\cdots,\frac{1}{2})}(z;b_{A,S}) \equiv 
\begin{pmatrix}
z{\bf 1}_M & \\
b_{A,S}&{\bf 1}_M
\end{pmatrix}
=
\begin{pmatrix}
z {\bf 1}_M & \\
 & {\bf 1}_M
\end{pmatrix}
\begin{pmatrix}
{\bf 1}_M & \\
b_{A,S} & {\bf 1}_M
\end{pmatrix} \ .
\label{eq:local_k=1_111patch}
\eeq
We denote the patch described by the above moduli matrix the 
$(\frac{1}{2},\cdots,\frac{1}{2})$-patch of the manifold 
$G'_{\rm C+F}/U(M)$. 
The complex parameters in the $M\times M$ matrix $b_{A,S}$ are the
(local) inhomogeneous coordinates of ${\cal M}_{\rm ori}$.
Indeed, the moduli matrix has $\frac{M(M\mp1)}{2}+1$ complex
parameters which is in fact the dimension of the moduli space as will be
demonstrated in Sec.~\ref{sec:dim-moduli}. 
This in turn implies that, in the present case, the moduli space for
the local vortex is {entirely generated by a $G'$ orbit}, except for the
position moduli.

By a similar argument we find $2^M$ patches, starting from the
special points $\vec{\tilde\mu} = (\pm \frac{1}{2},\cdots,\pm\frac{1}{2})$
given in Eq.~(\ref{eq:special_2M}).
Indeed, this can easily be done by means of permutations, e.g.
\beq
H_0^{(}\overbrace{{}^{-\frac{1}{2},\cdots,-\frac{1}{2}}}^{r}{}^{,}
\overbrace{{}^{\frac{1}{2},\cdots,\frac{1}{2}}}^{M-r}{}^{)}(z;b_{A,S}) 
= P_r^{-1} H_0^{(\frac{1}{2},\cdots,\frac{1}{2})}(z;b_{A,S}) P_r\ ,
\label{eq:perm_local_k=1}
\eeq
where the permutation matrix is
\beq
P_r \equiv
\begin{pmatrix}
{\bf 0}_r & & \epsilon \, {\bf 1}_r & \\
& {\bf 1}_{M-r} & & {\bf 0}_{M-r}\\
{\bf 1}_r & & {\bf 0}_r & \\
& {\bf 0}_{M-r} & & {\bf 1}_{M-r}
\end{pmatrix}\ , \quad
P_r^{\rm T} J P_r = J \ .
\label{eq:perm}
\eeq
One can easily check that the constraint 
\[\left[P_r^{-1} H_0^{(\frac{1}{2},\cdots,\frac{1}{2})}
  P_r\right]^{\rm T} J 
\left[P_r^{-1} H_0^{(\frac{1}{2},\cdots,\frac{1}{2})} P_r\right] = z J
\ ,
\] 
is indeed satisfied. 
The determinant of the permutation matrices is
\beq
\det P_r = (-\epsilon)^r \ .
\label{eq:sign_perm}
\eeq
Note that $P_r$ is an element of $G'$ iff $\det P_r=1$.

The problem now is to find the transition functions among the $2^M$
patches just found.  
As in the case of $U(N)$ vortices \cite{Eto:2004rz}, the transition
functions between the $(\frac{1}{2},\cdots,\frac{1}{2})$-patch 
and the
$(\underbrace{\hbox{$-\frac12,\cdots,-\frac12$}}_{r},
\underbrace{\hbox{$\frac{1}{2},\cdots,\frac{1}{2}$}}_{M-r})$-patch
are obtained by using the $V$-transformation
(\ref{eq:V-transf}):\vspace{-2em}
\beq
H_0^{(}\overbrace{{}^{-\frac{1}{2},\cdots,-\frac{1}{2}}}^{r}{}^{,}
\overbrace{{}^{\frac{1}{2},\cdots,\frac{1}{2}}}^{M-r}{}^{)}(z;b_{A,S}') 
= V(z)H_0^{(\frac{1}{2},\cdots,\frac{1}{2})}(z;b_{A,S})\ .
\label{eq:transit_local_k=1}
\eeq
By solving the above equation, one obtains the transition functions 
between the two patches having $\det P_r=1$ as
\beq
b_1' = \epsilon\, b_1^{-1}\ ,\quad
b_2' = b_1^{-1}b_2\ ,\quad
b_3' = b_3 + \epsilon \, b_2^{\rm T} b_1^{-1} b_2\ ,
\label{eq:tf_local_k1_b}
\eeq
where $b_{A,S}$ is decomposed to 
an $r$-by-$r$ matrix $b_1$, an-$r$ by-$(M-r)$ matrix $b_2$ and 
an $(M-r)$-by-$(M-r)$ matrix $b_3$ defined as follows
\beq
 b_{A,S} = 
\begin{pmatrix}
b_1 & b_2 \\
- \epsilon \, b_2^{\rm T} & b_3 
\end{pmatrix} \ ,\quad
b_{1,3}^{\rm T} = -\epsilon \, b_{1,3} \ ,
\eeq
and similarly for $b'_i$. 
The technical details will be postponed till the next section.
This derivation of the quotient space $G'/U(M)$ in the moduli 
matrix formalism, can be related to the ordinary derivation with 
$2M$ dimensional vector spaces which we call the orientation
vectors. See App.~\ref{orient1} for the details.

As shown in Eq.~(\ref{eq:sign_perm}), $\det P_r$ is always $+1$ in the
case of $G' = USp(2M)$, while both $+1$ and $-1$ are possible for
$G'=SO(2M)$. 
Hence, all $2^M$ patches can be connected for $G'=USp(2M)$. 
However, two patches which are related by the permutation $P_r$ with 
$\det P_r = -1$ are disconnected since such a permutation is not an
element of $SO(2M)$ but of $O(2M)$ and thus there does not exist any
transition function ($V$-transformation). 
Therefore, we conclude that the patches for $G'=SO(2M)$ are divided
into two disconnected parts according to the sign of 
$\det P_r = \pm 1$. In summary, the moduli space of the $k=1$ vortex
is 
\beq
{\cal M}_{USp(2M)} &=&{\mathbb C}\times{\cal M}^{\rm ori}_{USp(2M)}= {\mathbb C}\times \frac{USp(2M)}{U(M)}\ ,\\
{\cal M}_{SO(2M)} &=& {\mathbb C}\times{\cal M}^{\rm ori}_{SO(2M)}= \left({\mathbb C}\times \frac{SO(2M)}{U(M)}\right)_{+} \cup \left({\mathbb C}\times \frac{SO(2M)}{U(M)}\right)_{-} 
\label{separateM}
\eeq
with ${\mathbb C}$ being the position moduli.  {The doubling of the moduli space in the $SO(N)$ case reflects the 
presence of a $\mathbb Z_{2}$ topological  charge for the vortex (see Eq. (\ref{soparity})), so that ${\cal M}^{\rm ori}_{SO(2M),+}\cap{\cal M}^{\rm ori}_{SO(2M).-}=\emptyset$.}

Furthermore, the structure of these moduli spaces seems to be consistent
with the GNOW duality \cite{GNO}. The dual of $USp(2M)$ is the
$Spin(2M+1)$ group, with a single spinor representation of
multiplicity, $2^{M}$.  
In the case of $SO(2M)$, its GNOW dual is $Spin(2M)$, where the
smallest irreducible representations are the two spinor
representations of chirality $\pm$, each with multiplicity  
$2^{M-1}$. Actually, the quotient $SO(2M)/U(M)$ is just a space for a
pure spinor in $2M$ dimensions \cite{Berkovits:2004bw}. 
Finally, by embedding the vortex theory into an underlying theory with
a larger gauge group which breaks to the group $SO(2M)$ or to 
$USp(2M)$, what is found here for the vortex moduli and their
transformation properties can be translated into the properties of the
monopoles appearing at the ends, through the homotopy matching
argument \cite{ABEK,Duality}. These aspects will be further discussed
in a separate article \cite{GNOWnine}.

We have introduced the dual weight diagram $\vec{\tilde \mu}$ to
represent the special moduli matrices (representative vortex
solutions),  $H_0^{(\tilde\mu_1,\tilde\mu_2,\cdots,\tilde\mu_M)}(z)$
in Sec.~\ref{sec:special_mm}. 
Now we reinterpret them in a slightly different way. The lattice
points of the diagram can be thought of as a representation of the
patches of the space, where the origin of the local coordinates are
just given by these special points.   
For example, in the case of $G'=SO(2M),USp(2M)$, the lattice point
$\vec{\tilde\mu} = (\frac{1}{2},\cdots,\frac{1}{2})$ represents the
patch given in Eq.~(\ref{eq:local_k=1_111patch})\footnote{{This interpretation gives an intrinsic meaning to the special points . Furthermore, their number is related (in many cases equal) to the Euler character of the moduli space.}}. Next we link the
lattice points painted with the same color, namely the patches related
by the permutation $P_r$ with $\det P_r = +1$. 
The structure of the moduli space discussed above can easily be read
off from the dual weight diagram {obtained  this way}.

{The dual lattices formed by special points representatives of  connected patches  are  equal to lattices of irreducible
representations of the dual group. On the contrary, two
disconnected parts of the moduli space (see ${\cal M}_{SO(2M)} $ in
Eq.~(\ref{separateM})) nicely correspond to distinct irreducible 
representations (two spinor representations of opposite chiralities). In the case of composite vortices, we will find irreducible representations obtained by tensor compositions of the fundamental ones. This picture holds for all the explicit cases  we could check (low rank groups), and is an important hint of a ``semi-classical'' emergence of the GNOW duality from the vortex side.} 
  


\subsubsection{Examples:   $G'=SO(2),SO(4),SO(6)$ and $G'=USp(2),USp(4)$}

\begin{figure}[ht]
\begin{center}
\includegraphics[height=3cm]{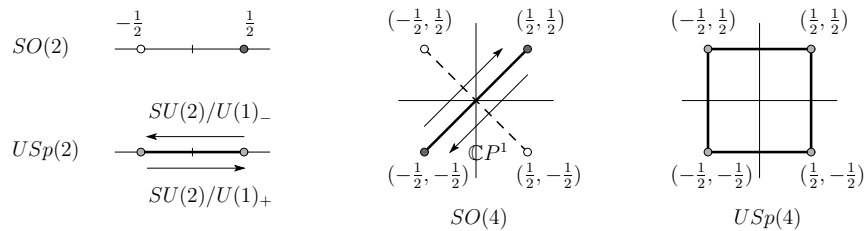}
\caption{{\small The moduli spaces of the $k=1$ local vortex.}}
\label{fig:local_k1_example}
\end{center}
\end{figure}

Let us illustrate the structure of the moduli spaces in some simple
cases, see Fig.~\ref{fig:local_k1_example}.
The $U(1) \times SO(2) \simeq U(1)_+ \times U(1)_-$ 
theory has two types of ANO vortices. One type is characterized by 
$\pi_1(U(1))$ and the other of $\pi_1(U(1)_-)$. They are
described by the following moduli matrices
\beq
H_0^{(\frac{1}{2})} = 
\begin{pmatrix}
z-z_1 & 0 \\
0 & 1
\end{pmatrix} \ ,\quad
H_0^{(-\frac{1}{2})} = 
\begin{pmatrix}
1 & 0 \\
0 & z-z_2
\end{pmatrix} \ .
\eeq

Because $USp(2) \simeq SU(2)$, the $G'=USp(2)$ vortex is indeed
identical to the $U(2)$ vortex which has been well-studied in the
literature. The orientational moduli are 
${\mathbb C}P^1 \simeq \frac{SU(2)}{U(1)}$. 
{Note that the special configurations 
$H_0^{(-\frac{1}{2})} = {\rm diag}(1,z)$ 
and $H_0^{(\frac{1}{2})} = {\rm diag}(z,1)$ are fixed points of the 
$U(1) \subset SU(2)$ group generated by $\sigma_3$: $U(1)={\rm diag}(e^{i\theta},e^{-i\theta})$.
One can} move from $H_0^{(-\frac{1}{2})}$ to $H_0^{(\frac{1}{2})}$ by using
$SU(2)/U(1)$ and vice versa \cite{Eto:2004rz}: 
\beq
\underbrace{
\begin{pmatrix}
1 & 0\\
0 & z
\end{pmatrix}}_{H_0^{(-\frac{1}{2})}}
\underbrace{
\begin{pmatrix}
1 & a\\
0 & 1
\end{pmatrix}}_{SU(2)/U(1)} 
=
\underbrace{
\begin{pmatrix}
0 & 1/a'\\
-a'& z
\end{pmatrix}}_{V\text{-transformation}}
\underbrace{
\begin{pmatrix}
z & 0\\
0 & 1
\end{pmatrix}}_{H_0^{(\frac{1}{2})}}
\underbrace{
\begin{pmatrix}
1 & 0 \\
a' & 1
\end{pmatrix}}_{SU(2)/U(1)}\ , \quad \text{with} \quad a a' = 1 \ .
\eeq
{The corresponding dual weight diagram, shown in the bottom-left of
Fig.~\ref{fig:local_k1_example}, represents  the fundamental
multiplet of the dual $SU(2)$ group. It can be also interpreted as the toric diagram of 
${\mathbb C}P^1$}. 

Next consider $G' = SO(4)$ vortices. We have two different vortices
which are characterized by the $\pi_{1}(SO(4)) = {\mathbb
  Z}_2$-parity. The orientational moduli again turn out to be
\beq
{\mathbb C}P^1 \simeq \frac{SO(4)}{U(2)} \simeq \frac{SU(2)\times SU(2)}{U(1) \times SU(2)} \simeq
\frac{SU(2)}{U(1)}\ .
\eeq
For instance, we find a similar relation between
$H_0^{(-\frac{1}{2},-\frac{1}{2})}$ and
$H_0^{(\frac{1}{2},\frac{1}{2})}$
\begin{align}
\underbrace{
\begin{pmatrix}
{\bf 1}_2 & \\
& z {\bf 1}_2
\end{pmatrix}}_{H_0^{(-\frac{1}{2},-\frac{1}{2})}}
\underbrace{
\begin{pmatrix}
{\bf 1}_2 & b_A\\
 & {\bf 1}_2
\end{pmatrix}}_{SO(4)/U(2)}
=
\underbrace{
\begin{pmatrix}
& b_A'{}^{-1}\\
- b_A' & z {\bf 1}_2
\end{pmatrix}}_{V\text{-transformation}}
\underbrace{
\begin{pmatrix}
z {\bf 1}_2 & \\
& {\bf 1}_2
\end{pmatrix}}_{H_0^{(\frac{1}{2},\frac{1}{2})}}
\underbrace{
\begin{pmatrix}
{\bf 1}_2 & \\
b'_A & {\bf 1}_2
\end{pmatrix}}_{SO(4)/U(2)}\ ,\quad \text{with} \  b_A b_A' = {\bf
  1}_2\ .
\end{align}
The two special points (the two sites of the dual weight diagram) are
again fixed points of the $U(1)$ symmetry, thus the dual weight
diagram can be thought of as the toric diagram for ${\mathbb C}P^1$. 
There are two ${\mathbb C}P^1$'s in this case, see
Fig.~\ref{fig:local_k1_example}.  Furthermore, the diagram can 
alternatively be thought of as representing the reducible 
$({\bf \frac{1}{2}}, {\bf 0})\oplus ({\bf 0}, {\bf \frac{1}{2}})$
representation of the spinor $Spin (4)$, which is the dual of $SO(4)$.

The diagram for the  $G'=USp(4)$ case consists of a single structure
where all the 4 points are connected 
\beq
{\cal M}^{\rm ori}_{USp(4)} =\frac{USp(4)}{U(2)}\ .
\label{eq:ms_usp(4)}
\eeq
This is {consistent with the interpretation of the diagram in Fig.~\ref{fig:local_k1_example} as being the weight lattice of the irreducible spinor representation  ${\bf 4}$ of 
$SO(5)$}, which is indeed the GNOW-dual of $USp(4)$ \cite{GNO}.

\begin{figure}[ht]
\begin{center}
\includegraphics[height=4cm]{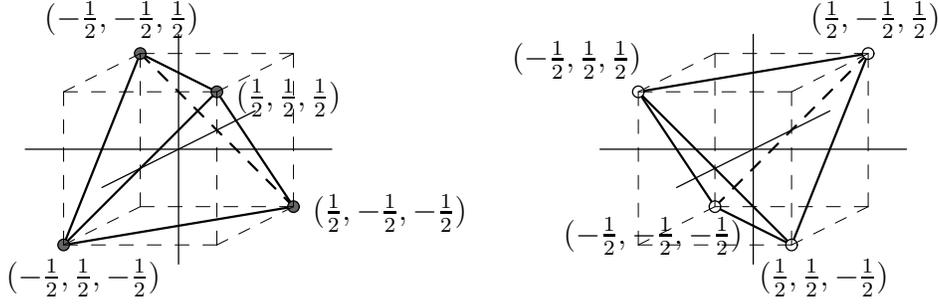}
\caption{{\small The moduli spaces of the $k=1$ local vortex in
    $G'=SO(6)$.}} 
\label{fig:k1_so6}
\end{center}
\end{figure}

The last example is $G'=SO(6)$ (see Fig.~\ref{fig:k1_so6}). This is
another neat example where the orientational moduli are a well-known
manifold and its dual weight diagram can be identified with a toric
diagram. The orientational moduli space is  
\beq
{\cal M}^{\rm ori}_{SO(6)}=\frac{SO(6)}{U(3)} \simeq \frac{SU(4)}{U(1)\times SU(3)} \simeq
     {\mathbb C}P^3 \ .
\eeq
The corresponding dual weight diagram is shown in
Fig.~\ref{fig:k1_so6}. 
There are two ${\mathbb C}P^3$'s similar to the case of $G'= SO(4)$. 
From the toric diagram, one can easily find the ${\mathbb C}P^1$ and
${\mathbb C}P^2$ subspaces which appear as edges and faces,
respectively.
Again these two separate parts of the moduli spaces can be {interpreted as} the two spinor representations, 
${\bf 4} \oplus {\bf 4}^{\bf *},$ of opposite
chiralities of the dual group 
\[ Spin(6) \sim SU(4)\ . \]

\subsection{The doubly-wound ($k=2$) local vortex in $G'=SO(2M)$ and $G'=USp(2M)$
  theories} 
\begin{figure}[htb]
\begin{center}
\includegraphics[height=10cm]{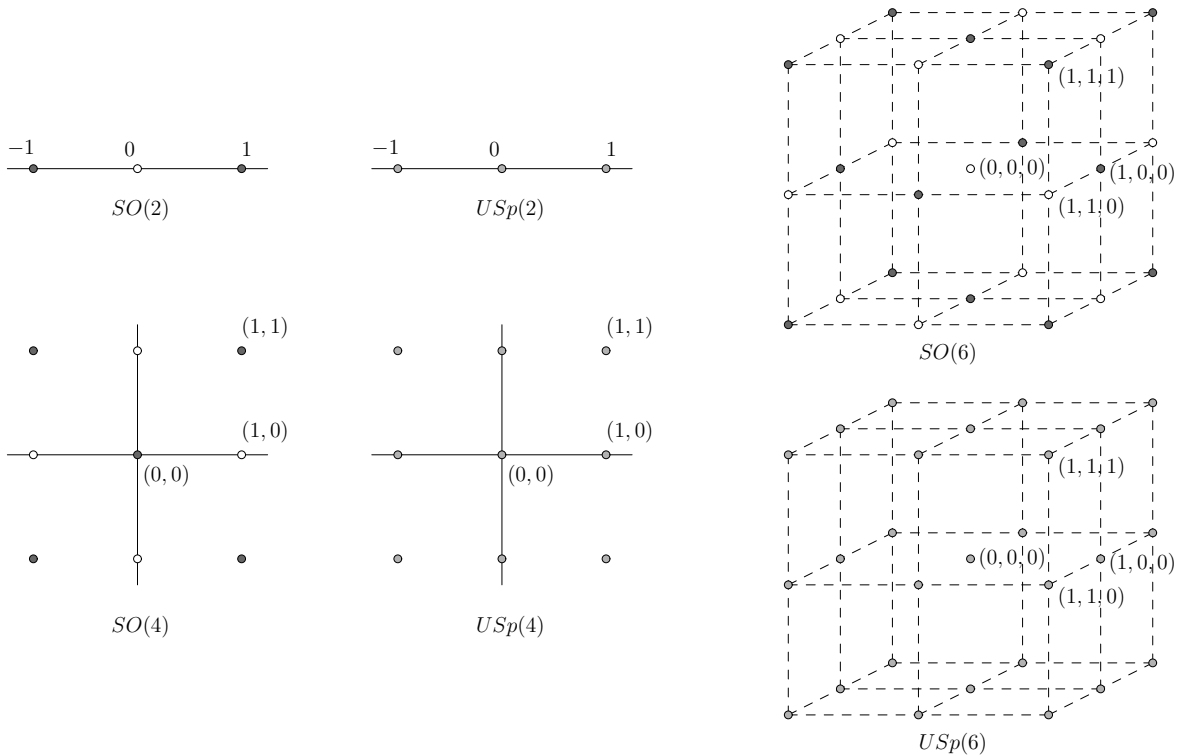}
\caption{The special points for the $k=2$ vortex.}
\label{fig:patches_k2}
\end{center}
\end{figure}

In the case of $G'=SO(2M), USp(2M)$ theories, the strong condition for
the $k=2$ vortices located at $z=z_1$ and $z=z_2$ is of the form 
\beq
H_0(z)^{\rm T} J H_0(z) = P(z) J\ ,\quad 
P(z)\equiv (z-z_1)(z-z_2)\ ,
\label{eq:strong_cond_k=2_so4}
\eeq
which can equivalently be parametrized as
\begin{eqnarray}
 P(z)=(z-z_0)^2-\delta\ ,\quad z_0=\frac{z_1+z_2}2\ ,\quad 
\delta=\left(\frac{z_1-z_2}2\right)^2\ .
\end{eqnarray}
Here $z_1$ and $z_2$ stand for the vortex positions which are where the
scalar field becomes zero, while $z_0$ and $\delta$ are the center of mass and  
the relative position (separation) of two vortices, respectively. 
Several examples of dual weight diagrams are given in
Fig.~\ref{fig:patches_k2}. 

We will now proceed to the doubly-wound ($k=2$) vortices in
$U(1)\times G'$ gauge theories, with $G'=SO(2M)$ or $USp(2M)$.
The $SU(N)_{\rm C+F}$-orbit structure of the moduli space of $k$
vortices in $U(N)$ gauge theory was studied in Ref.~\cite{Eto:2004ii}
using the K\"ahler quotient construction of Hanany and Tong \cite{HT}. 
Here we study the orbit structure of the moduli space of $k=2$
vortices for $G'=SO(2M)$ or $USp(2M)$ more systematically by using the
moduli matrix formalism.
Before going into the detail, let us recall the properties of the
$k=2$ ANO vortices in the usual Abelian-Higgs model.
They can be also studied using the moduli matrix which, in this case is
simply a holomorphic function in $z$, i.e.~a second-order polynomial:  
\beq
H_0^{\rm ANO}(z) = z^2 - \alpha z + \beta = (z-z_1)(z-z_2) \ ,
\eeq
with $\alpha = z_1 + z_2$ and $\beta = z_1z_2$.
Since these two vortices are indeed identical, we cannot distinguish
them. In fact, the moduli matrix is invariant under the exchange of
$z_1$ and $z_2$. Thus the corresponding moduli space is the symmetric
product of ${\mathbb C}$: 
\beq
{\cal M}_{\rm ANO}^{k=2} = \frac{{\mathbb C} \times 
{\mathbb C}}{{\mathfrak S}_2} \simeq {\mathbb C}^2/\mathbb Z_{2}\ .
\eeq

There is a nice property of the moduli matrix for the local vortices.
Suppose $H_0^{i}$ satisfies the strong condition for $k_i$ local
vortices, namely $(H_0^{i})^{\rm T} J H_0^{i} = P_i(z) J$ with a polynomial
function of the $k_i$-th power. 
Then the product of two matrices 
$H_0^{(i,j)} \equiv H_0^{i} H_0^{j}$ automatically satisfies the
strong condition for $k=k_i+k_j$ local vortices: 
$(H_0^{(i,j)})^{\rm T} J H_0^{(i,j)} = P_i(z)P_j(z) J$.
In this way we can construct the moduli matrices for the higher
winding number vortices from those with the lower winding numbers,
which was found in $U(N)$ vortices \cite{Eto:2006cx,Duality}.  
This feature implies that the moduli space for separated local
vortices can be constructed as a symmetric product of copies of those
of a single local vortex:
\beq
{\cal M}_{\rm sep}^{k} \simeq \frac{(\mathbb C \times{\cal M}_{\rm ori})^{k}}{{\mathfrak S}_k} 
\label{eq:sepmoduli}
\eeq
{The consideration above is valid when the component vortices are separated even for small vortex separations. When two or more vortex axes coalesce,  the symmetric product degenerates, and the topological structure of the moduli space undergoes a change. Thus the coincident case must
be treated more carefully.  We shall study the case of two coincident vortices in detail in the next Section.}

Our study of the moduli matrix in the present work is complete up to
$k=2$ vortices ($k=1$ for odd $SO$ groups). The problem of a
complete classification of the moduli matrix for the higher winding
number $(k\ge3)$ is left for future work.

The product of moduli matrices, especially for the $G'=SO(N)$ case,
gives us a natural understanding in the following sense.
The single $G'=SO(N)$ vortex has a ${\mathbb Z}_2$-parity $+1$ or
$-1$. They are physically distinct, hence the $k=2$ configuration
is expected to be classified into three categories by the 
${\mathbb Z}_2$-parity of the component vortex as 
$(Q_{{\mathbb Z}_2}^{(1)},Q_{{\mathbb Z}_2}^{(2)}) = (+1,+1),(+1,-1),(-1,-1)$.
The total ${\mathbb Z}_2$-parity of the configurations with
$(Q_{{\mathbb Z}_2}^{(1)},Q_{{\mathbb Z}_2}^{(2)}) = (+1,+1),(-1,-1)$
is $+1$ while that of 
$(Q_{{\mathbb Z}_2}^{(1)},Q_{{\mathbb Z}_2}^{(2)}) = (+1,-1)$ is $-1$.
Therefore, the former and the latter are disconnected.
An interesting question is whether 
$(Q_{{\mathbb Z}_2}^{(1)},Q_{{\mathbb Z}_2}^{(2)}) = (+1,+1)$  and
$(-1,-1)$ are connected or not.  {The naive answer would be yes,  because the two solutions represent two equivalent objects from the topological point of view. However, the true answer, as we will show, is subtler, and is different for the local and  semi-local cases. For the latter case, the two moduli spaces are smoothly connected and in fact are the same space. More interestingly, in the local case they represent two different spaces which intersect at some submanifold.  As we shall see, this result is compatible with the interpretation that weight lattices formed by connected special points are in correspondence with irreducible representations of the dual group\footnote{{The fact that there is no topology which can explain this disconnection somehow enforces our interpretation in terms of the dual group.}}
\cite{FGK}. } 

The patch structure for the $k=2$ local vortices in generic
$G'=SO(2M),USp(2M)$ theories is rather complex. 
In this subsection, we just present the result without details.  
The result will be discussed again when we shall consider the generic
configurations satisfying the weak condition (\ref{eq:cd_SO,USp}) in
Sec.~\ref{sec:semilocal}.  
The moduli matrix in a generic patch takes the form
\beq
H_0^{(}\overbrace{{}^{1,\cdots,1}}^{r}{}^{,}
\overbrace{{}^{0,\cdots,0}}^{M-r}{}^{)}(z) 
=
\begin{pmatrix}
P(z) {\bf 1}_r & 0 & 0 & 0\\
B_1(z) & (z-z_0){\bf 1}_{M-r} + \Gamma_{11} & 0 & \Gamma_{12}\\
A(z) & C_1 & {\bf 1}_r & C_2\\
B_2(z) & \Gamma_{21} & 0 & (z-z_0){\bf 1}_{M-r} + \Gamma_{22}
\end{pmatrix}\ ,\label{eq:k2Localgeneric}
\label{eq:mm_k2_generic}
\eeq
\beq
A(z) &=& a_{1;A,S}\, z + a_{0;A,S} + \lambda_{S,A}\ ,\\
\begin{pmatrix}
B_1(z)\\
B_2(z)
\end{pmatrix} 
&=& - \left((z-z_0){\bf 1}_{2(M-r)} + \Gamma\right)J_{2(M-r)}
\begin{pmatrix}
C_1^{\rm T}\\
C_2^{\rm T}
\end{pmatrix} \ ,\\
\Gamma &\equiv& 
\begin{pmatrix}
\Gamma_{11} & \Gamma_{12}\\
\Gamma_{21} & \Gamma_{22}
\end{pmatrix} \ ,
\eeq
where $a_{i;A,S}$ $(i=0,1)$ is an $r \times r$ constant (anti-)symmetric matrix,
$C_i$ is an $r \times (M-r)$ constant matrix and we have defined
\beq
\lambda_{S,A} \equiv -\frac{1}{2} (C_1,\ C_2) \, J_{2(M-r)}\, 
\begin{pmatrix}
C_1^{\rm T}\\
C_2^{\rm T}
\end{pmatrix} \ ,\quad
J_{2(M-r)} \equiv
\begin{pmatrix}
& {\bf 1}_{M-r}\\
\epsilon \, {\bf 1}_{M-r} & 
\end{pmatrix} \ .
\eeq
The strong condition is now translated into the following form
\beq
\Gamma^{\rm T} J_{2(M-r)} + J_{2(M-r)} \Gamma = 0\ ,\quad
\Gamma^2 = \delta\, {\bf 1}_{2(M-r)}\ ,\qquad(\Tr \;\Gamma = 0)\ .
\label{eq:re_strong_condition}
\eeq
Solutions to this condition for separated vortices are discussed in
App.~\ref{sec:orbit_even_1}. 
It is a hard task to study the moduli space collecting all the
patches, for generic $SO(2M)$ and $USp(2M)$. 
A complete analysis of the moduli space in several cases will be given
later.

Some of the moduli parameters in Eq.~(\ref{eq:mm_k2_generic}) 
are the Nambu-Goldstone (NG) modes associated with global symmetry
breaking and the rest are interpreted as so-called quasi-NG modes
\cite{quasi-NG}. 
The former is, for instance, the overall orientation of the two
vortices and the center of mass.  The relative separation between two local
vortices ($\mathbb C$) and some of the relative orientational modes are typical examples of the latter.  For two coincident vortices  the situation is subtler, but in general there will still be a set of NG modes generated by the $G'_{\rm C+F}$ symmetry, while the remaining modes are  quasi-NG modes. As we will see in the following, the number of the 
quasi-NG modes is $\left[\frac{M}{2}\right]$ or
$\left[\frac{M}{2}\right]-1$ for $SO(2M)$ and $M$ for $USp(2M)$, 
which was actually difficult to find  without using the moduli
matrix formalism.

\subsubsection{$G'_{\rm C+F}$-orbits for coincident  vortices\label{sec:orbit_even_2}} 

Let us now specialize to the case of the $k=2$ co-axial (axially
symmetric) vortices. 
The details of the analysis can be found in App.~\ref{app:details}. 
Consider a special moduli matrix 
\beq
H_0^{(}\overbrace{{}^{1,\cdots,1}}^{r}{}^{,}
\overbrace{{}^{0,\cdots,0}}^{M-r}{}^{)} 
= {\rm diag}\big(
\underbrace{z^2,\cdots,z^2}_r,
\underbrace{z,\cdots,z}_{M-r},
\underbrace{1,\cdots,1}_r,
\underbrace{z,\cdots,z}_{M-r}\big)\ . \label{eq:special_mm_k2} 
\eeq
Clearly, this vortex breaks the color-flavor symmetry as
\beq
SO(2M) \to U(r) \times SO(2(M-r))\ , \quad
USp(2M) \to U(r) \times USp(2(M-r))\ . \label{eq:color-flavor-breaking}
\eeq
Thus depending on $r~(r=0,1,\cdots,M)$, we have $M+1$ different
$G'_{\rm C+F}$ orbits. Each orbit reflects the NG modes associated
with the symmetry breaking. 
The different orbits are connected by the quasi-NG modes which are
unrelated to symmetry. 
The total space is stratified with $G'_{\rm C+F}$-orbits as leaves. 
To see this, let us consider the following moduli matrix 
(for $G' = SO(2M)$): 
\begin{align}
H_0 &=
\begin{pmatrix}
z^2{\bf 1}_{M-2} & & &\\
& z {\bf 1}_2 & &i\sigma_2 \lambda\\
& & {\bf 1}_{M-2} &\\
& & & z {\bf 1}_2
\end{pmatrix} 
= V^{-1}
\begin{pmatrix}
z^2{\bf 1}_{M-2} & & &\\
& z^2 {\bf 1}_2 & &\\
& & {\bf 1}_{M-2} &\\
& - i\sigma_2 \lambda^{-1}z & & {\bf 1}_2
\end{pmatrix} \ ,
\label{eq:mm_k2_example}\\
V &= 
\begin{pmatrix}
{\bf 1}_{M-2} & & &\\
& z {\bf 1}_2 & & - i\sigma_2 \lambda\\
& & {\bf 1}_{M-2} &\\
& -i\sigma_2 \lambda^{-1} & & {\bf 0}_2
\end{pmatrix} \in SO(2M)\ .
\end{align}
We can always take $\lambda$ to be non-negative and real
$\mathbb{R}_{>0}$ by means of the color-flavor rotation 
\begin{align}
H_0 \to U^{-1} H_0 U \ , \quad
U = 
\begin{pmatrix}
{\bf 1}_{M-2} & & & \\
& a{\bf 1}_2 & & \\
& & {\bf 1}_{M-2} & \\
& & & a^{-1} {\bf 1}_2
\end{pmatrix} \in SO(2M)\ .
\end{align}
In two limits $\lambda \to 0$ and $\lambda \to \infty$, 
the moduli matrix (\ref{eq:mm_k2_example}) reduces to 
the special matrix (\ref{eq:special_mm_k2}) with $r=M-2$ and $r=M$,
respectively. 
The orbit with intermediate values $0 < \lambda <\infty$ corresponds
to the symmetry breaking pattern 
\beq
\frac{SO(2M)}{U(M-2) \times USp(2)}\ .
\eeq
In fact, the moduli matrix (\ref{eq:mm_k2_example}) is left invariant
under the $USp(2) \in SO(2M)_{\rm C+F}$ transformations 
\beq
U = 
\begin{pmatrix}
{\bf 1}_{M-2} & & &\\
& g^{-1} & & \\
& & {\bf 1}_{M-2} \\
& & & g^{\rm T}
\end{pmatrix}
\in SO(2M)\  ,\quad g^{\rm T} (i\sigma_2) g = i\sigma_2 \ .
\eeq
Therefore, the quasi-NG mode $\lambda$ connects two different 
$SO(2M)_{\rm C+F}$ orbits:
\beq
\frac{SO(2M)}{U(M)}\times \mathbb Z_2
\ \overset{\lambda\to0}{\longleftarrow}\ 
\mathbb{R}_{>0} \times \frac{SO(2M)}{U(M-2) \times USp(2)} \times
\mathbb Z_2
\ \overset{\lambda \to \infty}{\longrightarrow}\ 
\frac{SO(2M)}{U(M-2) \times SO(4)}\ ,
\label{Z2introduc}
\eeq
where the $\mathbb Z_2$ factor indicates a permutation, $P^{-1}H_0P$
with $P\in O(2M)/SO(2M)$.  
This permutations does not belong to the $SO(2M)_{\rm C+F}$ symmetry,
nonetheless it generates a new moduli matrix solution. We thus see, as
explained before, how the moduli space of coincident vortices of
positive chirality is generically made of two disconnected parts. If
$M-r\not = 0$, such a permutation acts trivially  
or can be pulled back by an $SO(2M)$ rotation on $H_0$. At these
special points the two copies coalesce. Nonetheless we must interpret
the two spaces as defining two different composite states of
vortices: $(+1,+1)$ and a $(-1,-1)$. This interpretation is fully
consistent if one studies interactions in the range of validity of the
moduli  space approximation \cite{Manton:1981mp}. It is easy to
realize that, in this approximation, the chirality of each of the
component vortices is conserved: two composite state of vortices
$(+1,+1)$ and $(-1,-1)$ do not interact, even if their trajectory in
the moduli space passes through an intersection
submanifold\footnote{The question if (or how) these vortices interact
  beyond the moduli space approximation, and in particular at the
  quantum level, is an interesting open question. See also a comment
  related on this issue at the end of the Conclusion.}

At the intersection, the dimension of the manifold always reduces by
\[\left[{\rm dim}\;\mathbb R_{>0}-{\rm dim}\;USp(2)\right]-(-{\rm
  dim}\;SO(4))=4 \ . \]

This can easily be extended to the following moduli matrix, with 
$t,\alpha \in \mathbb{Z}_{\ge 0}$
\begin{align}
H_0=\left(
\begin{array}{ccc|ccccc}
z^2{\bf 1}_{t} & & & & &   \\
& z^2{\bf 1}_{2\alpha} & & & &   \\
&  & z{\bf 1}_{M-t-2\alpha} & & &  \\
\hline
0& & &  {\bf 1}_{t} & &  \\
& z \tilde \Lambda & & & {\bf 1}_{2\alpha} & \\
& &  0 & & & z{\bf 1}_{M-t-2\alpha} 
\end{array}
\right),\quad 
\tilde \Lambda=
\begin{pmatrix}
\tilde \lambda_{1} \tilde J_{2\tilde p_1}  & & \\
&\ddots &\\
&&\tilde \lambda_{s} \tilde J_{2\tilde p_s}
\end{pmatrix}\ ,
\label{eq:so2m_k2_hie}
\end{align}
where $\tilde J_{2\tilde p_i}$ is the invariant tensor of 
$USp(2\tilde p_i)$ and
\beq
\alpha = \sum_{i=1}^s \tilde p_i\ ,\quad  
t+2\alpha\le M \ , \quad
0 < \tilde \lambda_i < \tilde \lambda_{i+1} \ .
\eeq
An arbitrary patch (\ref{eq:k2Localgeneric}) with $\delta=0$ in the
$SO(2M)$ case, can be brought onto the above form as explained in
App.~\ref{app:details}. 
The set of numbers $(t,s,\tilde p_i)$ and the quasi-NG modes
$\lambda_i$ are, of course, independent of $r$ which indicates the
patch which we take as a starting point.

Note that this is invariant with respect to the group 
$\prod_{i=1}^s USp(2\tilde p_i) \in SO(2M)_{\rm C+F}$ 
\beq
U = \text{block-diag} \left({\bf 1}_{t}, g_{2\tilde
p_1}^{-1},\cdots,g_{2\tilde p_s}^{-1},{\bf 1}_{M-t-2\alpha},
{\bf 1}_{t}, g_{2\tilde p_1}^{\rm T},\cdots,g_{2\tilde p_s}^{\rm T},{\bf 1}_{M-t-2\alpha}
\right)\ ,
\eeq
with 
$g_{2\tilde p_i}^{\rm T} \tilde J_{2\tilde p_i} g_{2\tilde p_i} =  \tilde J_{2\tilde p_i}$. 
Therefore, the local structure of the $SO(2M)$-orbit has the form 
\beq
\mathbb{R}_{>0}^s 
\times \frac{O(2M)}{U(t)\times \prod_{i=1}^s USp(2\tilde p_i) 
\times O(2 u)}\ ,\quad
{\rm with\quad} t+u+2\sum_{i=1}^s \tilde p_i=M\ .
\label{eq:SOlocalstructure} 
\eeq
When we take the limit $\tilde \lambda_1 \to 0$, 
a subgroup $U(t) \times USp(2\tilde p_1)$ of the isotropy group gets
enhanced to $U(t+2\tilde p_1)$ and the orbit shrinks, thus the local 
structure around the new orbit is given by changing the indices in
Eq.~(\ref{eq:SOlocalstructure}) as follows
\begin{eqnarray}
 (s,t,u;\tilde p_1,\tilde p_2,\cdots, \tilde p_s)\quad 
\stackrel{\tilde \lambda_1\to 0}{\to}\quad 
(s',t',u';\tilde p_i')= 
(s-1,t+2\tilde p_1,u;\tilde p_2,\cdots, \tilde p_s) \ .
\end{eqnarray}
In the opposite limit where $\tilde \lambda_s \to \infty$,
another subgroup $USp(2\tilde p_s) \times SO(2u)$ of the isotropy 
group is getting enlarged to $SO(2u +4\tilde p_s)$, hence  
the local structure around this new orbit is obtained by 
\begin{eqnarray}
 (s,t,u;\tilde p_1,\cdots,\tilde p_{s-1}, \tilde p_s)\quad 
\stackrel{\tilde \lambda_s\to \infty}{\to}\quad 
(s'',t'',u'';\tilde p_i'')= 
(s-1,t,u+2 \tilde p_s;\tilde p_1,\cdots, \tilde p_{s-1})\ .
\end{eqnarray}
By choosing various $t,\tilde p_i$ and taking the limits 
$\tilde\lambda_i \to 0,\infty$, 
we can reach all the points of the moduli space. However, since these
transitions are always induced by the 
$2\tilde p_i \times 2\tilde p_i $ matrix $\tilde J_{2\tilde p_i}$,
the patches with only an even number of $z^2$'s in the diagonal element
are connected.  Analogously, the patches with an odd number of $z^2$'s
are mutually connected. 
Nevertheless, the former and latter remain disconnected and this of
course is just a consequence {of the different chiralities ($\mathbb{Z}_2$ topological factor). }

\begin{figure}[ht]
\begin{center}
\includegraphics[width=15cm]{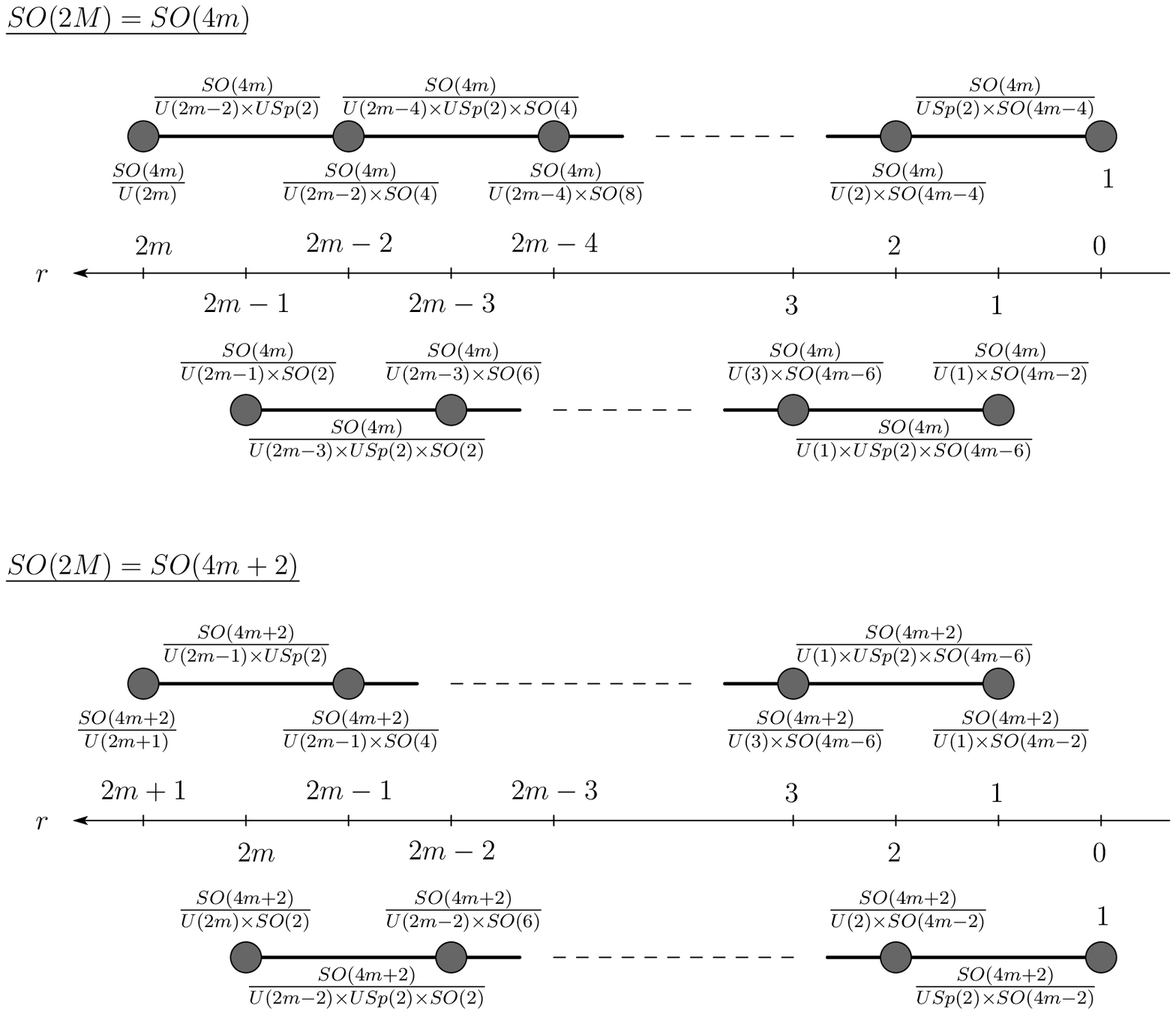}
\caption{{\small Sequences of the $k=2$ vortices in $SO(4m)$ and
    $SO(4m+2)$. The sites (circles) correspond to the special orbits
    of Eq.~(\ref{eq:mm_k2_example}) and the links connecting them
    denote the insertion of the minimal pieces $\tilde\lambda_i
    \tilde J_{2}$ such as in Eq.~(\ref{eq:so2m_k2_hie}).}}
\label{fig:sequence}
\end{center}
\end{figure}

For instance, by inserting a minimal extension, i.e.~the following
piece, $\tilde\lambda \tilde J_2$, the special orbits in
Eq.~(\ref{eq:special_mm_k2}) can sequentially be shifted as 
\begin{align}
{\rm diag}(z^2,\cdots,z^2,z^2,z^2,1,\cdots,1,1,1) &\to 
{\rm diag}(z^2,\cdots,z^2,z,z,1,\cdots,1,z,z) \to \cdots \non&\to 
{\rm diag}(z,\cdots,z,z\cdots,z) \ .  \nonumber
\end{align}
However, the connection pattern depends on whether 
$SO(2M)=SO(4m)$ or $SO(4m+2)$, see Fig.~\ref{fig:sequence}. 
At a generic point $(\tilde p_i = 1,\ s=m)$ 
where the color-flavor symmetry is maximally broken
the corresponding moduli spaces can locally be written as 
\beq
{\cal M}^{k=2,\,\text{ori}}_{SO(4m),+} &=& 
\mathbb{R}_{>0}^m \times \frac{SO(4m)}{USp(2)^m}
\times \mathbb Z_2\ ,\\
{\cal M}^{k=2,\,\text{ori}}_{SO(4m),-} &=& 
\mathbb{R}_{>0}^{m-1} \times \frac{SO(4m)}{U(1) \times USp(2)^{m-1}
  \times SO(2)}\ ,\\ 
{\cal M}^{k=2,\,\text{ori}}_{SO(4m+2),+} &=& 
\mathbb{R}_{>0}^m \times \frac{SO(4m+2)}{U(1)\times USp(2)^m}
\times \mathbb Z_2\ ,\\
{\cal M}^{k=2,\,\text{ori}}_{SO(4m+2),-} &=& 
\mathbb{R}_{>0}^{m} \times \frac{SO(4m+2)}{USp(2)^m \times SO(2)} \ .
\eeq
{The two copies of the moduli space, in the case of positive chirality, intersect at some submanifold if $M\neq1$}.
The dimensions of these moduli spaces are summarized as
\beq
{\rm dim}_{\mathbb C}
\left[ {\cal M}^{k=2,\,\text{ori}}_{SO(2M),\pm}\right] &=& 
M^2-M \ .
\eeq
Taking the vortex position into account, the complex dimension of the
full moduli space is $M^2 - M + 2$ which is nothing but twice the
dimension of the $k=1$ moduli space.

In the case of vortices in $USp(2M)$ theory, we can bring a generic 
moduli matrix onto the following form
\beq
H_0=\left(
\begin{array}{ccc|ccccc}
z^2{\bf 1}_{t} & & & & &   \\
& z^2{\bf 1}_{\beta} & & & &   \\
&  & z{\bf 1}_{M-t-\beta} & & &  \\
\hline
0& & &  {\bf 1}_{t} & &  \\
& z \tilde \Lambda & & & {\bf 1}_{\beta} & \\
& &  0 & & & z{\bf 1}_{M-t-\beta} 
\end{array}
\right),\quad 
\tilde \Lambda=\left(
\begin{array}{ccc}
\tilde \lambda_{1} {\bf 1}_{\tilde p_1}  & & \\
&\ddots &\\
&&\tilde \lambda_{s} {\bf 1}_{\tilde p_s}
\end{array}\right)\,
\label{eq:usp2m_k2_hie}
\eeq
with
\beq
\beta = \sum_{i=1}^s \tilde p_i,\quad  t+\beta\le M,\quad
0 < \tilde \lambda_i < \tilde \lambda_{i+1}.
\eeq
This matrix is invariant under 
$\left[ \prod_{i=1}^s O(\tilde p_i) \right] \in USp(2M)$
\beq
U = \text{block-diag}\left(
{\bf 1}_{t},g_{\tilde p_1}^{-1},\cdots,g_{\tilde p_s}^{-1},{\bf 1}_{M-t-\beta},
{\bf 1}_{t},g_{\tilde p_1}^{\rm T},\cdots,g_{\tilde p_s}^{\rm T},{\bf 1}_{M-t-\beta}
\right) \ ,
\eeq
with $g_{\tilde p_i}^{\rm T} g_{\tilde p_i} = {\bf 1}_{\tilde p_i}$. 
Therefore, the local structure around the $USp(2M)$ orbit is
given by
\beq
\mathbb{R}_{>0}^s \times
\frac{USp(2M)}{U(t) \times \left[ \prod_{i=1}^s O(\tilde p_i) \right]
\times USp(2u)}\ ,\quad 
{\rm with}\quad t+u+\sum_{i=1}^s\tilde p_i=M\ . 
\label{eq:k2SOevenorbit}
\eeq
In the limit $\tilde \lambda_1 \to 0$, the local structures of 
the orbit changes according to
\begin{eqnarray}
 (s,t,u;\tilde p_1,\tilde p_2,\cdots, \tilde p_s)\quad 
\stackrel{\tilde \lambda_1\to 0}{\to}\quad 
(s',t',u';\tilde p_i')= 
(s-1,t+\tilde p_1,u;\tilde p_2,\cdots, \tilde p_s)\ .
\end{eqnarray}
On the other hand, in the opposite limit $\tilde \lambda_s \to \infty$, 
the local structure of the orbit becomes 
\begin{eqnarray}
 (s,t,u;\tilde p_1,\cdots,\tilde p_{s-1}, \tilde p_s)\quad 
\stackrel{\tilde \lambda_s\to \infty}{\to}\quad 
(s'',t'',u'';\tilde p_i'')= 
(s-1,t,u+ \tilde p_s;\tilde p_1,\cdots, \tilde p_{s-1})\ .
\end{eqnarray}

Since the minimal insertion is a real positive number $\tilde
\lambda$, all the special orbits are connected, contrary to the case
of the $SO(2M)$ vortices. This is consistent with the fact that 
there is no $\mathbb{Z}_2$-parity in the $USp(2M)$ case.

At the most generic point where 
$0 < \tilde \lambda_1 < \cdots < \tilde \lambda_M$,
the color-flavor symmetry is broken down to the discrete subgroup 
$\mathbb{Z}_2^{M}$, 
\beq
\mathbb{R}_{>0}^M \times \frac{USp(2M)}{\mathbb{Z}_2^{M}} \ .
\eeq
We can read off the dimensions of moduli space for the $k=2$ co-axial 
local $USp(2M)$ vortices from this 
\beq
{\rm dim}_{\mathbb C} \left[{\cal M}^{k=2,\text{ori}}_{USp(2M)}\right]
= \frac{M}{2} + \frac{2M(2M+1)}{4} = M^2 + M \ .
\eeq

\subsubsection{Examples: $G'=SO(2),SO(4)$ and $G'=USp(2),USp(4)$}

\subsubsection*{$k=2$ local vortices for $G'=SO(2),USp(2)$
  \label{sec:k2SO2USp2}} 

\begin{figure}[ht]
\begin{center}
\includegraphics[height=2cm]{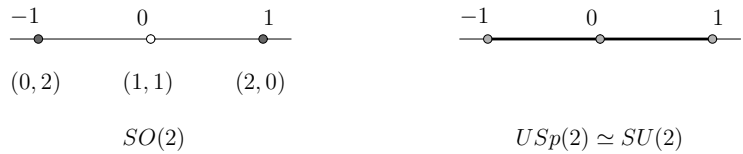}
\caption{{\small The $k=2$ local vortices for $G'=SO(2),USp(2)$.}}
\label{fig:so2_k2}
\end{center}
\end{figure}

Let us first consider the $G'=SO(2)$ theory. 
Although there is no ${\mathbb Z}_2$-parity due to the
fact that $\pi_1(SO(2)) = {\mathbb Z}$, there are nevertheless two
distinct classes of vortices characterized by 
$\pi_1(U(1)_+)$ and $\pi_1(U(1)_-)$ with 
$U(1)\times SO(2) \simeq U(1)_+ \times U(1)_-$.
Thus there are three possible $k=2$ configurations.
$(\pi_1(U(1)_+),\pi_1(U(1)_-)) = \{(2,0),(0,2),(1,1)\}$, see
Fig.~\ref{fig:so2_k2}. 
The corresponding moduli matrices are given by
\beq
H_0^{(+1)} = 
\begin{pmatrix}
P(z) & 0 \\
0 & 1
\end{pmatrix} \ ,\quad 
H_0^{(-1)} = 
\begin{pmatrix}
1 & 0 \\
0 & P(z)
\end{pmatrix} \ ,\quad
H_0^{(0)} = 
\begin{pmatrix}
z-z_1 & 0 \\
0 & z-z_2
\end{pmatrix} \ .
\eeq
Clearly, $z_1$ and $z_2$ are not distinguishable in the first two
matrices while they are in the third matrix. This reflects the fact
that the configuration consists of two identical vortices and two
different vortices, in the two respective cases. 
Therefore, the moduli space is made of three disconnected pieces 
\beq
{\cal M}^{k=2}_{SO(2)} = 
{\cal M}^{(2,0)}_{SO(2)} \cup 
{\cal M}^{(0,2)}_{SO(2)} \cup {\cal M}^{(1,1)}_{SO(2)}\ ,
\eeq
where these spaces are defined by
\beq
{\cal M}^{(2,0)}_{SO(2)} 
&=& \left({\cal M}^{(1,0)}_{SO(2)} \times {\cal M}^{(1,0)}_{SO(2)}\right)
    /{\mathfrak S}_2
= ({\mathbb C} \times {\mathbb C}) 
    / {\mathfrak S}_2 = {\mathbb C}^2/\mathbb Z_{2} \ ,\\
{\cal M}^{(0,2)}_{SO(2)} 
 &=& \left({\cal M}^{(0,1)}_{SO(2)} \times {\cal M}^{(0,1)}_{SO(2)}\right)
    /{\mathfrak S}_2
= ({\mathbb C} \times {\mathbb C}) / {\mathfrak S}_2 = 
 {\mathbb C}^2/\mathbb Z_{2} \ ,\\
{\cal M}^{(1,1)}_{SO(2)} &=& {\cal M}^{(1,0)}_{SO(2)} \times {\cal M}^{(0,1)}_{SO(2)} 
= {\mathbb C}^2 \ .
\eeq
{The $\mathbb Z_{2}$ factor gives rise to crucial differences in the interactions between these vortices}. 
For instance, a head-on collision of two identical local vortices 
in ${\cal M}^{(2,0)}_{SO(2)}$ or ${\cal M}^{(0,2)}_{SO(2)}$ 
leads to a 90 degree scattering, while such a collision of the two
different local vortices living in ${\cal M}^{(1,1)}_{SO(2)}$  would
be transparent, which yields opposite results for the 
reconnection of two colliding vortex-strings \cite{Hashimoto}. {Again, this result is a consequence of the fact that   
vortices with different chiralities must be considered as different, and non-interacting objects}.

The next example is $G'=USp(2)$. As was noted earlier the vortices in
the $G'=USp(2)$ theory are the ones thoroughly studied due to
$USp(2)=SU(2)$.  
The moduli spaces including the patches and the transition functions
for the $k=2$ vortices, in terms of the moduli matrix, are given in 
Ref.~\cite{Eto:2006cx,Duality}.  We shall not repeat the discussion
here. The result is \cite{Eto:2006cx,Duality}
\begin{align}
&{\cal M}_{SU(2)}^{k=2,{\rm separated}} \simeq ({\mathbb C} \times {\mathbb C}P^1)^2/{\mathfrak S}_2,\nonumber \\
&{\cal M}_{SU(2)}^{k=2,{\rm coincident}} \simeq 
\mathbb C\times W{\mathbb C}P^2_{(2,1,1)} \simeq 
\mathbb C\times{\mathbb C}P^2/{\mathbb Z}_2\ . \label{eq:SU(2)moduli}
\end{align}
The dual weight diagram for this case is shown in
Fig.~\ref{fig:so2_k2}.

\subsubsection*{$k=2$ local vortices for $G'=SO(4)$}
\label{sec:k2_local_so4}

Let us now consider $G'=SO(4)$.
\begin{figure}[ht]
\begin{center}
\includegraphics[height=4cm]{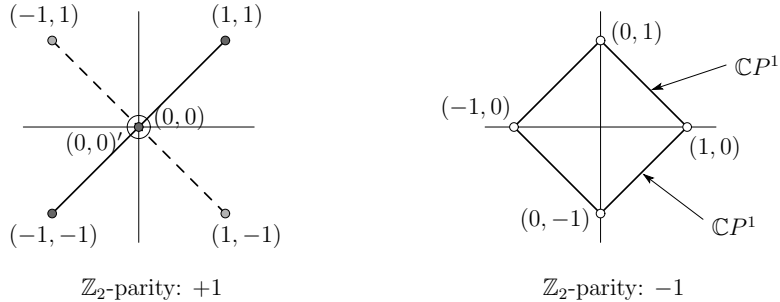}
\caption{{\small The patches of the $k=2$ local vortices in $G'=SO(4)$.}}
\label{fig:so4_k2}
\end{center}
\end{figure}
As can be seen from Fig.~\ref{fig:patches_k2}, there are 9 special
points in the entire moduli space. Five out of them have 
$Q_{{\mathbb Z}_2}=+1$, and the other four have 
$Q_{{\mathbb Z}_2}=-1$.

Note that the isomorphism  
$SO(4)\simeq [SU(2)_+\times SU(2)_-]/\mathbb Z_2$ can indeed be
complexified as  
\begin{eqnarray}
 SO(4)^{\mathbb C}\simeq [SL(2,\mathbb C)_+\times SL(2,\mathbb C)_-]
/\mathbb Z_2\ , \nn {}
[U(1)\times SO(4)]^{\mathbb C}/\mathbb{Z}_2
\simeq [GL(2,\mathbb C)_+\times GL(2,\mathbb C)_-]/\mathbb C^*\ .
\end{eqnarray}
In fact, an arbitrary matrix $X$ satisfying 
$X^{\rm T}J X\propto J$ can always be rewritten as
\begin{eqnarray}
 X&=&\sigma^{-1}(A \otimes B)\sigma=f_+(A) f_-(B)=f_-(B)f_+(A),\nn
&& f_+(A)=\sigma^{-1}(A\otimes {\bf 1}_2)\sigma,\quad 
f_-(B)=\sigma^{-1}({\bf 1}_2 \otimes B)\sigma, \quad
\sigma=\left(
\begin{array}{ccc}
 {\bf 1}_2& & \\
&&1\\ &-1&
\end{array}\right) 
\end{eqnarray}
where $A,B\in GL(2,\mathbb C)$ and 
$f_\pm$ define maps from $GL(2,\mathbb C)_\pm$ to 
$[U(1)\times SO(4)]^{\mathbb C}/\mathbb{Z}_2$.  
The elements of $GL(2,\mathbb C)_\pm$, $f_\pm(A)$, are related by
the odd parity permutation
\begin{eqnarray}
 P^{-1}f_{\pm}(A)P= f_{\mp}(A)\ ,\quad  
P = 
\begin{pmatrix}
1 & & & \\
& & & 1\\
& & 1 & \\
& 1 & & 
\end{pmatrix} \ ,\quad (\det P=-1)\ .\label{eq:so4perm}
\end{eqnarray}
Fixed points of this permutation are given by $A\propto {\bf 1}_2$. 
This complexified isomorphism tells us that a moduli matrix for
$G'=SO(4)$ obeying the strong condition can always be decomposed to a
couple of the moduli matrices for  
$G'=SU(2)$ which have been well-studied.
This fact simplifies the analysis of the moduli space in the present
case. For instance, $f_\pm$ are maps from the moduli matrix for 
$k=1, G'=SU(2)$ to those of $k=1,SO(4)$ with the parity 
$Q_{{\mathbb Z}_2}=\pm 1$, since 
$f_+({\rm diag}(z,1))={\rm diag}(z,z,1,1)$.

Consider first  the  $Q_{{\mathbb Z}_2}=+1$ patches.
There are corresponding patches
of the four special points 
$\vec{\tilde\mu}=(\pm 1,\pm 1),(\pm 1,\mp 1)$.
The $(1,1)$-patch is explicitly given by the moduli matrix
\beq
H_0^{(1,1)} &=& 
\begin{pmatrix}
z^2 + b_1 z + b_2& & & \\
& z^2 + b_1 z + b_2 & & \\
& -b_3 z - b_4& 1 & \\
b_3 z + b_4& & & 1
\end{pmatrix} \ ,
\eeq
with $(z-z_1)(z-z_2)=z^2 + b_1 z + b_2 $.
The rest of the patches $H_0^{(1,-1)}, H_0^{(-1,1)},H_0^{(-1,-1)}$ can
be obtained by appropriate permutations of $H_0^{(1,1)}$. 
Note that the special point $\vec{\tilde\mu}=(0,0)$ of the moduli
space has two different vicinities which we call the $(0,0)_+$-patch and 
the $(0,0)_-$-patch, that is, the point $\vec{\tilde\mu}=(0,0)$ is on an 
intersection of two submanifolds.
In fact, we find that the two different matrices
\begin{align}
H_0^{(0,0)_+} = 
\begin{pmatrix}
z-a_1& & & a_4\\
& z-a_1 & -a_4& \\
& a_3& z-a_2 & \\
-a_3& & & z-a_2
\end{pmatrix} \ , \quad
H_0^{(0,0)_-} = 
\begin{pmatrix}
z-a_1'& a_4'& & \\
-a_3'& z-a_2' & & \\
& & z-a_2' & a_3'\\
& & -a_4' & z-a_1'
\end{pmatrix} \ ,
\label{eq:00patch_k2_so4}
\end{align}
with
\beq
(z-z_1)(z-z_2) = (z-a_1)(z-a_2) + a_3a_4 = (z-a_1')(z-a_2') + a_3'a_4'  \;,  
\eeq
are connected at the points where $a_3=a_4=a_3'=a_4'=0$ and 
$a_1=a_2=a_1'=a_2'$ only.
As mentioned, these concrete expressions for the patches can be
obtained by the maps from those of the $G'=SU(2)$ case as follows 
\begin{eqnarray}
 H_0^{(0,0)_+}=f_+(h^{(1,1)}(a_i))\ ,\quad 
H_0^{(1,1)}=f_+(h^{(2,0)}(b_i))\ , \quad 
H_0^{(-1,-1)}=f_+(h^{(0,2)}(c_i))\ ,\label{eq:eventhree}\\
H_0^{(0,0)_-}=f_-(h^{(1,1)}(a_i'))\ ,\quad 
H_0^{(1,-1)}=f_-(h^{(2,0)}(b_i'))\ , \quad 
H_0^{(-1,1)}=f_-(h^{(0,2)}(c_i'))\ , \label{eq:oddthree}
\end{eqnarray}
where $h^{(*,*)}(a_i)$ are the moduli matrices for $G'=SU(2),k=2$,
\begin{eqnarray}
&&h^{(1,1)}(a_i) =
\begin{pmatrix}
z-a_1 & a_4\\
-a_3& z-a_2
\end{pmatrix} \ ,\nn
&& h^{(2,0)}(b_i) =
\begin{pmatrix}
z^2 + b_1 z + b_2 & 0\\
b_3 z + b_4& 1 
\end{pmatrix} \ ,\quad 
h^{(0,2)}(c_i) =
\begin{pmatrix}
1 & c_3 z + c_4\\
0 & z^2 + c_1 z + c_2
\end{pmatrix} \ .
\end{eqnarray}
The transition functions among these patches are given by the
$V$-transformation (\ref{eq:V-transf}) with  
$V(z)=f_+(V_+(z))f_-(V_-(z))$ where $V_\pm(z)$ are those of
$G'=SU(2)$, i.e.~they are exactly the same as in the $SU(2)$ case
\cite{Eto:2005yh,Eto:2006pg,Eto:2006cx}.
Now, connectedness of the patches is manifest since we know the moduli
space for $G'=SU(2)$ is indeed simply connected. 
The three patches in Eq.~(\ref{eq:eventhree}) compose a submanifold 
${\cal M}_{SO(4),++}^{k=2}$ and Eq.~(\ref{eq:oddthree}) composes 
${\cal M}_{SO(4),--}^{k=2}$. 
The moduli space with $Q_{{\mathbb Z}_2}=+1$, therefore, can be
expressed as
\begin{eqnarray}
{\cal M}_{SO(4),+}^{k=2}\simeq 
{\cal M}_{SO(4),++}^{k=2}\cup  {\cal M}_{SO(4),--}^{k=2} \ ,\quad
 {\cal M}_{SO(4),++}^{k=2}\simeq {\cal M}_{SO(4),--}^{k=2}\simeq
{\cal M}_{SU(2)}^{k=2}\ ,\label{eq:k=2_local_+1}
\end{eqnarray} 
where ${\cal M}_{SU(2)}^{k=2}$ is shown in Eq.~(\ref{eq:SU(2)moduli}).
As we have shown, these two submanifolds intersect at the fixed point
of the permutation (\ref{eq:so4perm}) in the $(0,0)_+$-patch and the
$(0,0)_-$-patch
\begin{eqnarray}
 {\cal M}_{SO(4),++}^{k=2}\cap {\cal M}_{SO(4),--}^{k=2}= \mathbb
 C \ , 
\end{eqnarray} 
where $\mathbb C$ describes the position of the two coincident local
vortices, $a_1=a_2=a_1'=a_2'$. Note that by comparing the right panel
of Fig.~\ref{fig:so2_k2} and the left panel of Fig.~\ref{fig:so4_k2}
(with a $\pm 45$ degrees rotation), it is easily seen that the 
$k=2$, $U(2)$ moduli spaces are embedded in that of the $SO(4)$
theory. 

Let us next study the transition functions among the 
$Q_{{\mathbb Z}_2}=-1$ patches,  
$(1,0)$-$(0,1)$-$(-1,0)$-$(0,-1).$  
The general form of the moduli matrix in the $(1,0)$-patch is:
\beq
H_0^{(1,0)}=f_+(h^{(1,0)}(z_1,d_1))f_-(h^{(1,0)}(z_2,d_2)) =
\begin{pmatrix}
(z-z_1)(z-z_2)& & & \\
-d_2(z-z_1)& z-z_1 & & \\
-d_1d_2& d_1& 1 & d_2\\
-d_1(z-z_2)& & & z-z_2
\end{pmatrix} \ ,\quad \label{andthis1}
\eeq
while the other three are 
\begin{eqnarray}
 H_0^{(0,1)}=f_+(h^{(1,0)}(z_1,d_1))f_-(h^{(0,1)}(z_2,e_2)) \ ,\nn
H_0^{(0,-1)}=f_+(h^{(0,1)}(z_1,e_1))f_-(h^{(1,0)}(z_2,d_2)) \ ,\nn
H_0^{(-1,0)}=f_+(h^{(0,1)}(z_1,e_1))f_-(h^{(0,1)}(z_2,e_2)) \ ,
\end{eqnarray}
where $h^{(1,0)}$ and $h^{(0,1)}$ are the two patches of 
${\cal M}^{k=1}_{SU(2)}\simeq {\mathbb C} \times {\mathbb C}P^1$,
\begin{eqnarray}
 h^{(1,0)}(z_0,b)=
\begin{pmatrix}
 z-z_0 & \\ -b&1
\end{pmatrix}\ ,\quad 
h^{(0,1)}(z_0,b')=
\begin{pmatrix}
 1 & -b'\\ &z-z_0
\end{pmatrix}\ .\label{eq:SU2k1}
\end{eqnarray}
Hence, we can conclude that the moduli space of the $k=2$ local
vortices with $Q_{{\mathbb Z}_2}=-1$ is 
\beq
{\cal M}^{k=2}_{SO(4),-} \simeq ({\cal M}^{k=1}_{SU(2)})^2\simeq
\left({\mathbb C} \times {\mathbb C}P^1\right)^2 \ .
\label{eq:k=2_local_-1}
\eeq
This can be also understood from the dual weight diagrams
in Figs.~\ref{fig:local_k1_example} and \ref{fig:so4_k2}.

\begin{figure}[ht]
\begin{center}
\includegraphics[height=6.5cm]{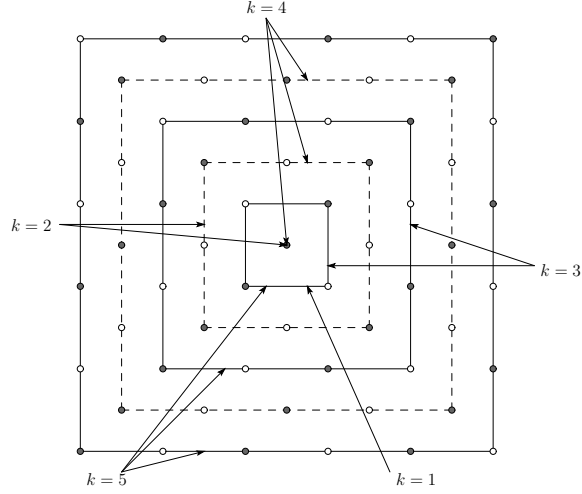}
\caption{{\small The dual weight lattice for $k=1,2,3,4,5$ vortices in
    $G'=SO(4)$.}} 
\label{fig:so4}
\end{center}
\end{figure}

The difference between the moduli spaces in
Eq.~(\ref{eq:k=2_local_+1}) and Eq.~(\ref{eq:k=2_local_-1}) can be
understood as follows. Recall that there exist two kinds of minimal
vortices in $G'=SO(2M)$ theory, namely one for $SU(2)_+$ with
$Q_{{\mathbb Z}_2} = +1$ and another for $SU(2)_-$ with 
$Q_{{\mathbb Z}_2} = -1$, see Fig.~\ref{fig:local_k1_example}. 
We can then choose two vortices with either the same or a different 
${\mathbb Z}_2$-parity in composing the $k=2$ vortex. 
Two vortices with the same parity can be regarded as physically
identical, while those with different parities are distinct. 
In the case of two identical vortices, the moduli space should be a
symmetric product, namely given by Eq.~(\ref{eq:k=2_local_+1}). 
Since the total parity $Q_{{\mathbb Z}_2}^{k=2} = +1$ can be made of   
$(Q_{{\mathbb Z}_2}^{(1)},Q_{{\mathbb Z}_2}^{(2)})=(+1,+1)$ and 
$(-1,-1)$, one finds two copies, as in Eq.~(\ref{eq:k=2_local_+1}).
In contrast, there is only one possibility for 
$Q_{{\mathbb Z}_2}^{k=2} = -1$, namely
$(Q_{{\mathbb Z}_2}^{(1)},Q_{{\mathbb Z}_2}^{(2)})=(+1,-1)$.
The dual weight diagrams are thus quite useful.
As a further illustration, let us show the diagrams for some
higher-winding vortices with $G'=SO(4)$ in Fig.~\ref{fig:so4}, without
going into any detail. 

\subsubsection*{$k=2$ local vortices for $G'=USp(4)$}

Consider now the $k=2$ local vortices for $G'=USp(4)$.
Since the moduli for a single ($k=1$) local vortex requires four
parameters, we expect that the $k=2$ configurations need eight.
The moduli matrices including the special points as the origin are of
the form 
\beq
H_0^{(0,0)} &=& (z-z_0) {\bf 1}_4 + A\ ,\\
H_0^{(1,0)} &=& 
\begin{pmatrix}
P(z) & 0 & 0 & 0\\
b_3 b_6 - b_4(z-z_0+b_5) & z-z_0+b_5 & 0 & b_6\\
b_1 z + b_2 & b_3 & 1 & b_4 \\
-b_4 b_7 + b_3(z-z_0-b_5) & b_7 & 0 & z-z_0 - b_5
\end{pmatrix} \ , 
\eeq
with $P(z) = (z-z_0)^2 - \delta = (z-z_0)^2 - (b_5^2 + b_6 b_7)$ and
\beq
H_0^{(1,1)} = 
\begin{pmatrix}
P(z) & 0 & 0 & 0 \\
0 & P(z) & 0 & 0 \\
c_3 z + c_4 & c_5 z + c_6 & 1 & 0 \\
c_5 z + c_6 & c_7 z + c_8 & 0 & 1
\end{pmatrix} \ ,
\eeq
where $P(z) = z^2 + c_1 z + c_2$.
All other patches are connected and can be obtained by suitable permutations.
The moduli matrices $H_0^{(1,1)}, H^{(1,0)}$ depend on eight free
parameters, as expected. The strong condition is already solved by
them, and thus these patches are $\mathbb C^8$.
The moduli matrix $H_0^{(0,0)}$ has however more complicated form. The
strong condition turns out to be: 
\beq
A^{\rm T} J + J A = 0 \ ,\quad A^2 = \delta{\bf 1}_4 \ .
\eeq
The first condition tells that $A$ takes a value in the algebra of
$USp(4)$, so 
\beq
A = 
\begin{pmatrix}
- \frac{a_{12} - a_{34}}{2} & a_{35} & a_{13} & a_{15}\\
-a_{45} & -\frac{a_{12} + a_{34}}{2} & a_{15} & -a_{14}\\
a_{24} & a_{25} & \frac{a_{12} - a_{34}}{2} & a_{45}\\
a_{25} & -a_{23} & -a_{35} & \frac{a_{12} + a_{34}}{2}
\end{pmatrix} \ .
\eeq
Now $A$ has $10$ parameters. The second set of constraints comes from  
imposing the Pl\"ucker condition on $a_{ij} = - a_{ji}$ ($i,j=1,2,3,4,5$)
\beq
a_{ij}a_{kl} - a_{ik}a_{jl} + a_{il}a_{jk} = 0 \ .
\eeq
Note that the number of linearly independent conditions is three,
hence seven parameters out of ten in the matrix $A$ are linearly
independent. Those together with $z_0$, yield eight degrees of
freedom, indeed as expected.
In this patch, $\delta$ depends on $a_{ij}$ as follows
\beq
\delta = \frac{1}{4}\left(a_{12}-a_{34}\right)^2 + a_{13}a_{24} -
a_{35}a_{45} + a_{15}a_{25}\ . 
\eeq
{Thus the patch $H^{(0,0)}$  is
expressed as}
\begin{eqnarray}
\{H_0^{(0,0)} \}&\simeq& 
\mathbb C\times \frac{\{ B | B: 2\times 5 ~{\rm matrix}\}}{SL(2,\mathbb C)}\simeq\mathbb C\times \left(\mathbb C^{*} \rtimes \frac{\{ B | B: 2\times 5 ~{\rm matrix}\}}{GL(2,\mathbb C)}\right)
\simeq\nonumber \\
&\simeq&\mathbb C\times \left(\mathbb C \rtimes \frac{\{ B | B: 2\times 5 ~{\rm matrix \ of \ rank} \ 2\}}{GL(2,\mathbb C)}\right)
= {\mathbb C} \times 
\left({\mathbb C} \rtimes Gr_{5,2}\right) \ .
\end{eqnarray}
{The last term in the bracket is a cone whose base space is  a $U(1)$ fibration of $Gr_{5,2}$. The tip of this cone corresponds to the origin of the patch, where $a_{ij}=0$, which is thus a conical singularity in the moduli space. Notice that this is a true singularity of the classical metric on the moduli space. It comes out by applying the strong condition on a smooth set of coordinates \cite{Hashimoto}. It is an interesting open problem how this singularity affects the interactions of vortices. }
The transition functions between these patches are easily obtained,
for instance, by requiring that $V(z)=H^{(1,1)} (H^{1,0})^{-1}$ be
regular with respect to $z$
\begin{align}
c_1&=-2 z_0\ , \quad 
&c_2&=z_0^2 - b_5^2 - b_6 b_7\ , \quad
&c_3&= b_1 + \frac{b_4^2}{b_6}\ , \quad 
&c_4&= b_2 - \frac1{b_6}(b_3 b_4 b_6 -b_4^2 (b_5 - z_0))\ ,\nn
c_5&= -\frac{b_4}{b_6}\ , \quad 
&c_6&= b_3 - \frac{b_4}{b_6} (b_5 - z_0)\ , \quad 
&c_7&= \frac1{b_6} \ , \quad 
&c_8&=\frac1{b_6}(b_5 - z_0)\ .
\end{align}
The parameters in $H^{(1,0)}$ are transformed to 
$a_{ij}=B_{1i}B_{2j}-B_{2i}B_{1j}$ of $H^{(0,0)}$ as
\begin{eqnarray}
B \simeq \frac1{\sqrt{b_1}} 
\begin{pmatrix}
 1&b_3^2-b_1 b_7 &0 &-b_2-z_0 b_1+b_3 b_4 +b_1 b_5 & -b_3\\
0 &-b_2-z_0 b_1-b_3 b_4 -b_1 b_5&1&-b_4^2-b_1 b_6 &b_4
\end{pmatrix} \ .
\end{eqnarray}

\subsection{The $k=1$ local vortex for $G' = SO(2M+1)$}


Let us now consider the vortex solutions of the $G' = SO(2M+1)$
theory. The strong condition for the $k=1$ local vortex positioned at
the origin in $G'=SO(2M+1)$ is given by Eq.~(\ref{eq:strong_cond_k=1})
with $n_0=1$. It is very similar to the condition
Eq.~(\ref{eq:strong_cond_k=2_so4}) for the $k=2$ coincident vortices
($z_1=z_2=0$) in $G'=SO(2M)$ 
\beq
H_0^{\rm T} J H_0 = z^2 J \ .
\label{eq:strong_cond_so(odd)_k=1}
\eeq
This implies that the complexity of a single local $SO(2M+1)$ vortex
is almost the same as in the case of the $k=2$ co-axial $SO(2M)$
vortices. 
Indeed, the corresponding dual weight diagrams, see
Figs.~\ref{fig:patches_k1} and \ref{fig:patches_k2}, for instance,  
are the same.

\begin{figure}[ht]
\begin{center}
\includegraphics[height=4cm]{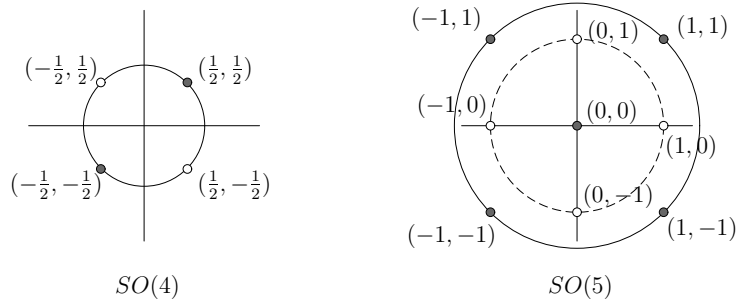}
\caption{{\small Comparison between the single (minimum-winding)
    vortices in $G'=SO(4)$ and $G'=SO(5)$ theories.}} 
\label{fig:so4_so5_k1}
\end{center}
\end{figure}

If however we restrict ourselves to the case of the minimal vortex,
there is a startling difference between the case of  $SO(2M)$ and that
of $SO(2M+1)$. Consider the dual weight diagrams in these two types of
theories. In the case of the $SO(2M)$ theory, all the weight vectors
have the same length $|\tilde{\vec{\mu}}|^2=M/4$, whereas those for
the $SO(2M+1)$ local vortices have different lengths
$|\tilde{\vec{\mu}}|^2$ from $0$ to $M$, see Fig.~\ref{fig:so4_so5_k1}
for $SO(4)$ and $SO(5)$. 
The $M-1$ dimensional sphere represents an orbit of 
$G_{\rm C+F}'=SO(2M)$ or $G_{\rm C+F}'=SO(2M+1)$ which is nothing but
the internal orientation moduli. In the case of $G'=SO(2M)$, the
single vortex has only one orbit, hence the moduli space consists of
the position ${\mathbb C}$ and the broken color-flavor symmetry
$SO(2M)/U(M)$. 
On the other hand, in the case of $G'=SO(2M+1)$, there exist multiple
orbits corresponding to the NG modes, and furthermore the quasi-NG modes
connecting them. For concreteness, let us consider the following
moduli matrix 
\beq
H_0^{(}\overbrace{{}^{1,\cdots,1}}^{r}{}^{,}
\overbrace{{}^{0,\cdots,0}}^{M-r}{}^{)}(z)
= {\rm diag} \big( \underbrace{z^2,\cdots,z^2}_{r}, \underbrace{z,\cdots,z}_{M-r},
\underbrace{1,\cdots,1}_{r}, \underbrace{z,\cdots,z}_{M-r},z\big) \ ,
\label{eq:sp_point_so(2M+1)}
\eeq
where $r$ takes on integer values from $0$ to $M$. We now act with the
color-flavor symmetry $G_{\rm C+F}'=SO(2M+1)$ on the moduli matrix 
from the right. Hence, the $U(r)$ subgroup in $SO(2M+1)$ can be
absorbed by the $V$-transformation (\ref{eq:V-transf}):
\beq
U_0 = 
\begin{pmatrix}
g^{-1} & & & &\\
& {\bf 1}_{M-r} & & &\\
& & g^{\rm T} & & \\
& & & {\bf 1}_{M-r} & \\
& & & & 1
\end{pmatrix} \in U(r) \subset SO(2M+1)\ ,\qquad g \in U(r) \ .
\eeq
The other subgroup $SO(2(M-r)+1) \subset SO(2M+1)$ can be also
absorbed by a $V$-transformation. 
Thus the orbit including the special point
(\ref{eq:sp_point_so(2M+1)}) is \cite{FGK}
\beq
{\cal M}_{\rm ori}^r = \frac{SO(2M+1)}{U(r) \times SO(2(M-r)+1)} \ .
\label{eq:special_orbit_odd}
\eeq
The orbit continuously connects the special points corresponding to 
the dual weight vectors of the same lengths, see
Fig.~\ref{fig:so4_so5_k1}. 
Although the internal moduli spaces (\ref{eq:special_orbit_odd})
with different $r$'s are not connected by the action of $SO(2M+1)$;
these are indeed connected by quasi-NG modes. 

The complete moduli space for the $k=1$, $SO(2M+1)$ vortex is very
similar to that of $k=2$ co-axial $SO(2M)$ vortices which have been
studied in Sec.~\ref{sec:orbit_even_2}. 
A generic solution to the strong condition
(\ref{eq:strong_cond_so(odd)_k=1}) is given by 
\begin{align}
H_0^{(}\overbrace{{}^{1,\cdots,1}}^{r}{}^{,}
\overbrace{{}^{0,\cdots,0}}^{M-r}{}^{)}(z) 
=
\begin{pmatrix}
(z-z_0)^2 {\bf 1}_r & 0 & 0 & 0\\
B_1(z) & (z-z_0){\bf 1}_{M-r} + \Gamma_{11} & 0 & \Gamma_{12}\\
A(z) & C_1 & {\bf 1}_r & C_2\\
B_2(z) & \Gamma_{21} & 0 & (z-z_0){\bf 1}_{M-r+1} + \Gamma_{22}
\end{pmatrix} \ ,
\end{align}
\beq
A(z) &\equiv& a_{1;A}\, z + a_{0;A} + \lambda_{S} \ ,\\
\begin{pmatrix}
B_1(z)\\
B_2(z)
\end{pmatrix}
&=& - \left((z-z_0){\bf 1}_{2(M-r)+1} + \Gamma\right)J_{2(M-r)+1}
\begin{pmatrix}
C_1^{\rm T}\\
C_2^{\rm T}
\end{pmatrix} \ ,\\
\Gamma &\equiv& 
\begin{pmatrix}
\Gamma_{11} & \Gamma_{12}\\
\Gamma_{21} & \Gamma_{22}
\end{pmatrix} \ ,
\eeq
where $a_{i;A}$ $(i=0,1)$ are $r \times r$ constant anti-symmetric
matrices, $C_1$ is an $r \times (M-r)$ constant matrix, 
$C_2$ is an $r \times (M-r+1)$ constant matrix, 
and we have defined
\beq
\lambda_{S} \equiv -\frac{1}{2} (C_1,\ C_2) \, J_{2(M-r)+1}\, 
\begin{pmatrix}
C_1^{\rm T}\\
C_2^{\rm T}
\end{pmatrix} \ ,\quad
J_{2(M-r)+1} \equiv
\begin{pmatrix}
& {\bf 1}_{M-r} & \\
{\bf 1}_{M-r} & &\\
& & 1
\end{pmatrix} \ .
\eeq
The strong condition is now translated into the following form 
\beq
\Gamma^{\rm T} J_{2(M-r)+1} + J_{2(M-r)+1} \Gamma = 0 \ ,\quad
\Gamma^2 = 0 \ .
\label{eq:st_cond_soodd}
\eeq

\begin{figure}[ht]
\begin{center}
\includegraphics[width=15cm]{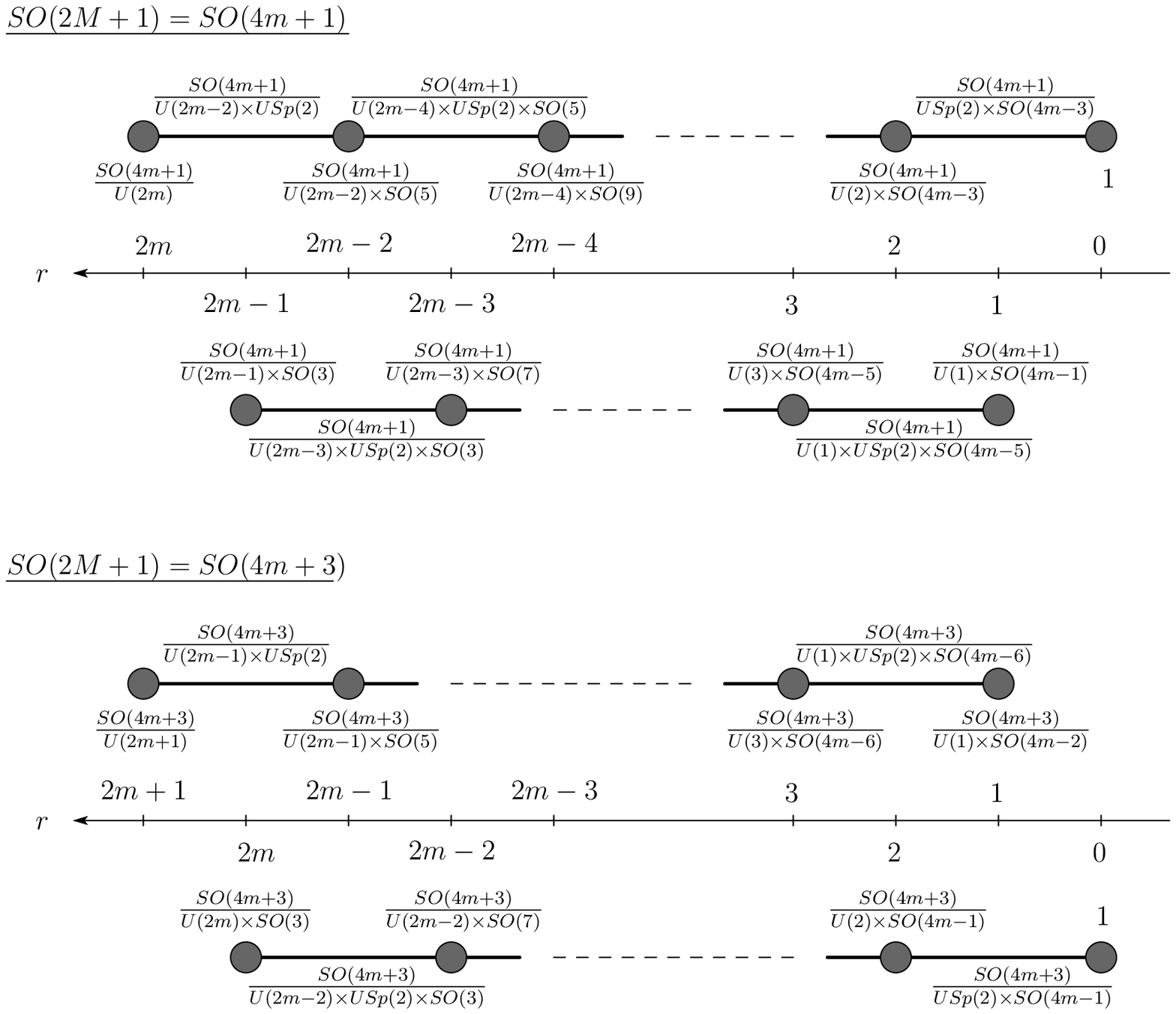}
\caption{{\small Sequences of the $k=1$ vortices for $SO(4m+1)$ and
    for $SO(4m+3)$ theories.}} 
\label{fig:sequence_odd}
\end{center}
\end{figure}

All moduli parameters are included in $a_{i;A},C_i,\Gamma$.
As in the case of $k=2$ co-axial $G'=SO(2M)$ vortices  (see
App.~\ref{app:details}), $a_{0;A}$ and $C_i$ can be removed by an
appropriate color-flavor rotation and $\Gamma$ satisfying the strong
condition (\ref{eq:st_cond_soodd}) can be written 
as (up to $SO(2M+1)_{\rm C+F}$ rotations)
\beq
\Gamma \simeq 
\left(
\begin{array}{cc|cc|c}
& & \Lambda & &\\
& & & {\bf 0}_{M-r-2\gamma}&\\
\hline
{\bf 0}_{2\gamma} & & &&\\
& {\bf 0}_{M-r-2\gamma} & &&\\
\hline
& & & & 0
\end{array}
\right),\quad
\Lambda \equiv i \sigma_2 \otimes {\rm diag}
\left(\lambda_1{\bf 1}_{p_1},\cdots,\lambda_q{\bf 1}_{p_q}\right) \ , 
\eeq
with $\lambda_i > \lambda_{i+1} > 0$ and $2\gamma~(<2(M-r)+1)$ being
the rank of $\Gamma$ ($\gamma = \sum_{i=1}^q p_i$).
By making use of the $V$-transformation and the $SO(2M+1)_{\rm C+F}$
symmetry, we finally obtain the following moduli matrix
\beq
H_0 =
\left(
\begin{array}{cc|cc||cc|cc||c}
z^2{\bf 1}_{r-2\alpha} & 0 & 0 & 0 & 0 & 0 & 0 & 0 & 0\\
0 & z^2{\bf 1}_{2\alpha} & 0 & 0 & 0 & 0 & 0 & 0 & 0\\
\hline
0 & 0 & z^2{\bf 1}_{2\gamma} & 0 & 0 & 0 & 0 & 0 & 0\\
0 & 0 & 0 & z{\bf 1}_{M-r-2\gamma} & 0 & 0 & 0 & 0 & 0\\
\hline
\hline
0 & 0 & 0 & 0 & {\bf 1}_{r-2\alpha} & 0 & 0 & 0 & 0\\
0 & \Lambda'\,z & 0 & 0 & 0 & {\bf 1}_{2\alpha} & 0 & 0 & 0\\
\hline
0 & 0 & \Lambda^{-1} z & 0 & 0 & 0 & {\bf 1}_{2\gamma} & 0 & 0\\
0 & 0 & 0 & 0 & 0 & 0 & 0 & z{\bf 1}_{M-r-2\gamma} & 0\\
\hline\hline
0 & 0 & 0 & 0 & 0 & 0 & 0 & 0 & z\\
\end{array}
\right)\ ,\label{eq:SOoddflavorfixed}
\eeq 
where we have diagonalized $a_{1;A}$ as
\beq
a_{1;A} = u \Lambda' u^{\rm T}\ ,\quad
\Lambda ' \equiv i\sigma_2 \otimes{\rm diag} 
\left(\lambda'_1{\bf 1}_{p'_1},\cdots,
\lambda'_{q'}{\bf 1}_{p'_{q'}}\right)\ , 
\quad u \in U(2\alpha)\ ,
\eeq
with $2\alpha$ being the rank of $a_{1;A}$ and 
$2\alpha = 2 \sum_{i=1}^{q'} p'_i$.
Let us now rearrange the eigenvalues $\{\lambda_i^{-1},\lambda'_i\}$ as
\beq
{\rm diag}(\Lambda',\Lambda^{-1}) \to i \sigma_2 \otimes 
{\rm diag} \left(
\tilde \lambda_1 {\bf 1}_{\tilde p_1},\cdots, \tilde \lambda_s {\bf 1}_{\tilde p_s}
\right) \ ,\quad
\tilde \lambda_a > \tilde \lambda_{a+1} > 0 \ ,
\eeq
and redefine $t\equiv r-2\alpha,\  u\equiv M-r-2\gamma$ with the
constraint:
\begin{eqnarray}
 s,t,u\in \mathbb Z_{\ge 0}\ ,\quad \tilde p_i \in \mathbb Z_{>0} \ ,\quad 
t+u+2\sum_{i=1}^s  \tilde p_i=M \ , \label{eq:SOoddconstraint}
\end{eqnarray}
such that the $r$-dependence in the form of
Eq.~(\ref{eq:SOoddflavorfixed}) disappears. 
We conclude that the moduli space of vortices is (apart from the
center of mass position):  
\beq
{\cal M}^{k=1, {\rm ori}}_{SO(2M+1)} &=& 
\bigcup_{\{t,u, \tilde p_i|\,{\rm Eq.\,}(\ref{eq:SOoddconstraint}) \}}
\mathbb{R}_{>0}^{s} \times {\cal O}_{t,u,\tilde p_i} \ ,\\
{\cal O}_{t,u,\tilde p_i} &=& 
\frac{SO(2M+1)}{U(t) \times SO(2 u+1) \times \prod_{a=1}^{s}
  USp(2\tilde p_a)}\ .
\eeq
Note that there does not appear any $\mathbb Z_2$ factor contrary to
the $SO(2M)$ case since 
\[P={\rm diag}(1,\cdots,1,-1)\in O(2M+1)/SO(2M+1) \]
acts trivially on $H_0$ in Eq.~(\ref{eq:SOoddflavorfixed}).
The special orbits in Eq.~(\ref{eq:special_orbit_odd}) are obtained
simply by choosing $s=0$. A sequence of the moduli space is given in
Fig.~\ref{fig:sequence_odd}. 
At the most generic points, the moduli spaces are locally of the form 
\beq
{\cal M}^{k=1, {\rm ori}}_{SO(4m+1),+} &=& 
\mathbb{R}_{>0}^m \times \frac{SO(4m+1)}{USp(2)^m} \ ,\\
{\cal M}^{k=1, {\rm ori}}_{SO(4m+1),-} &=& 
\mathbb{R}_{>0}^{m-1} \times \frac{SO(4m+1)}{U(1) \times USp(2)^{m-1}
  \times SO(3)} \ ,\\ 
{\cal M}^{k=1, {\rm ori}}_{SO(4m+3),+} &=& 
\mathbb{R}_{>0}^m \times \frac{SO(4m+3)}{U(1)\times USp(2)^m} \ ,\\
{\cal M}^{k=1, {\rm ori}}_{SO(4m+3),-} &=& 
\mathbb{R}_{>0}^{m} \times \frac{SO(4m+3)}{USp(2)^m \times SO(3)} \ .
\eeq
The dimensions of the moduli spaces are then summarized as
\beq
{\rm dim}_{\mathbb C}\left[ {\cal M}^{k=1, {\rm ori}}_{SO(2M+1),+} \right] &=&
M^2 \ ,\\
{\rm dim}_{\mathbb C}\left[ {\cal M}^{k=1, {\rm ori}}_{SO(2M+1),-} \right] &=&
M^2-1 \ .
\eeq

\subsubsection{Examples: $G'=SO(3),SO(5)$}

\subsubsection*{$k=1$ local vortex for $G'=SO(3)$}

Let us discuss the simplest example, {\it viz.} $G'=SO(3)$. In this model
there are two patches having $Q_{{\mathbb Z}_2}=+1$. 
The moduli matrices take the respective forms
\beq
H_0^{(1)} = f_3(h^{(1,0)}(0,a))=
\begin{pmatrix}
z^2 & 0 & 0\\
- a^2 & 1 & \sqrt{2} a\\
-\sqrt{2} a z & 0 & z
\end{pmatrix} \ ,\quad
H_0^{(-1)} = f_3(h^{(0,1)}(0,b)) \ .
\eeq
where $h^{(*,*)}(z_0,a)$ are the two patches (\ref{eq:SU2k1}) of
${\cal M}_{SU(2)}^{k=1}$ and the map $f_3$ is defined by 
\begin{eqnarray}
f_3: A=
\begin{pmatrix}
 c & d\\
e & f
\end{pmatrix} \in GL(2,\mathbb C)\to f_3(A)=
\begin{pmatrix}
 c^2& -d^2& \sqrt{2}c d\\
-e^2&f^2 &-\sqrt{2} e f\\
\sqrt{2} c e &-\sqrt{2} d f&
c f+d e
\end{pmatrix} \ , 
\end{eqnarray}
and expresses the isomorphism 
$GL(2,\mathbb C)/\mathbb Z_2\simeq [U(1)\times SO(3)]^\mathbb C$.
On the other hand, there exists just a single patch with 
$Q_{{\mathbb Z}_2}=-1$.  
This ``patch'' actually contains only a point
\beq
H_0^{(0)} =f_3(\sqrt{z}{\bf 1}_2)= 
z {\bf 1}_3 \ .
\eeq
This vortex does not break the color-flavor symmetry $G'_{\rm
  C+F}=SO(3)$: it is an Abelian vortex i.e.~not having any
orientational moduli.  
Hence, the moduli spaces ${\cal M}_{SO(3),\pm}^{k=1}$ are
\begin{eqnarray}
 {\cal M}_{SO(3),+}^{k=1}\simeq {\cal M}_{SU(2)}^{k=1}\simeq 
\mathbb C\times \mathbb CP^1\ ,\quad
{\cal M}_{SO(3),-}^{k=1}\simeq \mathbb C \ .
\end{eqnarray}
Note that $f_3$ always maps the moduli matrix of $G'=SU(2)$ 
to that of $G'=SO(3)$ with $Q_{{\mathbb Z}_2}=+1$.

\begin{figure}[ht]
\begin{center}
\includegraphics[height=4cm]{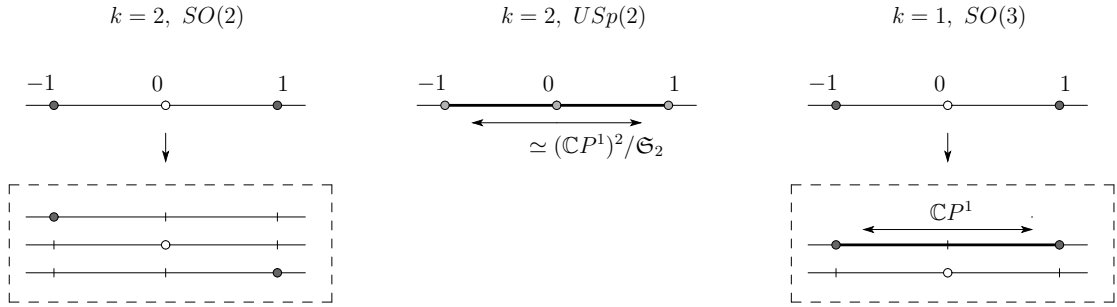}
\caption{{\small $k=1$ $SO(3)$ and $k=2$ $SO(2),USp(2)$.}}
\label{fig:k1_so3}
\end{center}
\end{figure}

We have seen very similar dual weight diagrams for $k=2$,
$SO(2),USp(2)$ and $k=1$, $SO(3)$ vortices. 
All of them consist of three sites on a straight line. However, when
the connectedness is taken into account, they are quite
different, see Fig.~\ref{fig:k1_so3}.  
The three points are isolated in the $SO(2)$ case while
they are all connected in the $USp(2)$ case. 
In the case of $SO(3)$, they split into two diagrams. One is a singlet
and the other has two sites mutually connected, which describes
${\mathbb C}P^1$.

\subsubsection*{$k=1$ local vortex for $G'=SO(5)$}

Finally, we move on to the second simplest case of odd $SO$ vortices: $G'=SO(5)$. Let us first list all the patches, starting with
those having $Q_{{\mathbb Z}_2} = +1$:
\beq
H_0^{(0,0)} &=& z{\bf 1}_5 + A \ ,\\
H_0^{(1,1)} &=& 
\begin{pmatrix}
z^2 & 0 & 0 & 0 & 0\\
0 & z^2 & 0 & 0 & 0\\
- c_3^2 & -c_1 z + c_2 - c_3 c_4 & 1 & 0 & \sqrt{2} c_3\\
c_1 z - c_2 - c_3 c_4 & - c_4^2 & 0 & 1 & \sqrt{2} c_4 \\
-\sqrt{2} c_3 z & -\sqrt{2} c_4 z & 0 & 0 & z
\end{pmatrix} \ ,
\eeq
where
\beq
A = 
\begin{pmatrix}
-a_1 a_2-a_3a_4 & - a_4^2 & 0 & a_1^2 & \sqrt{2} a_1 a_4\\
a_3^2 & -a_1a_2+a_3a_4 & -a_1^2 & 0 & -\sqrt{2} a_1 a_3\\
0 & a_2^2 & a_1a_2+a_3a_4 & -a_3^2 & \sqrt{2} a_2 a_3\\
-a_2^2 & 0 & a_4^2 & a_1a_2-a_3a_4 & \sqrt{2} a_2 a_4\\
-\sqrt{2} a_2 a_3 & - \sqrt{2} a_2 a_4 & - \sqrt{2} a_1 a_4 & \sqrt{2} a_1 a_3 & 0
\end{pmatrix} \ .
\eeq
The patches $H_0^{(1,-1)}$, $H_0^{(-1,1)}$ and $H_0^{(-1,-1)}$
can be obtained from $H_0^{(1,1)}$ by the permutations
(\ref{eq:perm}). 
This means that the four patches 
$\{ H_0^{(1,1)}, H_0^{(1,-1)}, H_0^{(-1,1)}, H_0^{(-1,-1)}\}$ are on
an $SO(5)$ orbit and they are certainly connected.
{By the general discussion in the previous Section, we know that also $H^{(0,0)}$ and all the other four
patches are connected. This can be seen explicitly by studying the transition functions among all these patches}:
\begin{align}
H_0^{(1,1)}&=V^{(1,1),(0,0)}H_0^{(0,0)}, \\
V^{(1,1),(0,0)}&=
\begin{pmatrix}
z+\frac{c_2+c_3c_4}{c_1} & \frac{c_4^2}{c_1} & 0 & -\frac{1}{c_1} & -\frac{\sqrt{2}c_4}{c_1}\\
-\frac{c_3^2}{c_1} & z+\frac{c_2-c_3c_4}{c_1} & \frac{1}{c_1} & 0 & \frac{\sqrt{2}c_3}{c_1}\\
0 & -c_1 & 0 & 0 & 0 \\
c_1 & 0 & 0 & 0 & 0 \\
-\sqrt{2}c_3 & -\sqrt{2}c_4 & 0 & 0 & 1
\end{pmatrix}\ ,\quad 
\left\{
\begin{array}{ll}
a_1=\pm \frac{1}{\sqrt{c_1}} \\
a_i=\pm \frac{c_i}{\sqrt{c_1}} \quad (i=2,3,4)\ ,
\end{array}
\right.
\end{align}
where the same sign has to be chosen for all the transition
functions. 
This means that the moduli space for the minimal vortex with
$Q_{{\mathbb Z}_2}=+1$ in $G'=SO(5)$ is
\beq
{\cal M}_{SO(5),+}^{k=1} = \mathbb C \times W{\mathbb C}P^4_{(2,1,1,1,1)}
\simeq \mathbb C \times {\mathbb C}P^4/{\mathbb Z}_2 \ ,
\eeq
where the subscript $(2,1,1,1,1)$ denotes the $U(1)^{\mathbb C}$
charges. The weighted complex projective space 
$W{\mathbb C}P^4_{(2,1,1,1,1)}$ is defined by the following
equivalence relation among five complex parameters $\phi_i$ (i.e.~the
homogeneous coordinates) 
\beq
(\phi_1,\phi_2,\phi_3,\phi_4,\phi_5) \sim 
(\lambda^2 \phi_1,\lambda \phi_2,\lambda \phi_3,\lambda \phi_4,\lambda
\phi_5) \ ,\quad
\lambda \in {\mathbb C}^* \ .
\eeq

On the other hand, the patches corresponding to $Q_{{\mathbb Z}_2}=-1$
take the form 
\beq
H_0^{(1,0)} = 
\begin{pmatrix}
z^2 & 0 & 0 & 0 & 0 \\
-b_1 z & z & 0 & 0 & 0\\
-b_1b_2 - b_3^2 & b_2 & 1 & b_1 & \sqrt{2}b_3\\
-b_2 z & 0 & 0 & z & 0\\
-\sqrt{2}b_3 z & 0 & 0 & 0 & z
\end{pmatrix} \ .
\eeq
The remaining patches $H_0^{(-1,0)}$, $H_0^{(0,1)}$ and $H_0^{(0,-1)}$
are obtained by permutations (\ref{eq:perm}) from $H_0^{(1,0)}$.  
Since all of them are on an $SO(5)$ orbit, the moduli space of the
$k=1$ vortices with $Q_{{\mathbb Z}_2} = -1$ is 
\beq
{\cal M}_{SO(5),-}^{k=1} =\mathbb C \times
\frac{SO(5)}{U(1) \times SO(3)}
\simeq {\cal M}_{USp(4)}^{k=1}=\mathbb C \times\frac{USp(4)}{U(2)} \ .
\eeq
The following $V$-transformation from the $(1,0)$-patch to the
$(-1,0)$-patch is 
\begin{align}
H^{(-1,0)}(z) &= V^{(-1,0),(1,0)}(z) H^{(1,0)}(z) \ , \\
V^{(-1,0),(1,0)} &=
\begin{pmatrix}
0 & 0 & -\frac{1}{2}\Xi & 0 & 0 \\
0 & \frac{{c'}^2}{\Xi} & a' z & -\frac{2{a'}^2}{\Xi} &
  -\frac{2a' c'}{\Xi} \\
-\frac{2}{\Xi} & -\frac{2b' z}{\Xi} & z^2 & -\frac{2a' z}{\Xi} &
  -\frac{2c' z}{\Xi} \\
0 & -\frac{2{b'}^2}{\Xi} & b' z & \frac{{c'}^2}{\Xi} & -\frac{2b' c'}{\Xi} \\
0 & -\frac{2 b' c'}{\Xi} & c' z & -\frac{2a' c'}{\Xi} & 
  1-\frac{2{c'}^2}{\Xi} 
\end{pmatrix} \ , \\
\Xi &\equiv 2a' b' + {c'}^2 \ .
\end{align}
The transition functions are as follows
\begin{align}
a = -\frac{2a'}{\Xi} \ , \quad 
b = -\frac{2b'}{\Xi} \ , \quad
c = \frac{2c'}{\Xi} \ .
\end{align}

\section{Semi-local vortices}
\label{sec:semilocal}

We now turn to the more general type of solutions, by relaxing
the strong condition (\ref{eq:strong_cond}). 
Namely, we shall make use of only the weak condition
(\ref{eq:cd_SO,USp}) to define our vortices. 
As will be seen shortly, this leads to a larger class of solutions:
the so-called semi-local vortices.

All our vortices including the semi-local ones being BPS saturated,  
can be analyzed by using the moduli matrix $H_0(z)$.  The latter 
has the general properties: 
\begin{itemize}
\item 
it is an $\NC \times \NF$ complex matrix;
\item
all of its elements are polynomials in $z$.
The algorithm given in Ref.~\cite{Baptista:2008ex}
implies that it is sufficient to consider only polynomials as
holomorphic functions;
\item 
it is defined only up to the $V$-equivalence relation,
Eq.~(\ref{eq:V-transf}); 
\item
it is subject to the weak condition, Eq.~(\ref{eq:cd_SO,USp}).
\end{itemize}
The moduli parameters $\phi^i$ for a BPS vortex solution emerge as
coefficients in $H_0(z)$ and thus the moduli space of the solutions is 
defined by the above properties only.
Of course, all the matrices which we found in 
Sec.~\ref{sec:local vortex} for the local vortices satisfy these
conditions {\it a fortiori}. 
Specifically, one can easily check that the special point 
$H_0^{(\tilde\mu_1,\cdots,\tilde\mu_M)}$ in Eq.~(\ref{eq:special_2M})
satisfies the weak condition.

In the strong coupling limit $e,g\rightarrow \infty$,
the master equations (\ref{master1}) and (\ref{master2}) 
are exactly solved by $\Omega'=\Omega'_{\infty}, \omega=\omega_\infty$ 
in Eq.~(\ref{eq:lumpsolution}) and the energy density and K\"ahler
potential for the effective action for 
the vortices (lumps) are given by \cite{Eto:2008qw}
\begin{eqnarray}
\mathcal{E} = 2\partial \bar\partial {\cal K} \ , \quad 
K(\phi^i,\phi^{i*})=\int d^2x\  {\cal K} \ ,\quad 
{\cal K}=\xi \log{\rm Tr}\left[\sqrt{I_{G'} I_{G'}^\dag}\right] \ ,
\end{eqnarray}
with the $G'$-invariant $I_{G'}=H_0^{\rm T}(z)JH_0(z)$. 
Even in the case of finite gauge couplings, these are considered to
be good approximations when $m_{e,g}L\gg 1$ where $L$ is the typical
distance from the core of the vortices. 
By substituting a typical form of $H_0(z)$ into the above formula, one
can obtain multiple peaks in the energy profile even for a minimal
winding vortex  $(k=1)$. We call these interesting multi-peak solutions  
{\it fractional vortices}.  These will be discussed  in a separate
paper \cite{Fractional}.  
Before explicitly studying the semi-local vortices, let us first solve some
 technical problems left out from the previous section.

\subsection{Dimension of the moduli space\label{sec:dim-moduli}}

The index theorem discussed in Appendix~\ref{sec:index_th} tells us that
our moduli space has dimension: 
\beq 
{\rm dim}_{\mathbb{C}}\left(\mathcal{M}_{G',k}\right) = 
\frac{k N^2}{n_0} = \nu \, N^{2}\ . 
\label{eq:dim_moduli_space}
\eeq
This dimension should  {coincide} with that of the space spanned by
the moduli in $H_0(z)$, if the master equations have a unique solution
for a given $H_0(z)$. It is easy to confirm this by considering the
vicinity of a special point of the moduli space.

Let us find the general form of $H_{0}$ in the vicinity of the special
point (\ref{eq:special_2M}) by perturbing $H_{0}$. For definiteness,
let us consider the perturbation around $H_0^{(\frac{k}{2},\cdots,\frac{k}{2})}$: 
\beq
H_0^{(\frac{k}{2},\cdots,\frac{k}{2})} + \delta H_0 = 
\begin{pmatrix}
z^k{\bf 1}_M & \\
& {\bf 1}_M
\end{pmatrix}
+
\begin{pmatrix}
\delta A(z) & \delta C(z)\\
\delta B(z) & \delta D(z)
\end{pmatrix} \ ,
\eeq
where $\delta A(z),\delta B(z),\delta C(z)$ and $\delta D(z)$ are
$M\times M$ matrices whose elements are holomorphic functions of $z$
with small (infinitesimal) coefficients\footnote{{Notice that here we are considering fluctuations around a $k$-vortex configuration with even parity. The generalization to the odd case is discussed at the end of the Section.}}. Not all of the fluctuations
are independent though: we must fix them uniquely by using the
$V$-equivalence (\ref{eq:V-transf}). The infinitesimal
$V$-transformation satisfies the condition $\delta V^{\rm T}(z)J+J\delta 
V(z)=0$ which just represents the algebra of 
$SO(2M,{\mathbb C}),USp(2M,{\mathbb C})$ and can be expressed as 
\beq
\delta V(z) =
\begin{pmatrix}
\delta L(z) & \delta N_{A,S}(z)\\
\delta M_{A,S}(z) & - \delta L^{\rm T}(z)
\end{pmatrix} \ .
\eeq
Again $\delta L(z),\delta M_{A,S}(z)$ and $\delta N_{A,S}(z)$ are
$M\times M$ matrices whose elements are holomorphic in $z$ and their
coefficients are infinitesimally small. Acting with this infinitesimal 
$V$-transformation on the moduli matrix
\beq
\delta V(z) H_0^{(\frac{k}{2},\cdots,\frac{k}{2})} + \delta H_0 \simeq
\begin{pmatrix}
z^k \delta L(z) & \delta N_{A,S}(z)\\
z^k \delta M_{A,S}(z) & - \delta L^{\rm T}(z)
\end{pmatrix}
+
\begin{pmatrix}
\delta A(z) & \delta C(z)\\
\delta B(z) & \delta D(z)
\end{pmatrix} \ ,
\eeq
we can set $\delta D(z) \to 0$, $\delta C \to \delta C_{S,A}(z)$ and
$\delta B(z) \to \delta B_{S,A}(z) + \delta b(z)$ yielding:
\beq
\delta H_0 = 
\begin{pmatrix}
\delta A(z) & \delta C_{S,A}(z)\\
\delta B_{S,A}(z) + \delta b(z) & 0
\end{pmatrix} \ .
\eeq
Note that we have adopted the notation that $\delta X(z)$ stands for a
general polynomial function while $\delta x(z)$ denotes a holomorphic
function whose degree is less than the vortex number $k$. 
Now the $V$-transformation is completely fixed, and one can determine
the true degrees of freedom of the fluctuations. The infinitesimal
form of the weak condition (\ref{eq:cd_SO,USp}) is 
\[\delta H_0^{\rm T}(z) J H_0(z) 
+ H_0(z) J \delta H_0(z) = {\cal O}(z^{k-1}) \ . \]
This leads to 
$\delta A \to \delta a(z)$, $\delta C_{S,A}(z) \to \delta c_{S,A}(z)$,
$\delta B_{S,A}(z) \to 0$ and $\delta b(z) \to \delta b_{A,S}(z)$: 
\beq
\delta H_0(z) = 
\begin{pmatrix}
\delta a(z) & \delta c_{S,A}(z)\\
\delta b_{A,S}(z) & 0
\end{pmatrix} \ .
\label{eq:fluc}
\eeq
These are good coordinates in the vicinity of the special point 
\[ H_0^{(\frac{k}{2},\cdots,\frac{k}{2})} = 
{\rm diag}\left(z^k,\cdots,z^k,1,\cdots,1\right) \ . \]
Of course, this is a only local description but it is sufficient for
counting the dimensions of the moduli space. The complex dimension
is the number of the complex parameters in the fluctuations 
\beq
{\rm dim}_{\mathbb C} 
{\cal M}_{SO(2M),USp(2M)}^{k\text{-semi-local}} = 2 k M^2 \ .
\eeq
In order to restrict the solutions to the local vortices, one further
imposes the following conditions: 
\beq
\delta a(z) \to \delta P(z) {\bf 1}_M \ ,\quad
\delta c_{S,A}(z) \to 0 \ ,
\eeq
with an arbitrary polynomial $\delta P(z)$ of order $(k-1)$.
This leads to the dimension of the $k$ local vortex moduli: 
\beq
{\rm dim}_{\mathbb C} {\cal M}_{SO(2M),+}^{k\text{-local}} &=&
k \left(1 + \frac{M(M-1)}{2}\right) \ ,\\
{\rm dim}_{\mathbb C} {\cal M}_{USp(2M)}^{k\text{-local}} &=&
k \left(1 + \frac{M(M+1)}{2}\right) \ .
\eeq
In a similar way, one can count the dimension in the vicinity of the
special point of positive chirality $(k,\cdots,k)$\footnote{Around other special points this strategy may not
  work in the local case. Other special points may sit  on an intersection of two different submanifolds and
  one cannot make a distinction between the fluctuations among them. It is possible, in any case, to identify, case by case, a special point which does not lie on an intersection. However, one might sometimes need to include quadratic fluctuations, in order to implement correctly the strong condition.}  for the $SO(2M+1)$ case and
obtain
\beq
{\rm dim}_{\mathbb C} {\cal M}_{SO(2M+1),+}^{k\text{-semi-local}} = 
k \left(2M+1\right)^2\ ,\quad
{\rm dim}_{\mathbb C} {\cal M}_{SO(2M+1),+}^{k\text{-local}} = 
k \left(M^2+1\right) \ .
\eeq
{Notice that these results can be considered as a non-trivial consistency check for the moduli matrix formalism. In fact, by physical arguments, we always expect the following relation among the dimensions of the moduli spaces:}
\beq
{\rm dim}_{\mathbb C} {\cal M}_{k}=k\, {\rm dim}_{\mathbb C} {\cal M}_{k=1},
\eeq 
{which is valid both for the local and semi-local case. This relations can be readily used to generalize the above equations to the other cases, including special points with odd chirality.} 
\subsection{The $k=1$ semi-local vortex in $G'=SO(2M),USp(2M)$
  theories} 


Let us study the minimal-winding semi-local vortex in this section. 
The $k=1$ vortex is special in the sense that all the fluctuations in
Eq.~(\ref{eq:fluc}) can actually be promoted to finite parameters.
Namely, the $H_0^{(\frac{1}{2},\cdots,\frac{1}{2})}$-patch is obtained by just replacing
the small fluctuation $\delta a(z),\delta b_{A,S}(z),\delta
c_{S,A}(z)$ by finite constant parameters $A,B_{A,S},C_{S,A}$,
respectively:
\beq
H_0^{(\frac{1}{2},\cdots,\frac{1}{2})}(z)
=
\begin{pmatrix}
z{\bf 1}_M + A & C_{S,A}\\
B_{A,S} & {\bf 1}_M
\end{pmatrix} \ .
\label{suchasthis}\eeq
One can verify that this indeed satisfies the weak condition
(\ref{eq:cd_SO,USp}) for $k=1$. 
Notice that the above matrix can also be rewritten as
\beq
H_0^{(\frac{1}{2},\cdots,\frac{1}{2})}(z) = 
\tilde U_C 
\begin{pmatrix}
z{\bf 1}_M + \tilde A & \\
&{\bf 1}_M
\end{pmatrix}
U_B \ ,
\label{eq:patche_sl_k=1}
\eeq
where we have defined
\beq
\tilde A \equiv A - C_{S,A}B_{A,S} \ ,\quad
U_B \equiv 
\begin{pmatrix}
{\bf 1}_M & \\
B_{A,S} & {\bf 1}_M
\end{pmatrix} \ ,\quad
\tilde U_C \equiv 
\begin{pmatrix}
{\bf 1}_M & C_{S,A}\\
 & {\bf 1}_M
\end{pmatrix} \ .
\eeq
When $A$ is proportional to the unit matrix and $C_{S,A}$ is zero,
that is, corresponding to a local vortex
(\ref{eq:local_k=1_111patch}), 
$U_{B}$ corresponds to the Nambu-Goldstone modes associated with
the symmetry breaking $G'_{\rm C+F} \to U(M)$. 
It is remarkable that this is not always the case for general
semi-local configurations since a non-vanishing $\tilde A$ and
$C_{S,A}$ break $U(M)$ further down.
{In general, the symmetry breaking is $G'_{\rm C+F} \to \mathbb Z_{n_{0}}$}.

Let us next consider the transition functions between two different
patches. As we did for the local vortices in 
Sec.~\ref{sec:local vortex}, the other patches can be obtained as in
Eq.~(\ref{eq:perm_local_k=1}), {\it i.e.} via the permutation matrix $P_r$
defined in Eq.~(\ref{eq:perm}). 
Transition functions are always obtained by means of the
$V$-transformations as in Eq.~(\ref{eq:V-transf}) 
\beq
H_0'(z) = V(z) H_0(z)\ ,\quad
V(z) \equiv V_e~V'(z) \ ,\quad
V_e \in {\mathbb C}^*\ ,\ 
V'(z) \in G'{}^{\mathbb C}\ .
\label{eq:trans_fnc_sl_k=1}
\eeq

For example, consider two patches, $H_0^{(\frac{1}{2},\cdots,\frac{1}{2})}(z)$ given
by Eq.~(\ref{eq:patche_sl_k=1}) and 
\beq
H_0^{(}\overbrace{{}^{-\frac{1}{2},\cdots,-\frac{1}{2}}}^{r}{}^{,}
\overbrace{{}^{\frac{1}{2},\cdots,\frac{1}{2}}}^{M-r}{}^{)}(z) 
&=& P_r^{-1} H_0^{(\frac{1}{2},\cdots,\frac{1}{2})}{}'(z) P_r \ ,\\
H_0^{(\frac{1}{2},\cdots,\frac{1}{2})}{}'(z) &=&
\tilde U_{C'} 
\begin{pmatrix}
z{\bf 1}_M + \tilde A' & \\
&{\bf 1}_M
\end{pmatrix} 
U_{B'} \ .
\eeq
The equation (\ref{eq:trans_fnc_sl_k=1}) in this case reads
\beq
\begin{pmatrix}
z {\bf 1}_M + \tilde A' & \\
& {\bf 1}_M
\end{pmatrix}
U_{B'} P_r U_{-B} = \tilde{U}_{-C'} P_r V \tilde{U}_{C}
\begin{pmatrix}
z {\bf 1}_M + \tilde A & \\
& {\bf 1}_M
\end{pmatrix} \ .
\eeq
The transition functions will be determined by this condition together with
\[(U_{B'}P_rU_{-B})^{\rm T}J(U_{B'}P_rU_{-B})=J \ , \quad
{\rm and } \quad (P_rV)^{\rm T} J (P_rV) = J \ . \]
The solution to these conditions are of the form 
\beq
U_{B'}P_rU_{-B} = 
\begin{pmatrix}
a & a\, d_{\!A,S}\\
0 & \left(a^{-1}\right)^{\rm T}
\end{pmatrix}\ ,\quad
\tilde{U}_{-C'} P_r V \tilde{U}_{C} = 
\begin{pmatrix}
a & (z\mathbf{1}_M + \tilde A') \, a \, d_{\!A,S}\\
0 & \left(a^{-1}\right)^{\rm T}
\end{pmatrix} \ ,
\label{eq:tf_sl_k1_last}
\eeq
with $a \in GL(M,{\mathbb C})$ and $d_{A,S}$ is an $M\times M$
(anti)symmetric matrix and 
\beq
\tilde A' &=& a \,\tilde A \, a^{-1}\ ,\\
C_{S,A}' &=& a \left[ C_{S,A} - \frac{1}{2}\left(\tilde A \, d_{\!A,S} 
- d_{\!A,S}\,\tilde A^{\rm T}\right)\right] a^{\rm T} \ .
\eeq
Notice that $\Tr \tilde A$ is invariant.
The final step is to determine $a,d_{A,S}$ and the transition function
for $B'_{A,S}$ by investigating the concrete form of $U_{B}$
\beq
U_{B} = 
\begin{pmatrix}
{\bf 1}_r & & & \\
& {\bf 1}_{M-r} & &\\
b_1 & b_2 & {\bf 1}_r & \\
-\epsilon \, b_2^{\rm T} & b_3 & & {\bf 1}_{M-r}
\end{pmatrix} \ , \quad 
b_{1,3}^{\rm T} = -\epsilon\,  b_{1,3} \ ,
\eeq
and analogously for $U_{B'}$.
Plugging this into the left hand side of the first equation in
(\ref{eq:tf_sl_k1_last}), one obtains the following result:
\beq
a = 
\begin{pmatrix}
- \epsilon\, b_1 & - \epsilon \,  b_2\\
0 & {\bf 1}_{M-r}
\end{pmatrix} \ , \quad
d_{A,S} = 
\begin{pmatrix}
-b_1^{-1} & \\
& {\bf 0}_{M-r}
\end{pmatrix} \ .
\eeq
The transition functions between $B_{A,S}$ and $B'_{A,S}$ are indeed
the same as those of the local vortex in Eq.~(\ref{eq:tf_local_k1_b}) 
\beq
b_1' = \epsilon \,  b_1^{-1} \ ,\quad
b_2' = b_1^{-1}b_2 \ ,\quad
b_3' = b_3 + \epsilon \,  b_2^{\rm T} b_1^{-1} b_2 \ .
\eeq

We again observe an important result from the first equation in
(\ref{eq:tf_sl_k1_last}). It tells us that 
\beq
\det P_r = +1\ ,
\eeq
thus there exist two copies of the moduli space, which are
disconnected even in the larger space including the semi-local
vortices, in the case of $G'=SO(2M)$. It is of course due to the 
${\mathbb Z}_2$ parity (see Sec.~\ref{sec:Z_2}). 
As in the case of the local vortices in $G'=SO(2M)$ theory discussed
earlier, the patches with different ${\mathbb Z}_2$-parity are
disconnected. 

\subsubsection{Example: $G' = SO(4)$ \label{example1}} 

Let us give an example in the $G'=SO(4)$ theory.
The patches with ${\mathbb Z}_2$-parity $+1$ are
\beq
H_0^{(\frac{1}{2},\frac{1}{2})} &=& 
\begin{pmatrix}
z + a & b & e & f\\
c & z + d & f & g\\
0 & i & 1 & 0\\
-i & 0 & 0 & 1
\end{pmatrix} \ ,\\
H_0^{(-\frac{1}{2},-\frac{1}{2})} &=&
\begin{pmatrix}
1 & 0 & 0 & i' \\
0 & 1 & -i' & 0 \\
e' & f' & z + a' & b'\\
f' & g' & c' & z + d'
\end{pmatrix} \ .
\eeq
These patches are connected by the $V$-transformation
(\ref{eq:V-transf}) $H_0^{(-\frac{1}{2},-\frac{1}{2})} = V^{(-\frac{1}{2},-\frac{1}{2}),(\frac{1}{2},\frac{1}{2})} H_0^{(\frac{1}{2},\frac{1}{2})}$,
\beq
V^{(-\frac{1}{2},-\frac{1}{2}),(\frac{1}{2},\frac{1}{2})} = 
\begin{pmatrix}
0 & 0 & 0 & i' \\
0 & 0 & -i' & 0 \\
0 & \frac{1}{i'} & z + \frac{a'+d'}{2} & 0\\
-\frac{1}{i'} & 0 & 0 & z + \frac{a'+d'}{2}
\end{pmatrix} \ ,
\eeq
The explicit form of the transition function (the relation between the
primed and unprimed parameters) is given in Eq.~(\ref{eq:tfk1}).

{There are two more patches for the vortex with ${\mathbb Z}_2$-parity
$-1$ and are described by the moduli matrices}
\beq
H_0^{(\frac{1}{2},-\frac{1}{2})} &=& 
\begin{pmatrix}
z+a'' & f'' & e'' & b''\\
-i'' & 1 & 0 & 0 \\
0 & 0 & 1 & i''\\
c'' & g'' & f'' & z + d''
\end{pmatrix} \\ 
H_0^{(-\frac{1}{2},\frac{1}{2})} &=& 
\begin{pmatrix}
 1 &  i'''& 0 & 0\\
  f''' & z+d'''  & b'''   & e''' \\
g'''    & c'''  & z+a'''  & f''' \\
  0  & 0 & -i'''  & 1
\end{pmatrix} \ .
\eeq
{These two patches are connected in the same way as the two with positive chirality. In fact they define another copy of the same space. In agreement with the general results found above, neither one of the
even patches: $H_0^{(-\frac{1}{2},-\frac{1}{2})}$,
$H_0^{(\frac{1}{2},\frac{1}{2})}$, is connected with one of the odd,  
$H_0^{(\frac{1}{2},-\frac{1}{2})}$ and $H_0^{-(\frac{1}{2},\frac{1}{2})}$. 
One can easily see that there does not exist any
$V$-transformation connecting them. 
One may construct a holomorphic matrix $X(z)$ which satisfies, for
example,  
$H_0^{(\frac{1}{2},-\frac{1}{2})} = X(z)
H_0^{(\frac{1}{2},\frac{1}{2})}$, however, violating the condition 
$X(z) \in SO(4,{\mathbb C})$. }

\subsection{The $k=2$ semi-local vortices}

\label{sec:semi-local-general-arguments}

Consider now the patches associated with the $k=2$ (doubly-wound) vortices.
Let us begin with infinitesimal fluctuations around the special point
\beq
H_0^{(}\overbrace{{}^{1,\cdots,1}}^{r}{}^{,}
\overbrace{{}^{0,\cdots,0}}^{M-r}{}^{)}
= 
\begin{pmatrix}
z^2{\bf 1}_r & & & \\
& z {\bf 1}_{M-r} & & \\
& & {\bf 1}_r & \\
& & & z{\bf 1}_{M-r}
\end{pmatrix}
\to H_0^{(1,\cdots,1,0,\cdots,0)} +  \delta H_0(z) \ .
\eeq
In order to get rid of the unphysical degrees of freedom in the
fluctuations $\delta H_0$, let us consider an infinitesimal
$V$-transformation (\ref{eq:V-transf})
\beq
\delta V =
\begin{pmatrix}
\delta K_{11} & \delta M_{11} & \delta K_{12;A,S} & \delta M_{12}\\
\delta L_{11} & \delta N_{11} & - \epsilon\, \delta M_{12}^{\rm T} & \delta N_{12;A,S}\\
\delta K_{21;A,S} & \delta M_{21} & - \delta K_{11}^{\rm T} & - \delta
L_{11}^{\rm T}\\
- \epsilon \, \delta M_{21}^{\rm T} & \delta N_{21;A,S} & - \delta
M_{11}^{\rm T} & - \delta N_{11}^{\rm T}
\end{pmatrix} \ .
\eeq
Acting with the $V$-transformation on the perturbed moduli matrix, we
find 
\beq
\delta H_0 \sim \delta H_0 + \delta V H_0^{(1,\cdots,1,0,\cdots,0)} \ .
\eeq
Since the explicit form of $\delta V H_0^{(1,\cdots,1,0,\cdots,0)}$ is 
\beq
\delta V H_0^{(1,\cdots,1,0,\cdots,0)} =
\begin{pmatrix}
z^2 \delta K_{11} & z \delta M_{11} & \delta K_{12;A,S} & z \delta M_{12}\\
z^2 \delta L_{11} & z \delta N_{11} & - \epsilon \, \delta M_{12}^{\rm T} & z \delta N_{12;A,S}\\
z^2 \delta K_{21;A,S} & z \delta M_{21} & - \delta K_{11}^{\rm T} & -
z \delta L_{11}^{\rm T}\\
- z^2 \epsilon \, \delta M_{21}^{\rm T} & z \delta N_{21;A,S} & - \delta
M_{11}^{\rm T} & - z \delta N_{11}^{\rm T}
\end{pmatrix} \ ,  \nonumber
\eeq
the physical degrees of freedom in the fluctuations can be expressed as
\beq
\delta H_0 =
\begin{pmatrix}
\delta A_{11} & \delta C_{11} & \delta A_{12;S,A} & \delta C_{12}\\
\delta B_{11} & \delta D_{11} & 0 & \delta D_{12;S,A} + \delta d_{12;A,S}\\
\delta A_{21;S,A} + \delta a_{21;A,S}^{(1)} z + \delta a_{21;A,S}^{(0)} & \delta c_{21} & 0 & \delta c_{22}\\
\delta B_{21} & \delta D_{21;S,A} + \delta d_{21;A,S} & 0 & \delta d_{22}
\end{pmatrix} \ , \nonumber
\eeq
where $\delta X$ denotes a generic holomorphic polynomial and $\delta
x$ stands for a constant matrix.
The infinitesimal version of the weak condition (\ref{eq:cd_SO,USp})  
\beq
\delta H_0^{\rm T}(z) J H_0(z) + H_0(z) J \delta H_0(z) = {\cal O}(z) \ ,
\eeq
turns out to be equivalent to the following conditions
\begin{align}
\left\{ \delta D_{11},~\delta D_{21;S,A},~\delta D_{12;S,A} \right\} &= 
{\cal O}(1)\ , \nonumber \\
\left\{ \delta A_{11},~\delta C_{11},~\delta A_{12;S,A},~\delta C_{12}
\right\} 
&= {\cal O}(z)\ ,  \nonumber \\
\delta A_{21;S,A} = 0 \ ,\quad
\delta B_{11} = - \delta c_{22}^{\rm T} z + \delta b_{11} \ ,\quad 
\delta B_{21} &= - \epsilon \, \delta c_{21}^{\rm T} z + \delta b_{21} \ .
\end{align}
We thus find the generic form of the fluctuations in the vicinity of
the special point $H_0^{(1,\cdots,1,0\cdots,0)}$ as 
\beq
\delta H_0
=
\begin{pmatrix}
\delta a_{11}^{(1)} z + \delta a_{11}^{(0)} & \delta c_{11}^{(1)} z + \delta c_{11}^{(0)}
& \delta a_{12;S,A}^{(1)} z + \delta a_{12;S,A}^{(0)} & \delta c_{12}^{(1)} z + \delta c_{12}^{(0)}\\
- \delta c_{22}^{\rm T} z + \delta b_{11} & \delta d_{11} & 0 & \delta d_{12}\\
\delta a_{21;A,S}^{(1)} z + \delta a_{21;A,S}^{(0)} & \delta c_{21} & 0 & \delta c_{22}\\
- \epsilon \, \delta c_{21}^{\rm T} z + \delta b_{21} & \delta d_{21} & 0 & \delta d_{22}
\end{pmatrix} \ .
\label{eq:fluc_k2}
\eeq
Let us count the dimensions of the moduli space. 
We have six matrices $\delta a_{ij}^{(\alpha)}$ of size $r\times r$,
two matrices $\delta b_{ij}$ of size $(M-r)\times r$,
six matrices $\delta c_{ij}^{(\alpha)}$ of size $r \times (M-r)$ and
four matrices $\delta d_{ij}$ of the size $(M-r)\times(M-r)$.
Thus summing up we obtain the correct dimension
\beq
{\rm dim}_{\mathbb C} \left[{\cal M}^{2\text{-semi-local}}_{SO(2M),USp(2M)}\right] = 4M^2 \ .
\eeq

The next task is to find the coordinate patches with {\it finite}
parameters (i.e.~large fluctuations). To this end, let us naively
promote all the small fluctuations in Eq.~(\ref{eq:fluc_k2}) to finite
parameters as $\delta x \to x$ (as was done in the case of the minimal
semi-local vortices) : 
\beq
H_0
=
\begin{pmatrix}
z^2{\bf 1}_r + a_{11}^{(1)} z + a_{11}^{(0)} & c_{11}^{(1)} z + c_{11}^{(0)}
& a_{12;S,A}^{(1)} z + a_{12;S,A}^{(0)} & c_{12}^{(1)} z + c_{12}^{(0)}\\
- c_{22}^{\rm T} z + b_{11} & z{\bf 1}_{M-r} + d_{11} & 0 & d_{12}\\
a_{21;A,S}^{(1)} z + a_{21;A,S}^{(0)} & c_{21} & {\bf 1}_r & c_{22}\\
- \epsilon \, c_{21}^{\rm T} z + b_{21} & d_{21} & 0 & z{\bf 1}_{M-r} + d_{22}
\end{pmatrix} \ .
\eeq
But such a procedure is inconsistent with the weak condition
(\ref{eq:cd_SO,USp}). 
Although $H_0^{\rm T} J H_0\big|_{{\cal O}(z^n)} = 0$ for $n\ge3$, the terms
of order ${\cal O}(z^2)$ turn out to be ($z^{2}$ times) 
\begin{align}
H_0^{\rm T} J H_0\big|_{{\cal O}(z^2)} = 
\begin{pmatrix}
- 2 \Lambda_{S,A} &
- a_{21;A,S}^{(1)} c_{11}^{(1)} & {\bf 1}_r - a_{21;A,S}^{(1)} a_{12;S,A}^{(1)} & - a_{21;A,S}^{(1)} c_{12}^{(1)}\\
c_{11}^{(1){\rm T}} a_{21;A,S}^{(1)} & 0 & 0 & {\bf 1}_{M-r}\\
\epsilon \, ({\bf 1}_r + a_{12;S,A}^{(1)} a_{21;A,S}^{(1)} ) & 0 & 0 & 0 \\
c_{12}^{(1){\rm T}} a_{21;A,S}^{(1)} & \epsilon \, {\bf 1}_{M-r} & 0 & 0
\end{pmatrix} \  ,
\label{sothat}
\end{align}
with
\beq
- 2 \Lambda_{S,A} \equiv 
a_{11}^{(1){\rm T}} a_{21;A,S}^{(1)} - a_{21;A,S}^{(1)} a_{11}^{(1)} +
c_{21}c_{22}^{\rm T} + \epsilon \, c_{22} c_{21}^{\rm T} \ .
\eeq
This must be $H_0^{\rm T} J H_0\big|_{{\cal O}(z^2)} = J$, i.e.~we have to
eliminate the undesired terms, such that Eq.~(\ref{sothat}) becomes
exactly equal to $ J$. To compensate the surplus terms, we add the
following extra term
\beq
H_0^{\rm extra} = 
\begin{pmatrix}
{\bf 0}_r & & & \\
& {\bf 0}_{M-r} & &\\
\Lambda_{S,A} & a_{21;A,S}^{(1)} c_{11}^{(1)} & a_{21;A,S}^{(1)} a_{12;S,A}^{(1)} & a_{21;A,S}^{(1)} c_{12}^{(1)}\\
& & & {\bf 0}_{M-r}
\end{pmatrix} \ .
\eeq
Finally we obtain the finite coordinate patch
\begin{align}
&H_0^{(}\overbrace{{}^{1,\cdots,1}}^{r}{}^{,}
\overbrace{{}^{0,\cdots,0}}^{M-r}{}^{)}(z)
=\nonumber\\
&\begin{pmatrix}
z^2{\bf 1}_r + a_{11}^{(1)} z + a_{11}^{(0)} & c_{11}^{(1)} z + c_{11}^{(0)}
& a_{12;S,A}^{(1)} z + a_{12;S,A}^{(0)} & c_{12}^{(1)} z + c_{12}^{(0)}\\
- c_{22}^{\rm T} z + b_{11} & z{\bf 1}_{M-r} + d_{11} & 0 & d_{12}\\
a_{21;A,S}^{(1)} z + a_{21;A,S}^{(0)} + \Lambda_{S,A} & c_{21} + a_{21;A,S}^{(1)} c_{11}^{(1)}
& {\bf 1}_r + a_{21;A,S}^{(1)} a_{12;S,A}^{(1)} & c_{22} + a_{21;A,S}^{(1)} c_{12}^{(1)}\\
- \epsilon \, c_{21}^{\rm T} z + b_{21} & d_{21} & 0 & z{\bf 1}_{M-r} + d_{22}
\end{pmatrix} \ .
\end{align}

All other patches can be obtained by making use of the permutation
(\ref{eq:perm}): 
\beq
H_0^{(}\overbrace{{}^{1,\cdots,1}}^{r}{}^{,}
\overbrace{{}^{0,\cdots,0}}^{M-r}{}^{)}(z) \to
P_{r'}^{-1} H_0^{(}\overbrace{{}^{1,\cdots,1}}^{r}{}^{,}
\overbrace{{}^{0,\cdots,0}}^{M-r}{}^{)}{}'(z)
P_{r'} \ .
\eeq
Since the transition functions between the different patches of the
$k=2$ semi-local vortices are rather complicated, we shall not discuss
them in this paper; we limit ourselves to showing just a few simple
examples below.

\subsubsection{$G'=SO(4)$\label{k2so4}}

As in the case of the $k=2$ local vortices discussed in
Sec.~\ref{sec:k2_local_so4}, at least nine patches are needed to
describe the $k=2$ semi-local vortices. 
They are divided into two disconnected parts as $9 = 5 + 4$ according
to the ${\mathbb Z}_2$-parity. 
The five matrices corresponding to 
$Q_{{\mathbb Z}_2}=+1$ are $\{ H_0^{(1,1)}$, $H_0^{(1,-1)}$,
$H_0^{(-1,1)}$, $H_0^{(-1,-1)}$, $H_0^{(0,0)}\}$ and the four matrices
with $Q_{{\mathbb Z}_2}=-1$ are $\{ H_0^{(1,0)}$, $H_0^{(-1,0)}$,
$H_0^{(0,1)}$, $H_0^{(0,-1)}\}$.

Let us start with the patches having $Q_{{\mathbb Z}_2}=+1$,
\beq
H_0^{(0,0)} &=& (z-z_0){\bf 1}_4 + D \ ,
\label{eq:00patch_k2_so4_sl}\\
H_0^{(1,1)} &=& 
\begin{pmatrix}
z^2 {\bf 1}_2 & \\
& {\bf 1}_2
\end{pmatrix}
+
\begin{pmatrix}
A_1 z + A_0 & C_{1S} z + C_{0S}\\
H_{1A} z + H_{0A} + \frac{1}{2}\left(H_{1A}A_1 - A_1^{\rm T}H_{1A}\right) & H_{1A} C_{1S}
\end{pmatrix} \ ,
\eeq
where $D$ is an arbitrary $4 \times 4$ matrix.
The other patches $\{ H_0^{(1,-1)}$, $H_0^{(-1,1)}$, $H_0^{(-1,-1)}\}$
can be obtained by the permutations (\ref{eq:perm}) of $H_0^{(1,1)}$. 

Now we can clearly see the difference between the local and semi-local
vortices. Let us consider the $(0,0)$-patch. The patches for the local
vortices are given in Eq.~(\ref{eq:00patch_k2_so4}) and those for the
semi-local vortices in Eq.~(\ref{eq:00patch_k2_so4_sl}). 
To avoid confusion, let us denote them by $(0,0)_{l+}$ and $(0,0)_{l-}$ 
for the former and $(0,0)_{sl}$ for the latter.
Clearly, the $(0,0)_{l+}$ and  $(0,0)_{l-}$ patches are unified into
the $(0,0)_{sl}$-patch when the strong condition is relaxed to the
weak one.

As explained in Sec.~\ref{sec:k2_local_so4}, the $(0,0)_{l+}$ patch
(with the $(1,1)$ and $(-1,-1)$ patches) and the $(0,0)_{l-}$-patch
(with the $(-1,1)$ and $(1,-1)$ patches) correspond to two possible
choices of the ${\mathbb Z}_2$-parities of the component vortices
$(Q_{{\mathbb Z}_2}^{(1)},Q_{{\mathbb Z}_2}^{(2)}) = (\pm1,\pm1)$. 
This reflects the fact that any product of the moduli matrices for
local vortices generates  automatically local vortices. It is tempting
to interpret the fact that the two spaces are disconnected  as meaning
that the  ${\mathbb Z}_2$-parity of each component vortex is
conserved. 
However, this is not the case for the semi-local vortices. Products of
moduli matrices satisfying the weak condition (\ref{eq:cd_SO,USp})
do not, in general,     satisfy it.
The ${\mathbb Z}_2$-parity of each vortex is therefore  not conserved
in the semi-local case. 

Let us examine the transition functions between the $(1,1)$ and
$(0,0)$-patches, explicitly.  
Notice, that we have already observed the connectedness between them,
as it was indeed present in the case of the local vortices. 
Our aim to express the following complicated results is completeness
of the calculations. 
Let us write down the moduli matrices as 
\begin{align}
H_0^{(1,1)} &=
\begin{pmatrix}
z^2 + a_1' z + a_0' & b_1' z + b_0' & e_1' z + e_0' & f_1' z + f_0'\\
c_1' z + c_0' & z^2 + d_1' z + d_0' & f_1' z + f_0' & g_1' z + g_0'\\
c_1' i_1' & i_1' z + i_0' - \frac{1}{2} a_1' i_1' + \frac{1}{2} d_1' i_1' & 1 + f_1' i_1' & g_1' i_1' \\
- i_1' z - i_0' - \frac{1}{2} a_1' i_1' + \frac{1}{2} d_1' i_1' & -b_1' i_1' & -e_1' i_1' & 1 - f_1' i_1'
\end{pmatrix} \ , \\
H_0^{(0,0)} &=
\begin{pmatrix}
z+a_0 & b_0 & c_0 & d_0 \\
e_0 & z+f_0 & g_0 & h_0 \\
i_0 & j_0 & z+k_0 & l_0 \\
m_0 & n_0 & o_0 & z+p_0
\end{pmatrix} \ .
\label{eq:1111_so4}
\end{align}
The transition functions are determined through a $V$-transformation
(\ref{eq:V-transf}) satisfying the relation 
$V^{(1,1),(0,0)} H_0^{(0,0)} = H_0^{(1,1)}$:
\beq
V^{(1,1),(0,0)} = 
\begin{pmatrix}
z + \frac{1}{2}a_1' + \frac{1}{2} d_1' - \frac{i_0'}{i_1'} & 0 & 0 & \frac{1}{i_1'}\\
0 & z + \frac{1}{2} a_1' + \frac{1}{2} d_1' - \frac{i_0'}{i_1'} & -\frac{1}{i_1'} & 0\\
0 & i_1' & 0 & 0\\
-i_1' & 0 & 0 & 0
\end{pmatrix} \ . \label{2200}
\eeq
The transition functions connecting the patches $H_0^{(0,0)}$ and $
H_0^{(1,1)}$ are thus given explicitly, see Eq.~(\ref{tf1}).

The transition functions between the $(1,-1)$ and $(0,0)$-patches can
be obtained by the permutation of the above $(1,1)$-$(0,0)$ system as 
\beq
P^{-1} H_0^{(1,1)} P = H_0^{(1,-1)} \ ,\quad 
P^{-1} H_0^{(0,0)} P = \tilde H_0^{(0,0)} \ ,\quad
P = 
\begin{pmatrix}
1 & 0 & 0 & 0 \\
0 & 0 & 0 & 1 \\
0 & 0 & 1 & 0 \\
0 & 1 & 0 & 0
\end{pmatrix} \ .
\eeq
Therefore, the transition functions are easily found as
\beq
V^{(1,-1),(0,0)} \tilde H_0^{(0,0)} = H_0^{(1,-1)} \ , \quad
V^{(1,-1),(0,0)} \equiv P^{-1} V^{(1,1),(0,0)} P \ .
\eeq
The transition functions between the $(1,1)$ and $(1,-1)$-patches 
can be obtained by combining two transition functions given above.

Let us next show the transition functions between the patches with
${\mathbb Z}_2$-parity $-1$. The explicit form of the moduli matrix is
given by 
\beq
H_0^{(1,0)} = 
\begin{pmatrix}
z^2 & & & \\
& z & & \\
& & 1 & \\
& & & z
\end{pmatrix}
+
\begin{pmatrix}
a_1 z + a_0 & b_1 z + b_0 & c_1 z + c_0 & d_1 z + d_0\\
-e_1 z + e_0 & f_0 & 0 & g_0\\
-e_1 i_1 & i_1 & 0 & e_1\\
-i_1 z + i_0 & j_1 & 0 & k_0
\end{pmatrix} \ . \label{eq:2101_so4}
\eeq
The $(-1,0)$-patch can be obtained by acting with the permutation
matrix on the $(1,1)$-patch as follows
\beq
H_0^{(-1,0)} = P^{-1} H_0^{(1,0)}{}' P \ ,\qquad
P = 
\begin{pmatrix}
0 & 0 & 1 & 0 \\
0 & 1 & 0 & 0 \\
1 & 0 & 0 & 0 \\
0 & 0 & 0 & 1
\end{pmatrix} \ . 
\label{perm}
\eeq
The transition functions between these two patches are obtained by
\beq
V^{(-1,0),(1,0)} H_0^{(1,0)} &=& H_0^{(-1,0)} \ ,\\
V^{(-1,0),(1,0)}  &=& 
\begin{pmatrix}
0 & 0 & -i_1' e_1' & 0 \\
0 & 0 & - e_1' z + e_0' & - \frac{e_1'}{i_1'}\\
- \frac{1}{e_1'i_1'} & \frac{1}{e_1'}\left(z- \frac{e_0'}{e_1'}\right) & \left(z- \frac{e_0'}{e_1'}\right)\left(z-
\frac{i_0'}{i_1'}\right) &
\frac{1}{i_1'}\left(z- \frac{i_0'}{i_1'}\right)\\
0 & - \frac{i_1'}{e_1'} & - i_1' z + i_0' & 0
\end{pmatrix} \ .  \label{oddk2}
\eeq
The other transition functions between all the other patches are
obtained through suitable permutations.

It can be shown that the patches with $Q_{{\mathbb Z}_2}=+1$ and
those with $Q_{{\mathbb Z}_2}=-1$ are indeed disconnected.  Let us
take the example of the two moduli matrices $H_0^{(0,0)}$ and
$H_0^{(1,0)}$. Assume that there exists a $V$-function such that  
\beq
V H_0^{(0,0)} = H_0^{(1,0)} \ . \label{transition}
\eeq
First we observe that $V$ is a matrix whose elements are all at most
of order $z$. This is due to $H_0^{(0,0)}$ having the term, 
$z {\bf 1}_4$ and the highest power of $V H_0^{(0,0)}$ should not
exceed 2 which is the highest degree of $H_0^{(1,0)}$.  
We can thus determine the linear term in $z$ of $V$ 
\beq
V = 
\begin{pmatrix}
1 & 0 & 0 & 0 \\
0 & 0 & 0 & 0 \\
0 & 0 & 0 & 0 \\
0 & 0 & 0 & 0
\end{pmatrix}
z + 
\begin{pmatrix}
v_{11} & v_{12} & v_{13} & v_{14} \\
v_{21} & v_{22} & v_{23} & v_{24} \\
v_{31} & v_{32} & v_{33} & v_{34} \\
v_{41} & v_{42} & v_{43} & v_{44}
\end{pmatrix} \ . 
\eeq
Furthermore, let us focus on the linear terms of $z$ in
Eq.~(\ref{transition}), i.e.,
\beq
\begin{pmatrix}
1 & 0 & 0 & 0 \\
0 & 0 & 0 & 0 \\
0 & 0 & 0 & 0 \\
0 & 0 & 0 & 0
\end{pmatrix}
D + 
\begin{pmatrix}
v_{11} & v_{12} & v_{13} & v_{14} \\
v_{21} & v_{22} & v_{23} & v_{24} \\
v_{31} & v_{32} & v_{33} & v_{34} \\
v_{41} & v_{42} & v_{43} & v_{44}
\end{pmatrix}
 = 
\begin{pmatrix}
a_1  & b_1 & c_1 & d_1\\
-e_1 & 1 & 0 & 0\\
0 & 0 & 0 & 0\\
-i_1 & 0 & 0 & 1
\end{pmatrix} \ .
\eeq
By comparison of the third row of both sides, we conclude that 
$\left( v_{31}, v_{32}, v_{33}, v_{34}\right) = \left(0, 0, 0,
0\right)$. 
However, $\det V = 0$ does not satisfy the requirement 
$V \in SO(4,\mathbb{C})$: hence these two patches are disconnected.

\subsection{The $k=1$ semi-local vortex for $G'=SO(2M+1)$}


The result of the index theorem (see App.~\ref{sec:index_th}) yields
that the real dimension is $2k(2M+1)^2$ for the moduli space in
$SO(2M+1)$. Following the technology explained in
Sec.~\ref{sec:semi-local-general-arguments}, it is straightforward to
extend the results to the case of $G'=SO(2M+1)$.
The moduli matrix for $k=1$ in the 
$(\overbrace{1,\cdots,1}^{r},\overbrace{0,\cdots,0}^{M-r})$-patch is
the most general semi-local moduli matrix and is given by 
\begin{align}
&H_0^{(}\overbrace{{}^{1,\cdots,1}}^{r}{}^{,}
\overbrace{{}^{0,\cdots,0}}^{M-r}{}^{)}(z)
=\nonumber\\
&\begin{pmatrix}
z^2{\bf 1}_r + a_{11}^{(1)} z + a_{11}^{(0)} & c_{11}^{(1)} z + c_{11}^{(0)}
& a_{12;S}^{(1)} z + a_{12;S}^{(0)} & c_{12}^{(1)} z 
+ c_{12}^{(0)} & e_{15}^{(1)}z + e_{15}^{(0)}\\
- c_{22}^{\rm T} z + b_{11} & z{\bf 1}_{M-r} + d_{11} & 0 
& d_{12} & e_{25}\\
a_{21;A}^{(1)} z + a_{21;A}^{(0)} + \Lambda_{S} & c_{21} + a_{21;A}^{(1)} c_{11}^{(1)}
& {\bf 1}_r + a_{21;A}^{(1)} a_{12;S}^{(1)} & c_{22} + 
a_{21;A}^{(1)} c_{12}^{(1)} & e_{35} + a_{21;A}^{(1)} e_{15}^{(1)} \\
- c_{21}^{\rm T} z + b_{21} & d_{21} & 0 & z{\bf 1}_{M-r} + d_{22} &
e_{45} \\ 
-e_{35}^{\rm T} z + e_{31}^{\rm T} & e_{32}^{\rm T} & 0 & 
e_{34}^{\rm T} & z + e_{55}
\end{pmatrix} \ ,
\end{align}
where we have defined
\beq -2\Lambda_{S} \equiv a_{11}^{(1){\rm T}}a_{21;A}^{(1)}
-a_{21;A}^{(1)} a_{11}^{(1)}
+c_{21} c_{22}^{\rm T}
+c_{22} c_{21}^{\rm T}
+e_{35} e_{35}^{\rm T} \ . 
\eeq

\subsubsection{$G'=SO(3)$}\label{sec:so3sl}

For $G'=SO(3), k=1$ there are 3 patches, {\it viz.} $(1)$, $(-1)$,
$(0)$. The moduli matrix for the $(0)$-patch is simply
\beq H_0^{(0)} = z\mathbf{1}_{3} + A \ , \eeq
where it is noteworthy to remark that the color$+$flavor symmetry is
unbroken. 

The moduli matrix for the $(1)$-patch is
\beq 
H_0^{(1)} =
\begin{pmatrix}
z^2 + z_1z + z_2 & a+fz & c+bz \\ -\frac{d^2}{2} & 1 & -d\\
e+dz & 0 & z - z_3 \end{pmatrix} \ , 
\eeq 
while the moduli matrix for the $(-1)$-patch is simply obtained by the
permutation 
\beq 
H_0^{(-1)} = P H_0^{(1)} P^{-1} \ , \qquad {\rm with} \ P =
\begin{pmatrix} 0&1&0\\1&0&0\\0&0&1 \end{pmatrix} \ . \eeq
The patches $(-1)$ and $(1)$ are connected by a $V$-transformation given by
\beq {H'}_0^{(1)} = V^{(1),(-1)}  {H}_0^{(-1)} \quad 
{\rm with~}\quad
 V^{(1),(-1)} = \begin{pmatrix}
\frac{(e'+d' z')^2}{{d'}^2} & -\frac{2}{{d'}^2} & -\frac{2(e'+d' z')}{{d'}^2}\\
-\frac{{d'}^2}{2} & 0 & 0 \\
e'+d' z' & 0 & -1 \end{pmatrix} \ , \eeq and the transition functions
can be found in the Appendix. 
The mass center of the system can be identified by taking the
coefficient of the $z^2$ term of ${\rm det} H_0$. It is given by:
$C.M.=-z_1'+z_3'+b'd'+d'{}^2f'/2=-z_1+z_3+b d+d^2 f/2$, which has a
form that is invariant under the change of patch.

The patches $(1)$ and $(0)$ are disconnected. This can be seen from
identifying the linear order of $V$ 
\beq H_0^{(1)} = V{H_0'}^{(0)} =
V\left(z\mathbf{1}_{3} + {A'}\right) \qquad \Rightarrow \qquad
V = z\ {\rm diag}(1,0,0) + V_{\rm const} \ . \eeq
Looking now at the linear order in $z$ of the equation
\beq \begin{pmatrix} z_1 & f & b \\ 0 & 0 & 0 \\ d & 0 & 1
\end{pmatrix} = \begin{pmatrix} 1&0&0\\0&0&0\\0&0&0
\end{pmatrix}{A'} + \begin{pmatrix} v_1 & v_2 & v_3\\ v_4 &
  v_5 & v_6\\ v_7 & v_8 & v_9 \end{pmatrix} \ , \eeq
which reveals that the second row of $V$ has to be zero, which takes
$V$ out of $SO(3,\mathbb{C})$ and the patches are thus disconnected.

\subsubsection{$G'=SO(5)$}\label{sec:so5sl}

For $SO(5)$ we have nine patches. The five having $\mathbb{Z}_2$
charge $+1$ are all connected and are described by the following moduli
matrices
\begin{align}
H^{(0,0)}(z) &= z\mathbf{1}_5 + 
\begin{pmatrix}
a'_1 & a'_2 & a'_3 & a'_4 & a'_5 \\
b'_1 & b'_2 & b'_3 & b'_4 & b'_5 \\
c'_1 & c'_2 & c'_3 & c'_4 & c'_5 \\
d'_1 & d'_2 & d'_3 & d'_4 & d'_5 \\
e'_1 & e'_2 & e'_3 & e'_4 & e'_5 
\end{pmatrix}
\ , \\
H^{(1,1)}(z) &= 
\begin{pmatrix}
z^2 + a_1 z+b_1 & a_2 z+b_2 & c_1 z+d_1 & c_2 z+d_2 & g_1 z+h_1 \\
a_3 z+b_3 & z^2 + a_4 z + b_4 & c_2 z+d_2 & c_3 z+d_3 & g_2 z+h_1\\
e a_3-\frac{i_1^2}{2} & e z + f - \frac{e\left(a_1-a_4\right)}{2} 
  -\frac{i_1 i_2}{2} & 1+e c_2 & e c_3 & i_1+e g_2 \\
-e z-f-\frac{e\left(a_1-a_4\right)}{2}-\frac{i_1 i_2}{2} & 
  -e a_2-\frac{i_2^2}{2} & -e c_1 & 1-e c_2 & i_2 - e g_1\\
-i_1 z+j_1 & -i_2 z+j_2 & 0 & 0 & z+y
\end{pmatrix},
\end{align}
with the rest being permutations of the latter.
The moduli matrix $(0,0)$-patch is connected to the $(1,1)$-patch by
the following $V$-transformation
\begin{align}
H^{(1,1)}(z) &= V^{(1,1),(0,0)}(z) H^{(0,0)}(z) \ , \\
V^{(1,1),(0,0)} &= 
\begin{pmatrix}
z + \frac{a_1+a_4}{2} - \frac{f}{e} - \frac{i_1 i_2}{2e} & 
  -\frac{i_2^2}{2e} & 0 & \frac{1}{e} & \frac{i_2}{e} \\
\frac{i_1^2}{2e} & z + \frac{a_1+a_4}{2} - \frac{f}{e} + 
  \frac{i_1 i_2}{2e} & -\frac{1}{e} & 0 & -\frac{i_1}{e} \\
0 & e & 0 & 0 & 0 \\
-e & 0 & 0 & 0 & 0\\
-i_1 & -i_2 & 0 & 0 & 1
\end{pmatrix}
\ ,
\end{align}
where the transition functions can be found in App.~\ref{app:trnsf}.
There are four patches having $\mathbb{Z}_2$-charge $-1$, which are
all connected. They are described by (and permutations of) the following
moduli matrix
\begin{align}
H^{(1,0)}(z) &= 
\begin{pmatrix}
z^2 + a_1 z + a_2 & c_1 z + c_0 & b_1 z+b_0 & d_1 z + d_0 & i_1 z +i_0 \\
f_0 - e_1 z & z+g_0 & 0 & g_1 & j_0 \\
-e_0 e_1 - \frac{j_1^2}{2} & e_0 & 1 & e_1 & j_1 \\
f_1 - e_0 z & g_2 & 0 & z+g_3 & j_2 \\
h_0 - j_1 z & h_1 & 0 & h_2 & z+k
\end{pmatrix} \ .
\end{align}
This patch is connected to $H^{(-1,0)}$ by the
following $V$-transformation
\begin{align}
H^{(-1,0)}(z) &= V^{(-1,0),(1,0)}(z) H^{(1,0)}(z) \ , \\
V^{(-1,0),(1,0)} &=
\begin{pmatrix}
0 & 0 & -\frac{1}{2}\Xi & 0 & 0 \\
0 & \frac{{j'}_1^2}{\Xi} & f'_0 - e'_1 z & -\frac{2{e'}_1^2}{\Xi} & 
  \frac{2e'_1 j'_1}{\Xi} \\
-\frac{2}{\Xi} & \frac{L_1(z)}{\Xi^2} & \frac{L_2(z)}{\Xi} 
  & \frac{L_3(z)}{\Xi^2} & \frac{L_4(z)}{\Xi^2} \\
0 & -\frac{2{e'}_0^2}{\Xi} & f'_1 - e'_0 z & \frac{{j'}_1^2}{\Xi} &
  \frac{2e'_0 j'_1}{\Xi} \\
0 & \frac{2e'_0 j'_1}{\Xi} & -h'_0 + j'_1 z & \frac{2e'_1 j'_1}{\Xi} 
  & 1 - \frac{2{j'}_1^2}{\Xi} 
\end{pmatrix} \ , \\
\Xi &\equiv 2e'_0 e'_1 + {j'}_1^2 \ , \\
\frac{1}{2}L_1(z) &\equiv f'_1{j'}_1^2 - 2{e'}_0^2\left(f'_0 - 
  e'_1 z\right) + e'_0 j'_1\left(j'_1 z - 2h'_0\right) \ , \\
L_2(z) &\equiv {h'}_0^2 - 2h'_0 j'_1 z + 
  2f'_0 \left(f'_1 - e'_0 z\right) + 
  z\left(2e'_0 e'_1 z + {j'}_1^2 z -2e'_1 f'_1\right) \ , \\
\frac{1}{2}L_3(z) &\equiv f'_0{j'}_1^2 - 
  2{e'}_1^2\left(f'_1-e'_0 z\right) + 
  e'_1 j'_1\left(-2 h'_0 + j'_1 z\right) \ , \\
\frac{1}{2}L_4(z) &\equiv j'_1\left(2 e'_1 f'_1 + 
  j'_1 \left(h'_0 - j'_1 z\right)\right) 
  -2 e'_0\left(e'_1\left(h'_0 + j'_1 z\right)-f'_0 j'_1\right) \ .
\end{align}
The patches of different chiralities are indeed disconnected, as we
expected from topological reasons.

\section{Conclusion and discussion}

In this paper we have analyzed the BPS vortices  appearing in $SO(N)\times U(1)$
and $USp(2N)\times U(1)$ gauge  {theories}. The 
concrete model which our analysis is based upon can be regarded as the
bosonic sector of the corresponding ${\cal N}=2$ gauge theories, but
many of our conclusions are valid on much more general grounds. A
short introduction to the {construction of} BPS vortices in a general gauge group has
already been given by some of us \cite{Short}.

It has been found that, in contrast to the vortices in  
$[SU(N)\times U(1)] / {\mathbb Z}_{N}\simeq U(N)$ models studied
extensively during the last several years, the vortex moduli in these
theories contain certain other moduli, generally known as semi-local
vortices, whose profile functions are characterized by their
asymptotic, power-like behavior, whereas the standard ANO vortices
(including their non-Abelian counterparts found in $U(N)$ theories)
have sharp, exponential cutoff to their transverse size. This is so
even with the minimal number of matter fields, sufficient for the
system to have a ``color-flavor-locked'' Higgs phase.  The difference
with the unitary gauge group case reflects the fact that, for a
given dimension, the number of gauge degrees of freedom is less here,
due to the fact that e.g., $SO(2N), USp(2N)$ groups constitute a 
strict subgroup $SU(2N)$. 

The existence of these semi-local extensions of the vortex moduli is
related to the existence of non-trivial vacuum moduli of the system,
and consequently, to the sigma model lumps which emerge in the strong
gauge coupling limit 
of our vortices \cite{Eto:2008qw,WVLump}. In this limit a vortex solution collapses to a  vacuum configuration everywhere on
the transverse plane.  It defines a map of a 2-cycle onto the moduli space of vacua, and is thus characterized by non-trivial elements of  $\pi_{2}(M_{vac})$. The existence of these semi-local moduli provides the vortex, even 
at finite coupling, with a very rich structure. In
this paper we tried to uncover their general properties, with the help of
concrete examples for the case of a few lower-rank groups.  An
interesting phenomenon concerns the emergence of fractional vortices,
where a certain multi-peaked vortex configuration appear, even if the
vortex, as a whole, has the topologically minimal winding allowed by
the stability.  These features will be discussed more extensively in a
separate article \cite{Fractional}.  

{Related to semi-local vortices is the issue of the non-normalizability of some of the moduli space parameters. 
In the case of $U(N)$ vortices 
this question was solved completely \cite{SemiL}, 
by using the general formula for the effective action of 
vortices in terms of the moduli matrix \cite{Eto:2006uw}. 
A part of this question was solved for a single vortex in $SO$ and $USp$
gauge theories in the lump limit \cite{Eto:2008qw}. Here we have refined our understanding of the non-normalizable modes, relating them as the moduli space parameters which live in a tangent bundle of the moduli space of vacua of the theory.  
}

We have determined the structure of the vortex moduli space, in some
cases identifying it with a well-known manifold, and determining the 
patches needed to cover the whole space. This has been done both
restricting to the local (ANO-like) vortices 
(Sec.~\ref{sec:local vortex}), and considering the full moduli space
(Sec.~\ref{sec:semilocal}).  The latter is closely
related to the issue of the sigma model lumps associated to the
non-trivial vacuum moduli in these theories \cite{Eto:2008qw}, as emphasized several times already. 

The study of the moduli space of local vortices
(Sec.~\ref{sec:local vortex}) is, on the other hand, deeply
related to the {nature of non-Abelian {\it
  monopoles}: i.e.,  to the issue of non-Abelian  (e.g.  GNOW) dualities}. Our results in this paper represent further
steps along the line of the work  \cite{ABEK,Duality}, even though here we have 
limited ourselves just to several examples and a few general
observations.  A more systematic discussion on this problem will be
presented elsewhere \cite{GNOWnine}.

Recently, some  non-BPS extensions of $U(N)$ vortices has been studied for
the local case \cite{Auzzi:2007wj} and for the semi-local case
\cite{Auzzi:2008wm} {with the aim of studying interactions and stability of non-BPS vortices}.  
A non-BPS extension of the $G'=SO,USp$ cases also remains as an open
problem. In connection with this, it is known that the $SO(2M)$ theory
admits a non-BPS ${\mathbb Z}_2$ vortex as 
$\pi_1 ({SO(2M)\times U(1)=\mathbb Z \times \mathbb Z_{2}}$
\cite{Benson:1994dp}, which has not been studied in this paper. {We limit ourselves to the consideration that such kind of lumps can, in principle,  mediate interactions between vortices of opposite chiralities, which, in the range of validity of the moduli space approximation \cite{Manton:1981mp}, are completely decoupled}.

\section*{Acknowledgement} 

We thank David Tong and Giacomo Marmorini for discussions. 
S.B.G~and W.V.~are grateful to Department of Physics, Keio University
and Department of Physics, Tokyo Institute of Technology for warm
hospitality where part of this work has been done.  
M.E., S.B.G.~and W.V.~are grateful to DAMTP, Cambridge University for
warm hospitality.
M.E.~is grateful to SISSA for warm hospitality. 

The work of M.E.~and K.O.~(T.F.) is supported by the Research 
Fellowships of the Japan Society for 
the Promotion of Science for Research Abroad (for Young
Scientists).
The work of M.N. is supported in part by Grant-in-Aid for Scientific
Research (No.~20740141) from the Ministry
of Education, Culture, Sports, Science and Technology-Japan.

The work of W.V is supported by Fondazione Angelo Della Riccia, Fondazione Svizzera, which supports young Italian
researchers working abroad Italy.

\appendix

\section{The index theorem}\label{sec:index_th}

We briefly discuss the dimension of the vortex moduli space along the
lines of Ref.~\cite{HT}, see also
Refs.~\cite{Callias:1977kg,Weinberg:1979er}.
In the following we will keep the gauge group completely generic with
a single overall $U(1)$ factor i.e.~$U(1)\times G'$.
Writing the BPS equations ($e=g$) with linear fluctuations
$\delta\!H,\delta\!\bar{A}$, we obtain 
\begin{align}
\bar{\D}\,\delta\!{H} &= - i\,\delta\!{\bar{A}}\,H \ , \\
\D\,\delta\!\bar{A} - \bar{\D}\,\delta\!A &= 
\frac{i e^2}{2}\tr\left\{\left(\delta\!H\,H^\dag + H\,{\delta\!H}^\dag\right)
t^\alpha\right\}t^\alpha \ , 
\end{align}
and the Gauss' law reads (with $\nu=0$)
\beq \tr\left[\left(\frac{2}{e^2}\D_\mu F^{\mu\nu} + i H(\D^\nu H)^\dag 
- i (\D^\nu H)H^\dag\right)t^\alpha\right] = 0 \ , \quad 
\forall\alpha \ , \eeq
which we use as a gauge fixing condition \cite{HT}
\beq \D\,\delta\!\bar{A}+\bar{\D}\,\delta\!{A} = 
\frac{i e^2}{2}\tr\left\{\left(\delta\!H\,H^\dag - 
H\,{\delta\!H}^\dag\right)t^\alpha\right\}t^\alpha \ . \eeq
A comment in store is that one might wonder why the Gauss law is not
already fulfilled by the fact that the solutions to the BPS equations
satisfy the Euler-Lagrange equations of the system. Fixing the gauge
can be done in many different ways, and instead of requiring the
fluctuations to be orthogonal to the gauge orbit, it
proves convenient to take a direction which corresponds to the time
direction of the Gauss law.  Even if there is no time dependence of the
fields in question, we promote these fluctuations as normal
fluctuations rendering the system better manageable. 
In other words, we constrain the a priori different directions of the
fluctuations to obey the linearized Gauss law.
This leads to the linear system
\begin{align}
\bar{\D}\,\delta\!{H} &= - i \, \delta\!{\bar{A}}\,H \ , \\
\D\,\delta\!{\bar{A}} &= 
\frac{i e^2}{2}\tr\left(\delta\!{H}\,H^\dag t^\alpha\right)t^\alpha \ .
\end{align}
First, we will introduce the following trick
\beq \delta\!{\bar{A}} = 2\;\tr\left(\delta\!{\bar{A}}\;t^\beta\right)t^\beta
\ , \eeq
which makes it possible to write the linear system conveniently as the 
following operator equation
\beq \Delta \begin{pmatrix}\delta\!{H} \\ \delta\!{\bar{A}} \end{pmatrix} 
= 0 \ , \eeq
with (taking $e^2=4$ for convenience)
\beq \Delta \equiv
\begin{pmatrix}
i\bar{\D} & - 2\;\tr\left(\;\circ\; t^\alpha\right)t^\alpha H \\ 
2\;\tr\left(\;\circ\; H^\dag t^\alpha\right)t^\alpha & i\D
\end{pmatrix} \ , \eeq
which has the adjoint operator
\beq \Delta^\dag =
\begin{pmatrix}
i\D & 2\;\tr\left(\;\circ\;t^\alpha\right)t^\alpha H \\ 
- 2\;\tr\left(\;\circ\; H^\dag t^\alpha\right)t^\alpha & i\bar{\D} 
\end{pmatrix} \ . \eeq
Let us start with showing that the operator $\Delta^\dag$ does
not have any zero-modes indeed. That is, the starting point for our
vanishing theorem is to take the complex norm $|X|^2 = \tr\, X X^\dag$
of the operator on a fluctuation
\begin{align} 
0 &= \int d^2x\ \left|\Delta^\dag
\begin{pmatrix}X\\Y\end{pmatrix}\right|^2 \\
&= \int d^2x\ \left[|\D X|^2 + |\bar{\D}Y|^2 + |Y H|^2 + 
\left|2\tr\left(X H^\dag t^\alpha\right)t^\alpha\right|^2
+ i\tr\partial\left(X H^\dag Y^\dag\right) 
- i\tr\bar{\partial}\left(Y H X^\dag\right)\right] \ , \nonumber
\end{align}
where the BPS equations have been used together with the fluctuation
$Y$ taking part of the algebra $Y=Y^\beta t^\beta$. This forces
$Y=0$. 
Here we assume the theory to be in the full Higgs phase.
We take the fluctuations to vanish at spatial infinity
($|z|\to\infty$), thus the boundary terms can be neglected and we can
think of the conditions
\beq \bar{\D}X^\dag = 0 \ , \quad \bar{D}Y = 0 \ , \quad
Y H = 0 \ , \quad \tr\left(t^\alpha H X^\dag\right) = 0 \ , \eeq
as the BPS equations and $F$-term conditions for an ${\cal N}=2$
($d=4$) 
theory with $Y$ being the adjoint scalar of the vector multiplet and
$X$ being anti-chiral fields with the superpotential
\beq W = \tr\left(Y H X^\dag\right) \ . \eeq
Recalling that this toy-theory is evaluated on the background
configuration where $H$ is the scalar fields of the vortex and
the gauge connection in the covariant derivative $\bar{A}$ is also
external fields determined by the background vortex configuration. 
The vortex configuration can always be rewritten by means of the
moduli matrix method yielding $H = S^{-1}H_0(z)$ which gives a
holomorphic description of the field $X^\dag \equiv \tilde H$ as
$\tilde{H} = \tilde{H}_0S$ with $S$ the complexified gauge fields of
the background configuration. It is now easy to show that the $F$-term
condition yields $\tr\left(t^\alpha H_0(z)\tilde{H}_0(z)\right) = 0$,
which in turn simplifies our problem to finding vacuum
configurations of this ${\cal N}=2$ theory,  which has the vacuum in
the Higgs phase almost 
everywhere. We utilize holomorphic invariants 
$I_{\mp}^i(H_0,\tilde{H}_0)$ having negative and positive
$U(1)$ charges, respectively. The boundary conditions for the
invariants are
\beq I_-^i = 0 \ , \quad I_+^i = \mathcal{O}\left(z^{n_i\nu}\right) \ , \eeq
with $\nu$ being the $U(1)$ winding. 
The key point now is to find independent invariants with positive
$U(1)$ charges which will reveal the possible existence of a non-zero
$\tilde{H}_0$. However, the contrary is important here:
\beq \textrm{\textit{iff there exist no independent $I_+^i$, then the 
 fluctuations $X^\dag$ must vanish.}} \nonumber \eeq
In our cases having $G=U(1)\times G'$ with $G' = U(N),SO(N),USp(2M)$
with a common $U(1)$ charge for all the fields it is an easy task to
show the non-existence of independent holomorphic invariants and the
theorem readily applies and completes the proof. We can now go on with
the calculation. 

Now let us calculate the following two operators $\Delta^\dag\Delta$
and $\Delta\Delta^\dag$
\begin{align}
\Delta^\dag\Delta &= -\mathbf{1}_2\partial\bar{\partial} + 
\begin{pmatrix}
\Gamma_1 + \frac{1}{2}B & L_1\\
L_2 & \Gamma_2 - \frac{1}{2}B^{\rm adj}
\end{pmatrix} \ , \\
\Delta\Delta^\dag &= -\mathbf{1}_2\partial\bar{\partial} + 
\begin{pmatrix}
\Gamma_1 & 0\\
0 & \Gamma_2 
\end{pmatrix} \ , 
\end{align}
where $B=F_{12}=-2[\D,\bar{\D}]$ and we have defined the following
operators 
\begin{align}
\Gamma_1 X &= -i A\bar{\partial}X - i(\bar{\partial}A)X - i
\bar{A}\partial X + \bar{A}A X 
+2\;\tr\left(X H^\dag t^\alpha\right)t^\alpha H \ , \\
\Gamma_2 Y &= -i\left[\bar{A},\partial Y\right] 
-i\left[\partial\bar{A},Y\right]
-i\left[A,\bar{\partial}Y\right]
+\left[A,\left[\bar{A},Y\right]\right] 
+2\;\tr\left(Y H H^\dag t^\alpha\right)t^\alpha \ , \\
L_1 Y &= -i Y \D H \ , \\
L_2 X &= i2\;\tr\left(X\bar{\D}H^\dag t^\alpha\right)t^\alpha \ ,
\end{align}
and the algebra of $Y$ has been used as well as the BPS equations. 

To calculate the index of $\Delta$ we can evaluate 
\beq \mathcal{I} = \lim_{M^2\to 0}\mathcal{I}(M^2) = \lim_{M^2\to 0}
\left[\Tr\left(\frac{M^2}{\Delta^\dag\Delta + M^2}\right) - 
\Tr\left(\frac{M^2}{\Delta\Delta^\dag + M^2}\right)\right] \ , \eeq
where $\Tr$ denotes a trace over states as well as over the matrices. 
Now as the eigenvalues of the operator $\Delta^\dag$ are all positive
definite, the index counts only the zero modes of the operator
$\Delta$. For well localized solutions (which go to zero faster than
$1/r$), the index is independent of $M^2$. For convenience we can
evaluate the index in the limit $M^2\to\infty$, thus we can expand and
obtain 
\beq \mathcal{I}(M^2) =
-M^2\Tr\left[\frac{1}{-\partial\bar{\partial}+M^2}
\begin{pmatrix}
\frac{1}{2}B & L_1 \\
L_2 & -\frac{1}{2}B^{\rm adj}
\end{pmatrix} 
\frac{1}{-\partial\bar{\partial}+M^2} + \ldots\right] \ , \eeq
where the ellipsis denote terms that vanish in taking the limit
$M^2\to\infty$. Tracing over the adjoint field strength gives zero. 
We can now evaluate the index as
\begin{align}
\mathcal{I} =&\  -\lim_{M^2\to\infty} M^2 \Tr
\int d^2 x\ \frac{1}{2}\tr(F_{12})
\left\langle x \left|
\left(-\partial\bar{\partial} + M^2\right)^{-2}
\right|x\right\rangle \ ,
\nonumber\\
=&\ -\lim_{M^2\to\infty} M^2 \sum_{1}^{\NF}\int d^2 x \ 
\frac{N}{2\sqrt{2N}}F_{12}^0
\int\frac{d^2 k}{(2\pi)^2}
\frac{1}{\left(\frac{1}{4}k^2 + M^2\right)^2} \ , \nonumber \\
=&\  \NF N\nu \ , 
\end{align}
where 
\beq \nu = -\frac{1}{2\pi\sqrt{2N}}\int d^2x\ F_{12}^0 = 
\frac{k}{n_0} \ . \eeq
Because of the vanishing theorem, the index gives exactly the number
of (complex) zero-modes for the BPS equations for the vortex.
Thus we obtain the same number of zero-modes as the number of moduli
parameters in the moduli matrix formalism. Note that the result is
obtained independently of the gauge group (however only valid when the
vanishing theorem applies) and the impact of the group is
simply encoded in  $\nu$. We also note  that our result reduces
to that of Ref.~\cite{HT} for $U(N)$ by recalling that 
$\nu = k/N$ in that case.

\section{The orientation vectors \label{orient1}}

We have considered the moduli matrix {\it per se} and studied the 
orientational moduli space of the local non-Abelian vortices. Our
result for $G'=SO(2M),USp(2M)$ is the quotient space given in
Eq.~(\ref{eq:orientation_space}). 
These spaces are well-known Hermitian symmetric spaces 
\cite{Hergason, Higashijima:2001vk}.
They can be embedded in the complex Grassmann space 
$G_{2M,M} \simeq SU(2M)/[SU(M) \times SU(M) \times U(1)]$
which is described by a $2M\times M$ complex matrix via a
$GL(M,{\mathbb C})$ equivalence relation 
\beq
Gr_{2M,M} \simeq
\Phi/\!\!/ GL(M,{\mathbb C}) = \{
\Phi \sim \Phi {\cal V}\}\ ,\quad {\cal V} \in GL(M,{\mathbb C}) \ .
\label{eq:Gr_2M,M}
\eeq
where the action of $GL(M,{\mathbb C})$ is free.  
In other words we require the rank of $\Phi$ to be $M$.
The embedding is defined by the constraint \cite{Higashijima:2001vk}
\beq
\Phi^{\rm T} J \Phi = 0\;,
\label{eq:isotoropy}
\eeq
where $J$ is given by Eq.~(\ref{invarianttensor}).

We can relate the matrix $\Phi$ to the orientation of the local
vortex as follows. 
Notice that the moduli matrix decreases its rank by $M$ at the
``vortex center'', $z=z_0$.
The orientational moduli can be extracted as $M$ linearly independent
$2M$-vectors orthogonal to $H_0(z=z_0)$ \cite{Eto:2005yh,Eto:2006pg} 
\beq
H_0(z=z_0) \,  \vec\phi_i = 0\ ,\qquad (i=1,2,\cdots,M)\ .
\label{eq:def_ori}
\eeq
Let us thus define a $2M \times M$ orientational matrix by putting 
$\vec\phi_i$ ($i=1,2,\ldots$) all together as
\beq
\Phi = \left(\vec\phi_1,\vec\phi_2,\cdots,\vec\phi_M\right)\ ,\qquad
H_0(z=z_0) \, \Phi = 0 \ .
\eeq
As $\Phi'$ given by  $\Phi' \equiv \Phi {\cal V}$ with 
${\cal V} \in GL(M,{\mathbb C})$  
-- which is just a change of the  basis -- 
satisfies the same equation (\ref{eq:def_ori}), $\Phi'$ represents the
same physical configuration as $\Phi$. 
This leads to the equivalence relation (\ref{eq:Gr_2M,M}) and to the
complex Grassmannian $Gr_{2M,M}$, as claimed. 
The isotropic condition (\ref{eq:isotoropy}) can be found as follows.
The strong condition (\ref{eq:strong_cond_k=1}) is written as
\beq
(H_0 \Phi)^{\rm T} J (H_0 \Phi) = z \Phi^{\rm T} J \Phi \ .
\eeq
Taking the derivative of this with respect to $z$, one obtains 
\beq
(\p H_0 \Phi)^{\rm T} J H_0 \Phi + (H_0 \Phi)^{\rm T} J \p H_0 \Phi =
\Phi^{\rm T} J\Phi \ .
\eeq
Evaluating this at $z =z_0$ one is led to the constraint
(\ref{eq:isotoropy}). 

The advantage of considering $\Phi$ instead of $H_0(z)$ is 
simplification of the calculation. 
In the rest of this subsection, one can completely forget the previous
argument of the moduli matrix. All the results derived from $H_0$
can be reproduced by $\Phi$ alone. 
Let us explain this by taking two examples: $SO(4)$ and $USp(4)$. 
Then $\Phi$ is a $4 \times 2$ matrix satisfying $\Phi^{\rm T} J \Phi = 0$.
Since $\Phi$ has rank 2, we can generally bring $\Phi$ onto the
following form by using $GL(2,{\mathbb C})$ 
\beq
\Phi_{SO(4)}^{(\frac{1}{2},\frac{1}{2})} = 
\begin{pmatrix}
1 & 0 \\
0 & 1\\
0 & -b\\
b & 0
\end{pmatrix} \ ,\quad
\Phi_{USp(4)}^{(\frac{1}{2},\frac{1}{2})} = 
\begin{pmatrix}
1 & 0 \\
0 & 1\\
a & b\\
b & c
\end{pmatrix} \ .
\eeq
Of course, further three patches 
$\{\Phi^{(-\frac{1}{2},\frac{1}{2})}$,$\Phi^{(\frac{1}{2},-\frac{1}{2})}$, 
$\Phi^{(-\frac{1}{2},-\frac{1}{2})}\}$  
are obtained by fixing $GL(2,{\mathbb C})$ in such a way that
the \{2-3 rows, 1-4 rows, 3-4 rows\} become the unit matrix,
respectively.

The transition functions among them are given through the
$GL(2,{\mathbb C})$. 
In the case of $G'=USp(4)$, the transition functions from the
$(\frac{1}{2},\frac{1}{2})$-patch to the 
$\{(-\frac{1}{2},\frac{1}{2})$,$(\frac{1}{2},-\frac{1}{2})$,
$(-\frac{1}{2},-\frac{1}{2})\}$-patches
are given by
\begin{align}
{\cal V}_{USp(4)}^{(\frac{1}{2},\frac{1}{2}) \to (-\frac{1}{2},\frac{1}{2})}
= 
\begin{pmatrix}
0 & 1\\
a & b
\end{pmatrix}^{-1},\ 
{\cal V}_{USp(4)}^{(\frac{1}{2},\frac{1}{2}) \to (\frac{1}{2},-\frac{1}{2})}
= 
\begin{pmatrix}
1 & 0\\
b & c
\end{pmatrix}^{-1},\ 
{\cal V}_{USp(4)}^{(\frac{1}{2},\frac{1}{2}) \to (-\frac{1}{2},-\frac{1}{2})}
= 
\begin{pmatrix}
a & b\\
b & c
\end{pmatrix}^{-1}.
\end{align}
When the inverse of ${\cal V}$ does not exist, such points are not
covered by two patches but only by one of them.
In the case of $G'=SO(2M)$, neither 
${\cal V}^{(-\frac{1}{2},\frac{1}{2}) \to (\frac{1}{2},\frac{1}{2})}$ nor
${\cal V}^{(\frac{1}{2},\frac{1}{2}) \to (-\frac{1}{2},\frac{1}{2})}$
have an inverse. Thus the $(\frac{1}{2},\frac{1}{2})$-patch is
disconnected from the $(-\frac{1}{2},\frac{1}{2})$-patch and the 
$(\frac{1}{2},-\frac{1}{2})$-patch.
It connects only with the $(-\frac{1}{2},-\frac{1}{2})$-patch and the
transition function is given by 
\beq
{\cal V}_{SO(4)}^{(\frac{1}{2},\frac{1}{2}) \to (-\frac{1}{2},-\frac{1}{2})}
= 
\begin{pmatrix}
0 & -b\\
b & 0
\end{pmatrix}^{-1}.
\eeq
Similarly, the $(-\frac{1}{2},\frac{1}{2})$-patch and the
$(\frac{1}{2},-\frac{1}{2})$-patch are connected. This is a
reinterpretation of the ${\mathbb Z}_2$-parity of the local vortex 
in the model with $G'=SO(4)$, see Fig.~\ref{fig:local_k1_example}.
An extension of this to the local vortex in $G'=SO(2M)$ is
straightforward.

\section{Some details}
\subsection{Spatially-separated vortices \label{sec:orbit_even_1}}

When the two vortices are separated, i.e.~$\delta \neq 0$,
the second equation of Eq.~(\ref{eq:re_strong_condition}) 
(together with $\Tr \, \Gamma = 0$) is solved by
\beq
\Gamma = o' \,  \Gamma_0\,  o'{}^{-1},\qquad
\Gamma_0  \equiv \sqrt \delta\,\left(
\begin{array}{cc}
{\bf 1}_{M-r} & \\
& - {\bf 1}_{M-r}
\end{array}
\right).
\eeq
There remains an arbitrariness under reshuffling the form, 
\beq
o'  \to o's,\quad \Gamma_0 \to s^{-1} \Gamma_0 s,\qquad
s \equiv
\left(
\begin{array}{cc}
u'_1 & \\
& u'_2
\end{array}
\right),
\eeq
where $u'_i \in GL(M-r,\mathbb{C})$.
Then the first condition in Eq.~(\ref{eq:re_strong_condition}) leads to
\beq
o'{}^{\rm T} J_{2(M-r)} \, o' = \left(
\begin{array}{cc}
0 & X\\
\epsilon \, X^{\rm T} & 0
\end{array}
\right) \sim J_{2(M-r)}\ ,
\eeq
where we have used the above-mentioned freedom to arrive at the last form for 
$J_{2(M-r)}$. 
The above relation means that $o'$ is an element of
$O(2(M-r))^{\mathbb C}$  ($USp(2(M-r))^{\mathbb C}$).
There exists still an unphysical transformation $u_1'{}^{\rm T} =
u_2'{}^{-1}\equiv u \in GL(M-r,\mathbb{C})$. 
Thus the solution of the strong condition (\ref{eq:re_strong_condition}) with $\delta \neq 0$ is given by
\beq
\Gamma \in \left\{
\begin{array}{lcl}
\left\{\mathbb{C}^* \times 
\left[\frac{O(2(M-r),\mathbb{C})}{U(M-r)}\right]^{\mathbb C}
\right\}/\mathbb Z_2&&\text{for}\quad G'=SO(2M)\ ,\\
 \left\{\mathbb{C}^* \times 
\left[\frac{USp(2(M-r),\mathbb{C})}{U(M-r)}\right]^{\mathbb C}
\right\}/\mathbb Z_2&& \text{for}\quad G'=USp(2M)\ ,
\end{array}
\right.
\eeq
with the first $\mathbb{C}^*$ factor being the relative distance 
$\sqrt{\delta}$. 
The $\mathbb Z_2$ factors in the denominators come about due to the fact
that a combination of a $\pi$-rotation in the $x^1$-$x^2$ space 
$\sqrt{\delta}\rightarrow -\sqrt{\delta}$ and 
a permutation $o'\rightarrow o'p$,  satisfying 
$p\Gamma_0p^{-1}=-\Gamma_0$  is an identity operation.

\subsection{Fixing NG modes for Sec. \ref{sec:orbit_even_2}\label{app:details}}

Let us go into a detailed investigation, in order to verify the
results in Sec.~\ref{sec:orbit_even_2}.
In the first place note that $a_{0;A,S}$ and $C_{1,2}$ 
are obviously NG modes when two vortices are coincident, namely $\delta = 0$. 
One can confirm this fact, 
for example, by considering an infinitesimal color-flavor $G'_{C+F}$
transformation accompanied by an appropriate $V$-transformation.
Therefore  any moduli matrices of the form (\ref{eq:k2Localgeneric}) 
can be always brought into the following
\begin{align}
H_0^{(}\overbrace{{}^{1,\cdots,1}}^{r}{}^{,}
\overbrace{{}^{0,\cdots,0}}^{M-r}{}^{)} &=&
\left(
\begin{array}{cccc}
(z-z_0)^2{\bf 1}_r & 0 & 0 & 0\\
0 & (z-z_0){\bf 1}_{M-r} + \Gamma_{11} & 0 & \Gamma_{12}\\
a_{1;A,S}\,z & 0 & {\bf 1}_r & 0\\
0 & \Gamma_{21} & 0 & (z-z_0){\bf 1}_{M-r} + \Gamma_{22}
\end{array}
\right)\ .
\label{eq:k2_patch_local}
\end{align}
For $\delta = 0$, the rank $2\gamma={\rm rank}(\Gamma)$ is less than 
$2\gamma < 2(M-r)$.
The first condition in Eq.~$(\ref{eq:re_strong_condition})$ states
that $\Gamma J_{2(M-r)}$ is anti-symmetric (symmetric), so that 
$\Gamma$ can be written as
\beq
\Gamma = \epsilon\,q \tilde J_{2\gamma} \, q^{\rm T} \, J_{2(M-r)}\ ,
\eeq
where $q$ is a $2(M-r)\times 2\gamma$ matrix whose rank is 
$2\gamma , \,(M-r\ge\gamma),$ and $\tilde J_{2\gamma}$ is the invariant tensor
of $\tilde G'_{2\gamma} = USp(2\gamma)$ for $G'=SO(2M)$ 
and $\tilde G'_{2\gamma} = SO(2\gamma)$ for $G'=USp(2M)$.
Then the second condition is translated into the following constraint
on $q$: 
\beq
A = 0 \ ,\quad A \equiv q^{\rm T} J_{2(M-r)} q \ .
\eeq
Note that the rank of $A=q^{\rm T} J_{2(M-r)} q$ is bounded as
\beq
4\gamma - 2(M-r) \le {\rm rank}(A) \le {\rm rank}(q) = 2\gamma \ .
\eeq
Therefore, $2\gamma \le M-r$ in the present case of ${\rm rank}(A) = 0$.
This last condition can be solved by
\beq
q = 
O \left(
\begin{array}{c}
g\\
{\bf 0}_{2(M-r-\gamma)\times 2\gamma}
\end{array}
\right)\ , \quad
g \in GL(2\gamma,\mathbb{C})\ ,\quad
O \in G'_{2(M-r)}\ .
\eeq
Thus we find
\beq
\Gamma = O \left(
\begin{array}{cc|cc}
g \tilde J_{2\gamma} g^{\rm T} &  & &\\
 & {\bf 0}_{M-r - 2\gamma} &  & \\
\hline
& & {\bf 0}_{2\gamma} & \\
& & & {\bf 0}_{M-r-2\gamma}
\end{array}
\right)
O^{\rm T} J_{2(M-r)}\ .
\eeq

In the case of $G'=SO(2M)$, we can bring the anti-symmetric matrix 
$g \tilde J_{2\gamma} g^{\rm T}$ onto a block-diagonal form as 
\beq
g \tilde J_{2\gamma} g^{\rm T} = u\, \Lambda \,u^{\rm T},\quad 
\Lambda \equiv i\sigma_2 \otimes {\rm diag}(\lambda_1{\bf 1}_{p_1},\lambda_2{\bf 1}_{p_2},\cdots,\lambda_q{\bf 1}_{p_q}),
\quad (\lambda_{i} > \lambda_{i+1} > 0),
\eeq
where $u \in U(2\gamma)$ and $2\sum_{i=1}^q p_i = 2\gamma$.
Thus we have found
\beq
\Gamma &=& 
O'
\left(
\begin{array}{cc|cc}
& & \Lambda & \\
& & & {\bf 0}_{M-r - 2\gamma}\\
\hline
{\bf 0}_{2\gamma} & & &\\
& {\bf 0}_{M-r-2\gamma} & &
\end{array}
\right)O'{}^{-1}\ ,\\
O' &\equiv& O\left(
\begin{array}{cccc}
u & & &\\
& {\bf 1}_{M-r-2\gamma} & & \\
& & \left(u^{\rm T}\right)^{-1} &\\
& & & {\bf 1}_{M-r-2\gamma}
\end{array}
\right) \in SO(2(M-r))\ .
\eeq
Similarly, the anti-symmetric tensor $a_{1;,A}$ can be brought onto a
diagonal form. 
Let ${\rm rank}(a_{1,A})= 2\alpha \le r$, then we obtain
\beq
a_{1;A} = 
\left(
\begin{array}{cc}
{\bf 0}_{r-\alpha} & \\
& u' \, \Lambda' \, u'{}^{\rm T}
\end{array}
\right) \ ,\quad
\Lambda' \equiv i\sigma_2 \otimes 
{\rm diag}(\lambda_1'{\bf 1}_{p'_1},\lambda_2'{\bf
  1}_{p'_2},\cdots,\lambda'_{q'}{\bf 1}_{p'_{q'}}) \ ,
\eeq
where $u' \in U(2\alpha)$, $2\sum_{i=1}^{q'} p'_i = 2\alpha$ and 
$\lambda'_i > \lambda'_{i+1} > 0$.
Finally, we arrive at the following expression
\beq
H_0 =
\left(
\begin{array}{cc|cc||cc|cc}
z^2{\bf 1}_{r-2\alpha} & 0 & 0 & 0 & 0 & 0 & 0 & 0\\
0 & z^2{\bf 1}_{2\alpha} & 0 & 0 & 0 & 0 & 0 & 0\\
\hline
0 & 0 & z{\bf 1}_{2\gamma} & 0 & 0 & 0 & \Lambda & 0\\
0 & 0 & 0 & z{\bf 1}_{M-r-2\gamma} & 0 & 0 & 0 & {\bf 0}_{M-r-2\gamma}\\
\hline
\hline
{\bf 0}_{r-2\alpha} & 0 & 0 & 0 & {\bf 1}_{r-2\alpha} & 0 & 0 & 0\\
0 & \Lambda'\,z & 0 & 0 & 0 & {\bf 1}_{2\alpha} & 0 & 0\\
\hline
0 & 0 & 0 & 0 & 0 & 0 & z{\bf 1}_{2\gamma} & 0\\
0 & 0 & 0 & 0 & 0 & 0 & 0 & z{\bf 1}_{M-r-2\gamma}
\end{array}
\right),
\eeq
where we have turned off the center of mass $z_0=0$.
One can return to the previous moduli matrix by using the color-flavor
symmetry  $H_0 \to U^{-1} H_0 U$ with
\beq
U \equiv \left(
\begin{array}{cc|c||cc|c}
{\bf 1}_{r-2\alpha} & & & & & \\
 & u'{}^{\rm T} & & & & \\
\hline
 & & O'{}^{-1} & & & \\
\hline
\hline
 & & & {\bf 1}_{r-2\alpha} & & \\
 & & & & u'{}^{-1} & \\
\hline
 & & & & & O'{}^{-1}
\end{array}
\right) \in SO(2M) \ .
\eeq
By making use of the $V$-transformation, one can bring this onto the
following form 
\beq
VH_0 &=&
\left(
\begin{array}{cc|cc||cc|cc}
z^2{\bf 1}_{r-2\alpha} & 0 & 0 & 0 & 0 & 0 & 0 & 0\\
0 & z^2{\bf 1}_{2\alpha} & 0 & 0 & 0 & 0 & 0 & 0\\
\hline
0 & 0 & z^2{\bf 1}_{2\gamma} & 0 & 0 & 0 & 0 & 0\\
0 & 0 & 0 & z{\bf 1}_{M-r-2\gamma} & 0 & 0 & 0 & 0\\
\hline
\hline
0 & 0 & 0 & 0 & {\bf 1}_{r-2\alpha} & 0 & 0 & 0\\
0 & \Lambda'\,z & 0 & 0 & 0 & {\bf 1}_{2\alpha} & 0 & 0\\
\hline
0 & 0 & \Lambda^{-1} z & 0 & 0 & 0 & {\bf 1}_{2\gamma} & 0\\
0 & 0 & 0 & 0 & 0 & 0 & 0 & z{\bf 1}_{M-r-2\gamma}
\end{array}
\right)\ ,\\
V &=&
\left(
\begin{array}{cc|cc||cc|cc}
{\bf 1}_{r-2\alpha} & 0 & 0 & 0 & 0 & 0 & 0 & 0\\
0 & {\bf 1}_{2\alpha} & 0 & 0 & 0 & 0 & 0 & 0\\
\hline
0 & 0 & z{\bf 1}_{2\gamma} & 0 & 0 & 0 & -\Lambda & 0\\
0 & 0 & 0 & {\bf 1}_{M-r-2\gamma} & 0 & 0 & 0 & 0\\
\hline
\hline
0 & 0 & 0 & 0 & {\bf 1}_{r-2\alpha} & 0 & 0 & 0\\
0 & 0 & 0 & 0 & 0 & {\bf 1}_{2\alpha} & 0 & 0\\
\hline
0 & 0 & \Lambda^{-1} & 0 & 0 & 0 & {\bf 0}_{2\gamma} & 0\\
0 & 0 & 0 & 0 & 0 & 0 & 0 & z{\bf 1}_{M-r-2\gamma}
\end{array}
\right) \ .
\eeq
where one can check that $V \in SO(2M,\mathbb{C})$ because 
$\Lambda^{\rm T} = - \Lambda$. 
We can rearrange the eigenvalues 
$\tilde \lambda_a = \{\lambda_i^{-1},\lambda'_j\}$ in such a way that 
\beq
{\rm diag}\left(
\Lambda'\,,\ 
\Lambda^{-1}
\right) = i\sigma_2 \otimes {\rm diag }\left(\tilde\lambda_1 {\bf
  1}_{\tilde p_1},\cdots,\tilde\lambda_s {\bf 1}_{\tilde p_s}\right)\ ,
\quad
\tilde\lambda_a > \tilde \lambda_{a+1} > 0\ ,
\eeq
hence the $G'_{\rm C+F} = SO(2M)$ orbit can easily be seen in
Eq.~(\ref{eq:k2SOevenorbit}).

The arguments for $G'=USp(2M)$ are analogous to those of $G'=SO(2M)$. 
A small difference is that $J_{2(M-r)}\Gamma$ and $a_{1;S}$ are now
symmetric. In the end, we obtain the moduli matrix on the following
form 
\beq
H_0 &=&
\left(
\begin{array}{ccc|ccc}
z^2{\bf 1}_{r-\beta} & 0 & 0 & 0 & 0 &  0\\
0 & z^2{\bf 1}_{\beta+\zeta} & 0 & 0 & 0 & 0\\
0 & 0 & z{\bf 1}_{M-r-\zeta} & 0 & 0 & 0\\
\hline
0 & 0 & 0 & {\bf 1}_{r-\beta} & 0 & 0\\
0 & \tilde \Lambda\,z & 0 & 0 & {\bf 1}_{\beta+\zeta} & 0\\
0 & 0 & 0 & 0 & 0 & z{\bf 1}_{M-r-\zeta}
\end{array}
\right)\ ,\\
\tilde \Lambda &=& {\rm diag}(\tilde \lambda_1 {\bf 1}_{\tilde p_1},
\cdots, \tilde \lambda_s {\bf 1}_{\tilde p_s}) \ ,
\eeq
with $\beta = {\rm rank}(\Gamma)$ and $\zeta = {\rm rank}(a_{1;S})$.

\section {Some transition functions}\label{app:trnsf}

Here we make a collection of some of the transition functions
discussed in the main text. 

\subsection {Example 1}

The transition functions between two ${\mathbb Z}_2$-parity $+1$
patches for the minimal semi-local vortices in $G'=SO(4)$ theory
of Sec.~\ref{example1}:  
\beq
\left\{
\begin{array}{l@{$\;$}l}
a &= - f' i' + \frac{a'+d'}{2}\ ,\\
b &= -g' i'\ ,\\
c &= e' i'\ ,\\
d &= f'i' + \frac{a'+d'}{2}\ ,\\
e &= -c' i' \ ,\\
f &= \frac{(a'-d')i'}{2}\ ,\\
g &= b' i'\ ,\\
i &= - \frac{1}{i'} \ .
\end{array}
\right. 
\label{eq:tfk1}
\eeq

\subsection {Example 2}
The transition functions between $H_0^{(0,0)} $ and $ H_0^{(1,1)}$ for
the $k=2$ semi-local vortices in $G'=SO(4)$ theory of
Sec.~\ref{k2so4}:  
\beq
\left\{
\begin{array}{l@{$\,$}l}
a_0 &= \frac{1}{2}a_1' - \frac{1}{2}d_1' + \frac{i_0'}{i_1'}\ ,\\
b_0 &= b_1' \ ,\\
c_0 &= e_1' \ ,\\
d_0 &= f_1' - \frac{1}{i_1'} \ ,\\
e_0 &= c_1'\ ,\\
f_0 &= - \frac{1}{2}a_1' + \frac{1}{2}d_1' + \frac{i_0'}{i_1'} \ ,\\
g_0 &= f_1' + \frac{1}{i_1'} \ ,\\
h_0 &= g_1' \ ,\\
i_0 &= - c_1' i_0' - c_0' i_1' + \frac{1}{2}a_1'c_1'i_1' +
\frac{1}{2}c_1'd_1'i_1' \ ,\\
j_0 &= a_1' i_0' - \frac{i_0^{'2}}{i_1'} - \frac{1}{4} a_1^{'2} i_1' -
d_0' i_1' + \frac{1}{4} d_1^{'2} i_1' \ ,\\
k_0 &= \frac{1}{2} a_1' + \frac{1}{2}d_1' - f_1' i_0' - f_0' i_1' - \frac{i_0'}{i_1'} + \frac{1}{2}a_1' f_1' i_1'
+ \frac{1}{2}d_1' f_1' i_1' \ ,\\
l_0 &= - g_1' i_0' - g_0' i_1' + \frac{1}{2} a_1' g_1' i_1' +
\frac{1}{2} d_1' g_1' i_1' \ ,\\
m_0 &= -d_1' i_0' + a_0' i_1' + \frac{i_0^{'2}}{i_1'} - \frac{1}{4}
a_1^{'2} i_1' + \frac{1}{4} d_1^{'2} i_1' \ ,\\
n_0 &= b_1' i_0' + b_0' i_1' - \frac{1}{2} a_1' b_1' i_1' - \frac{1}{2}
d_1' b_1' i_1' \ ,\\
o_0 &= e_1' i_0' + e_0' i_1' - \frac{1}{2} a_1' e_1' i_1' - \frac{1}{2}
d_1' e_1' i_1' \ ,\\
p_0 &= \frac{1}{2} a_1' + \frac{1}{2} d_1' + f_1' i_0' + f_0' i_1' - \frac{i_0'}{i_1'} - \frac{1}{2} a_1' f_1' i_1' -
\frac{1}{2} d_1' f_1' i_1' \ .
\end{array}
\right. \label{tf1}
\eeq
These transition functions are, of course, invertible.

\subsection{Example 3}

The transition functions between the patches with ${\mathbb
  Z}_2$-parity $-1$, viz.~$H_0^{(1,0)} $ and $ H_0^{(-1,0)}$, for the
$k=2$ semi-local vortices in $G'=SO(4)$ theory discussed in
Sec.~\ref{k2so4}, are
\beq
\left\{
\begin{array}{l@{$\,$}l}
a_1 &= - c_1'e_1'i_1' - \frac{e_0'}{e_1'} - \frac{i_0'}{i_1'} \ ,\\
a_0 &= - c_0' e_1' i_1' + \frac{e_0' i_0'}{e_1' i_1'} \ ,\\
b_1 &= - b_1' e_1' i_1' - f_0' i_1' - e_1'j_0' + \frac{e_0' i_1'}{e_1'}
\ ,\\
b_0 &= - b_0' e_1' i_1' - f_0' i_0' - e_0'j_0' + \frac{e_0' i_0'}{e_1'}
\ ,\\
c_1 &= - a_1' e_1' i_1' - e_0'i_1' - e_1' i_0' \ ,\\
c_0 &= - a_0' e_1' i_1' + e_0'i_0' \ ,\\
d_1 &= - d_1' e_1' i_1' - g_0'i_1' - e_1' k_0' - \frac{e_1'i_0'}{i_1'}
\ ,\\
d_0 &= - d_0' e_1' i_1' + g_0'i_0' + e_0' k_0' - \frac{e_0'i_0'}{i_1'} \,\\
e_1 &= - \frac{1}{i_1'} \ ,\\
e_0 &= - \frac{i_0'}{i_1^{'2}} \ ,\\
f_0 &= - \frac{i_0'}{i_1'} - \frac{e_1' j_0'}{i_1'} \ ,\\
g_0 &= - \frac{e_1' k_0'}{i_1'} - \frac{e_1' i_0'}{i_1^{'2}} \ ,\\
i_1 &= - \frac{1}{e_1'} \ ,\\
i_0 &= - \frac{e_0'}{e_1^{'2}} \ ,\\
j_0 &= - \frac{i_1' f_0'}{e_1'} - \frac{i_1'e_0'}{e_1^{'2}} \ ,\\
k_0 &= - \frac{g_0' i_1'}{e_1'} - \frac{e_0'}{e_1'} \ .
\end{array}
\right.
\eeq

\subsection{Example 4}

The transition functions between the patches $(-1)$ and $(1)$ for the
$k=1$ semi-local vortices in $G'=SO(3)$ theory discussed in
Sec.~\ref{sec:so3sl}, are 

\beq \left\{\begin{array}{l@{$\,$}l}
d &= -\frac{2}{d'} \ , \\
e &= -\frac{2e'}{{d'}^2} \ , \\
z_3 &= -\frac{2e'}{d'} - z'_3 \ , \\
f &= d' e' - \frac{1}{2}{d'}^2z'_1 \ ,\\
a &= \frac{1}{2}\left({e'}^2 -{d'}^2 z'_2\right) \ , \\
b &= -\frac{1}{2}b' {d'}^2 -e' - d' z'_3 \ , \\
c &= -\frac{1}{2}c' {d'}^2 - e'\left(\frac{e'}{d'}+z'_3\right) \ , \\
z_1 &= \frac{2e'}{d'} - \frac{1}{2}{d'}^2 f' \ , \\
z_2 &= \frac{{e'}^2}{{d'}^2} - \frac{1}{2}a' {d'}^2 \ .
\end{array}\right. \eeq

\subsection{Example 5}

The transition functions between the patches $(1,1)$ and $(0,0)$ for
the $k=1$ semi-local vortices in $G'=SO(5)$ theory discussed in
Sec.~\ref{sec:so5sl}, are 
\begin{align}
\left\{\begin{array}{l@{$\,$}l}
a'_1 &= \frac{a_1 - a_4}{2} + \frac{f}{e} + \frac{i_1 i_2}{2e} \ , \\
a'_2 &= a_2 + \frac{i_2^2}{2e} \ , \\
a'_3 &= c_1 \ , \\
a'_4 &= -\frac{1}{e} + c_2 \ , \\
a'_5 &= g_1 - \frac{i_2}{e} \ , \\
b'_1 &= a_3 - \frac{i_1^2}{2e} \ , \\
b'_2 &= -\frac{a_1 - a_4}{2} + \frac{f}{e} - \frac{i_1 i_2}{2e} \ , \\
b'_3 &= \frac{1}{e} + c_2 \ , \\
b'_4 &= c_3 \ , \\
b'_5 &= g_2 + \frac{i_1}{e} \ , \\
c'_1 &= -e b_3 + \frac{e a_3 \left(a_1 + a_4\right)}{2} - a_3 f
  -\frac{i_1\left(a_1 i_1 + a_3 i_2\right)}{2} - i_1 j_1 \ , \\
c'_2 &= - e b_4 - \frac{e\left(a_1^2-a_4^2\right)}{4} +a_1 f 
  -\frac{f^2}{e} - \frac{i_1\left(a_2 i_1+a_4 i_2\right)}{2} 
  -i_1 j_2 \ , \\
c'_3 &= -e d_2 +\frac{c_2 e\left(a_1+a_4\right)}{2}+\frac{a_1+a_4}{2} 
  -\frac{f}{e} - c_2 f - \frac{i_1\left(c_1 i_1+c_2 i_2\right)}{2} 
  -\frac{i_1 i_2}{2e} \ , \\
c'_4 &= -e d_3 + \frac{c_3 e\left(a_1+a_4\right)}{2} - c_3 f 
  +\frac{i_1^2}{2e} - \frac{i_1\left(c_2 i_1+c_3 i_2\right)}{2} \ , \\
c'_5 &= -e h_2 +\frac{g_2 e\left(a_1+a_4\right)}{2} - f g_2
  +\frac{i_1\left(a_1+a_4\right)}{2} - \frac{f i_1}{e}
  -\frac{i_1\left(g_1 i_1 + g_2 i_2\right)}{2} - i_1 y \ , \\
d'_1 &= e b_1 - \frac{e\left(a_1^2-a_4^2\right)}{4} - a_4 f
  +\frac{f^2}{e} - \frac{i_2\left(a_1 i_1 + a_3 i_2\right)}{2} 
  - i_2 j_1 \ , \\
d'_2 &= e b_2 - \frac{a_2 e\left(a_1+a_4\right)}{2} + a_2 f
  -\frac{i_2\left(a_2 i_1 + a_4 i_2\right)}{2} - i_2 j_2 \ , \\
d'_3 &= e d_1 - \frac{c_1 e\left(a_1+a_4\right)}{2} + c_1 f
  -\frac{i_2\left(c_1 i_1 + c_2 i_2\right)}{2} 
  -\frac{i_2^2}{2e} \ , \\
d'_4 &= e d_2 - \frac{c_2 e\left(a_1+a_4\right)}{2} 
  + \frac{a_1+a_4}{2} - \frac{f}{e} + c_2 f
  -\frac{i_2\left(c_2 i_1 + c_3 i_2\right)}{2} 
  +\frac{i_1 i_2}{2e} \ , \\
d'_5 &= e h_1 - \frac{g_1 e\left(a_1+a_4\right)}{2} + f g_1
  +\frac{i_2\left(a_1+a_4\right)}{2} - \frac{f i_2}{e} 
  -\frac{i_2\left(g_1 i_1 + g_2 i_2\right)}{2} 
  -i_2 y \ , \\
e'_1 &= j_1 +\frac{i_1\left(a_1 - a_4\right)}{2} +\frac{f i_1}{e}
  +a_3 i_2 \ , \\
e'_2 &= j_2 -\frac{i_2\left(a_1 - a_4\right)}{2} +\frac{f i_2}{e}
  +a_2 i_1 \ , \\
e'_3 &= c_1 i_1 + c_2 i_2 + \frac{i_2}{e} \ , \\
e'_4 &= c_2 i_1 + c_3 i_2 - \frac{i_1}{e} \ , \\
e'_5 &= y + g_1 i_1 + g_2 i_2 \ , 
\end{array}\right. 
\end{align}

\subsection{Example 6}

The transition functions between the patches $(-1,0)$ and $(1,0)$ for
the $k=1$ semi-local vortices in $G'=SO(5)$ theory discussed in
Sec.~\ref{sec:so5sl}, are 
\begin{align}
\left\{\begin{array}{l@{$\,$}l}
a_0 &= \frac{2f'_0 f'_1 + {h'}_0^2}{\Xi} - \frac{1}{2}b'_0\Xi \ , \\
a_1 &= -\frac{2\left(e'_0 f'_0 + e'_1 f'_1 + h'_0 j'_1\right)}{\Xi}
 - \frac{1}{2}b'_1\Xi \ , \\
b_0 &= f'_0 f'_1 + \frac{1}{2}{h'}_0^2 -\frac{1}{2}a'_0\Xi \ , \\
b_1 &= - e'_0 f'_0 - e'_1 f'_1 - h'_0 j'_1 -\frac{1}{2}a'_1\Xi\ , \\
c_0 &= f'_1 g'_0 + f'_0 g'_2 + h'_0 h'_1 
  +\frac{e'_0 \left(2f'_0 f'_1 + {h'}_0^2\right)}{\Xi} 
  - \frac{1}{2}c'_0 \Xi \ , \\
c_1 &= f'_1 - e'_0 g'_0 - e'_1 g'_2 - h'_1 j'_1 
  -\frac{2e'_0\left(e'_0 f'_0 + e'_1 f'_1 + h'_0 j'_1\right)}{\Xi} 
  -\frac{1}{2}c'_1\Xi \ , \\
d_0 &= f'_1 g'_1 + f'_0 g'_3 + h'_0 h'_2 
  +\frac{e'_1 \left(2 f'_0 f'_1 + {h'}_0^2\right)}{\Xi} 
  - \frac{1}{2}d'_0\Xi \ , \\
d_1 &= f'_0 - e'_0 g'_1 - e'_1 g'_3 - h'_2 j'_1 
  -\frac{2 e'_1\left(e'_0 f'_0 + e'_1 f'_1 + h'_0 j'_1\right)}{\Xi} 
  -\frac{1}{2}d'_1\Xi \ , \\
e_0 &= -\frac{2e'_0}{\Xi} \ , \\
e_1 &= -\frac{2e'_1}{\Xi} \ , \\
f_0 &= \frac{-4 e_1^{'2} f'_1 - 4 e'_1 h'_0 j'_1 + 2 f'_0
  j_1^{'2}}{\Xi^2} \ , \\ 
f_1 &= \frac{-4 e_0^{'2} f'_0 - 4 e'_0 h'_0 j'_1 + 2 f'_1
  j_1^{'2}}{\Xi^2} \ , \\ 
g_0 &= g'_0 - \frac{2 e'_1 \left(e'_0 g'_0 + e'_1 g'_2 + h'_1 j'_1
  \right)}{\Xi} + \frac{2 e'_0 \left(-2 e_1^{'2} f'_1 - 2 e'_1 h'_0
  j'_1 + f'_0 j_1^{'2}\right)}{\Xi^2} \ , \\  
g_1 &= \frac{4 e_1^{'3} \left(f'_1 + e'_0 g'_3\right) + g'_1 j_1^{'4} - 
 2 e_1^{'2} j'_1 \left(2 h'_0 + 2 e'_0 h'_2 + g'_3 j'_1\right) + 
 2 e'_1 j_1^{'2} \left(f'_0 + e'_0 g'_1 - h'_2 j'_1\right)}{\Xi^2}
\ , \\ 
g_2 &= \frac{-4 e_0^{'3} \left(f'_0 + e'_1 g'_0\right) + g'_2 j_1^{'4} - 
 2 e_0^{'2} j'_1 \left(2 h'_0 + 2 e'_1 h'_1 + g'_0 j'_1\right) + 
 2 e'_0 j_1^{'2} \left(f'_1 + e'_1 g'_2 - h'_1 j'_1\right)}{\Xi^2} \ ,
\\ 
g_3 &= g'_3 - \frac{4 e'_0 e'_1 \left(e'_0 f'_0 + e'_1 f'_1 + h'_0
  j'_1\right)}{\Xi^2} - \frac{2 \left(-e'_1 f'_1 + e_0^{'2} g'_1 +
  e'_0 e'_1 g'_3 + e'_0 h'_2 j'_1\right)}{\Xi} \ , \\
h_0 &= \frac{-4 e'_0 e'_1 h'_0 + 4 e'_0 f'_0 j'_1 + 4 e'_1 f'_1 j'_1 +
  2 h'_0 j_1^{'2}}{\Xi^2} \ , \\ 
h_1 &= \frac{4 e_0^{'2} \left(-e'_1 \left(h'_0 + e'_1 h'_1\right) +
  \left(f'_0 + e'_1 g'_0\right) j'_1\right) +  
 j_1^{'3} \left(2 e'_1 g'_2 + h'_1 j'_1\right) + 
 2 e'_0 j'_1 \left(2 e'_1 \left(f'_1 + e'_1 g'_2\right) + j'_1
 \left(h'_0 + g'_0 j'_1\right)\right)}{\Xi^2} \ , \\ 
h_2 &= \frac{-4 e'_0 e_1^{'2} \left(h'_0 + e'_0 h'_2\right) + 
 4 e'_1 \left(e'_1 f'_1 + e'_0 \left(f'_0 + e'_0 g'_1 + e'_1
 g'_3\right)\right) j'_1 +  
 2 e'_1 h'_0 j_1^{'2} + 2 \left(e'_0 g'_1 + e'_1 g'_3\right)
 j_1^{'3} + h'_2 j_1^{'4}}{\Xi^2} \ , \\ 
i_0 &= \frac{2 e_0^{'2} e_1^{'2} i'_0 - 2e'_0 e'_1 \left(f'_1 j'_0 -
  i'_0 j_1^{'2} + f'_0 j'_2 + h'_0 k'\right) +  j'_1 \left(-f'_1 j'_0
  j'_1 + \frac{1}{2} i'_0 j_1^{'3} - f'_0 \left(2 f'_1 + j'_1 j'_2\right) - 
    h'_0 \left(h'_0 + j'_1 k'\right)\right)}{\Xi} \ , \\
i_1 &= \frac{2e_0^{'2} e'_1 \left(e'_1 i'_1 + j'_0\right) + 
 j'_1 \left(2 e'_1 f'_1 +  h'_0 j'_1 + \frac{1}{2}i'_1 j_1^{'3} + 
  e'_1 j'_1 j'_2 + j_1^{'2} k'\right) + 
 e'_0 \left(j'_1 \left(2 f'_0 + j'_0 j'_1\right) + 2 e_1^{'2} j'_2 + 
    2 e'_1 \left(-h'_0 + j'_1 \left(i'_1 j'_1 +
    k'\right)\right)\right)}{\Xi} \ , \\ 
j_0 &= \frac{-j_1^{'3} \left(2 f'_0 + j'_0 j'_1\right) + 4 e'_0 e_1^{'3} j'_2 + 
 2 e_1^{'2} j'_1 \left(2 f'_1 + j'_1 j'_2 + 2 e'_0 k'\right) + 
 2 e'_1 j_1^{'2} \left(2 h'_0 - e'_0 j'_0 + j'_1 k'\right)}{\Xi^2} \ ,
\\
j_1 &= \frac{2j'_1}{\Xi} \ , \\
j_2 &= \frac{4 e_0^{'3} e'_1 j'_0 - j_1^{'3} \left(2 f'_1 + j'_1 j'_2\right) + 
 2 e_0^{'2} j'_1 \left(2 f'_0 + j'_0 j'_1 + 2 e'_1 k'\right) + 
 2 e'_0 j_1^{'2} \left(2 h'_0 - e'_1 j'_2 + j'_1 k'\right)}{\Xi^2} \ ,
\\
k &= \frac{-2 e'_0 j'_1 \left(-2 e'_1 h'_0 + 
 j'_1 \left(2 f'_0 + j'_0 j'_1\right) + 2 e_1^{'2} j'_2\right) + 
 4 e_0^{'2} e'_1 \left(-j'_0 j'_1 + e'_1 k'\right) - 
 j_1^{'2} \left(4 e'_1 f'_1 + 2 h'_0 j'_1 + 2 e'_1 j'_1 j'_2 +
 j_1^{'2} k'\right) }{\Xi^2} \ . 
\end{array}\right.
\end{align}


\clearpage

\begin{thebibliography}{100}

\bibitem{Abrikosov}
A.~A.~Abrikosov, 
 Sov.\ Phys.\ JETP {\bf 5}, 1174 (1957) 
[Zh.\ Eksp.\ Teor.\ Fiz.\ {\bf 32}, 1442 (1957) ].

\bibitem{Nielsen:1973cs}
H.~B.~Nielsen and P.~Olesen,
Nucl.\ Phys.\ B {\bf 61}, 45 (1973).


\bibitem{Achucarro}
  T.~Vachaspati and A.~Achucarro,
  Phys.\ Rev.\  D {\bf 44}, 3067 (1991);
  A.~Achucarro and T.~Vachaspati,
  Phys.\ Rept.\  {\bf 327}, 347 (2000);

\bibitem{Jeannerot:2003qv}
 R.~Jeannerot, J.~Rocher and M.~Sakellariadou,
  Phys.\ Rev.\ D {\bf 68}, 103514 (2003).
  
\bibitem{Lubensky} T.C. Lubensky. 
Physica A220,   99  (1995). 

\bibitem{Vilenkin}
A. Vilenkin and E. P. S. Shellard, ``Cosmic Strings
and Other Topological Defects'', Cambridge Univ. Press
(1994); M. B. Hindmarsh and T. W. B. Kibble, Rept. \
Prog. \  Phys. \ {\bf 58}, 477 (1995).

\bibitem{HT}
  A.~Hanany and D.~Tong,
  JHEP {\bf 0307}, 037 (2003)
  [arXiv:hep-th/0306150].

\bibitem{ABEKY}
  R.~Auzzi, S.~Bolognesi, J.~Evslin, K.~Konishi and A.~Yung,
  Nucl.\ Phys.\  B {\bf 673}, 187 (2003) 
  [arXiv:hep-th/0307287].

\bibitem{Isozumi:2004vg}
  Y.~Isozumi, M.~Nitta, K.~Ohashi and N.~Sakai,
  Phys.\ Rev.\  D {\bf 71}, 065018 (2005)
  [arXiv:hep-th/0405129].

\bibitem{Eto:2005yh}
M.~Eto, Y.~Isozumi, M.~Nitta, K.~Ohashi and N.~Sakai,
Phys.\ Rev.\ Lett.\ {\bf 96}, 161601 (2006) 
[arXiv:hep-th/0511088].

 \bibitem{Eto:2006pg}
M.~Eto, Y.~Isozumi, M.~Nitta, K.~Ohashi and N.~Sakai,
J.\ Phys.\ A {\bf 39}, R315 (2006) 
[arXiv:hep-th/0602170].

\bibitem{Isozumi:2004jc}
  Y.~Isozumi, M.~Nitta, K.~Ohashi and N.~Sakai,
  Phys.\ Rev.\ Lett.\  {\bf 93}, 161601 (2004)
  [arXiv:hep-th/0404198];
  Phys.\ Rev.\  D {\bf 70}, 125014 (2004)
  [arXiv:hep-th/0405194].

\bibitem{HashiTong}   
  K.~Hashimoto and D.~Tong,
  JCAP {\bf 0509}, 004 (2005)
  [arXiv:hep-th/0506022].

\bibitem{ASY}  
  R.~Auzzi, M.~Shifman and A.~Yung,
  Phys.\ Rev.\  D {\bf 73}, 105012 (2006)
  [Erratum-ibid.\  D {\bf 76}, 109901 (2007)]
  [arXiv:hep-th/0511150].

\bibitem{Eto:2006cx}
  M.~Eto, K.~Konishi, G.~Marmorini, M.~Nitta, K.~Ohashi, W.~Vinci and N.~Yokoi,
  Phys.\ Rev.\  D {\bf 74}, 065021 (2006)
  [arXiv:hep-th/0607070].

\bibitem{Duality}
 M. Eto, L. Ferretti,  K. Konishi, G. Marmorini, M. Nitta, K. Ohashi, W. Vinci,  N. Yokoi,
Nucl.Phys. {\bf B780}, 161-187 (2007) 
 [arXiv: hep-th/0611313].

\bibitem{SYSemi} 
  M.~Shifman and A.~Yung,
  Phys.\ Rev.\  D {\bf 73}, 125012 (2006)
  [arXiv:hep-th/0603134].

\bibitem{SemiL}    
M.~Eto, J.~Evslin, K.~Konishi, G.~Marmorini, M.~Nitta, K.~Ohashi, W.~Vinci, and N.~Yokoi,   
  Phys.\ Rev.\  D {\bf 76}, 105002 (2007)
  [arXiv:0704.2218 [hep-th]].

\bibitem{DKO}
  D.~Dorigoni, K.~Konishi and K.~Ohashi,
 Phys. Rev. D {\bf 79},  045011 (2009);  
  ``Non-Abelian Vortices without Dynamical Abelianization,''
  arXiv:0801.3284 [hep-th].

\bibitem{FK} 
  L. Ferretti,  K. Konishi 
  ``Duality and confinement in SO(N) gauge theories'', 
  FestSchrift, ``Sense of Beauty in Physics,'' in honor of the 70th birthday of A. Di Giacomo,
  Edizioni PLUS (University of Pisa Press),  2006; 

\bibitem{FGK}
  L.~Ferretti, S.~B.~Gudnason and K.~Konishi,
  Nucl.\ Phys.\  B {\bf 789}, 84 (2008)
  [arXiv:0706.3854 [hep-th]].

\bibitem{Eto:2008qw}
  M.~Eto, T.~Fujimori, S.~B.~Gudnason, M.~Nitta and K.~Ohashi,
  arXiv:0809.2014 [hep-th].

\bibitem{Short}
  M.~Eto, T.~Fujimori, S.~B.~Gudnason, K.~Konishi, M.~Nitta, K.~Ohashi and W.~Vinci,
  Phys.\ Lett.\  B {\bf 669}, 98 (2008)
  [arXiv:0802.1020 [hep-th]].

\bibitem{Baptista:2008ex}
  J.~M.~Baptista,
  arXiv:0810.3220 [hep-th].

\bibitem{Eto:2004rz}
  M.~Eto, Y.~Isozumi, M.~Nitta, K.~Ohashi and N.~Sakai,
  Phys.\ Rev.\  D {\bf 72}, 025011 (2005)
  [arXiv:hep-th/0412048].

\bibitem{ABEK}
  R.~Auzzi, S.~Bolognesi, J.~Evslin and K.~Konishi,
  Nucl.\ Phys.\  B {\bf 686}, 119 (2004)
  [arXiv:hep-th/0312233].

\bibitem{Eto:2004ii}
  M.~Eto, M.~Nitta and N.~Sakai,
  Nucl.\ Phys.\  B {\bf 701}, 247 (2004)
  [arXiv:hep-th/0405161].

\bibitem{Hashimoto} 
M.~Eto, K.~Hashimoto, G.~Marmorini, M.~Nitta, K.~Ohashi and
W.~Vinci, 
Phys. Rev. Lett. {\bf 98}, 091602 (2007) 
 [arXiv: hep-th/0609214]. 

\bibitem{Eto:2006uw}
  M.~Eto, Y.~Isozumi, M.~Nitta, K.~Ohashi and N.~Sakai,
  Phys.\ Rev.\  D {\bf 73}, 125008 (2006)
  [arXiv:hep-th/0602289].

\bibitem{Auzzi:2007wj}
  R.~Auzzi, M.~Eto and W.~Vinci,
  JHEP {\bf 0802}, 100 (2008)
  [arXiv:0711.0116 [hep-th]].
  R.~Auzzi, M.~Eto and W.~Vinci,
 JHEP {\bf 0711} (2007) 090
 [arXiv:0709.1910 [hep-th]].

\bibitem{Auzzi:2008wm}
  R.~Auzzi, M.~Eto, S.~B.~Gudnason, K.~Konishi and W.~Vinci,
  arXiv:0810.5679 [hep-th].
    
     
     
\bibitem{HT2}
  A.~Hanany and D.~Tong,
  JHEP {\bf 0404}, 066 (2004)
  [arXiv:hep-th/0403158].

\bibitem{SY}
  M.~Shifman and A.~Yung,
  Phys.\ Rev.\  D {\bf 66}, 045012 (2002)
  [arXiv:hep-th/0205025];
  
\bibitem{Shifman:2004dr}
  M.~Shifman and A.~Yung,
  Phys.\ Rev.\  D {\bf 70}, 045004 (2004)
  [arXiv:hep-th/0403149].

\bibitem{GSY}
  A.~Gorsky, M.~Shifman and A.~Yung,
  Phys.\ Rev.\  D {\bf 71}, 045010 (2005)
  [arXiv:hep-th/0412082].

\bibitem{Shifman:2005st}
  M.~Shifman and A.~Yung,
  Phys.\ Rev.\  D {\bf 72}, 085017 (2005)
  [arXiv:hep-th/0501211].

\bibitem{Tong:2005un}
D.~Tong,
arXiv:hep-th/0509216; 
  D.~Tong,
  ``Quantum Vortex Strings: A Review,''
  arXiv:0809.5060 [hep-th].

\bibitem{SYReview}  
  M.~Shifman and A.~Yung,
  Rev.\ Mod.\ Phys.\  {\bf 79}, 1139 (2007)
  [arXiv:hep-th/0703267].

\bibitem{Eto:2006mz}
  M.~Eto, T.~Fujimori, Y.~Isozumi, M.~Nitta, K.~Ohashi, K.~Ohta and N.~Sakai,
  Phys.\ Rev.\  D {\bf 73}, 085008 (2006)
  [arXiv:hep-th/0601181];
  M.~Eto, T.~Fujimori, M.~Nitta, K.~Ohashi, K.~Ohta and N.~Sakai,
  Nucl.\ Phys.\  B {\bf 788}, 120 (2008)
  [arXiv:hep-th/0703197].

\bibitem{Tong:2006pa}
  D.~Tong,
  JHEP {\bf 0612}, 051 (2006)
  [arXiv:hep-th/0610214].

\bibitem{Edalati:2007vk}
  M.~Edalati and D.~Tong,
  JHEP {\bf 0705}, 005 (2007)
  [arXiv:hep-th/0703045];
  D.~Tong,
  JHEP {\bf 0709}, 022 (2007)
  [arXiv:hep-th/0703235];
  M.~Shifman and A.~Yung,
  Phys.\ Rev.\  D {\bf 77}, 125016 (2008)
  [arXiv:0803.0158 [hep-th]];
  M.~Shifman and A.~Yung,
  Phys.\ Rev.\  D {\bf 77}, 125017 (2008)
  [arXiv:0803.0698 [hep-th]].

\bibitem{Aldrovandi:2007bn}
  L.~G.~Aldrovandi,
  Phys.\ Rev.\  D {\bf 76}, 085015 (2007)
  [arXiv:0706.0446 [hep-th]].
   
\bibitem{KonishiMN}  K.~ Konishi, 
``Vortices which do not Abelianize dynamically: 
Semi-classical origin of non-Abelian monopoles'',  Contribution to 
Continuous Advances in QCD 2008,  Minneapolis, May 2008.
   e-Print: arXiv:0809.1374 [hep-th]. 

\bibitem{Collie:2008mx}
  B.~Collie and D.~Tong,
  arXiv:0805.0602 [hep-th];
  B.~Collie,
  arXiv:0809.0394 [hep-th].


\bibitem{Dorey}  N. Dorey,
 JHEP {\bf 9811},  005 (1998) [arXiv:hep-th/9806056].


 \bibitem{GNO}   P.~Goddard, J.~Nuyts and D.~Olive,  Nucl.\ Phys.\ B {\bf 125}
(1977) 1;   F.~A.~Bais, Phys.\ Rev.\ D {\bf 18} (1978) 1206;   B.~J.~Schroers and  F.~A.~Bais, Nucl.\ Phys.\ B {\bf 512}  (1998) 250,  
hep-th/9708004;  Nucl.\ Phys.\ B {\bf 535} (1998) 197,  hep-th/9805163;   
E.~J.~Weinberg, Nucl.\ Phys.\ B {\bf167} (1980) 500;  
Nucl. Phys. B {\bf 203} (1982) 445. 

\bibitem{KS} K. Konishi and L. Spanu, 
Int. J. Mod. Phys. A18  (2003)  249,  
arXiv: hep-th/0106175. 

\bibitem{CDyons} A. Abouelsaood,  Nucl. Phys.  B {\bf 226},  309 (1983);  P. Nelson, A. Manohar,  Phys. Rev. Lett. {\bf  50},  943
(1983);  A. Balachandran, G. Marmo, M. Mukunda, J. Nilsson, E. Sudarshan, F. Zaccaria,    Phys. Rev. Lett. {\bf  50},  1553
(1983);  P. Nelson, S. Coleman,  Nucl. Phys.  B {\bf 227},  1  (1984);
  N. Dorey, C. Fraser, T.J. Hollowood,  M.A.C. Kneipp,  
[arXiv: hep-th/9512116];   Phys.Lett.  B {\bf 383}, 422  (1996)

\bibitem{Hindmarsh:1991jq}
M.~Hindmarsh,
Phys.\ Rev.\ Lett.\ {\bf 68} (1992) 1263;  
  Nucl.\ Phys.\  B {\bf 392} (1993) 461
  [arXiv:hep-ph/9206229];
A. Ach\'ucarro, K. Kuijken, L. Perivolaropoulos, T. Vachaspati,  
  Nucl.\ Phys.\  B {\bf 388} (1992) 435
  [arXiv:hep-ph/9206229].

\bibitem{Polyakov:1975yp}
  A.~M.~Polyakov and A.~A.~Belavin,
  JETP Lett.\  {\bf 22}, 245 (1975)
  [Pisma Zh.\ Eksp.\ Teor.\ Fiz.\  {\bf 22}, 503 (1975)].

\bibitem{Hergason} S. Hergason, ``Differential Geometry, Lie groups, and Symmetric Spaces'',  Academic Press, New York (1978). 

\bibitem{Higashijima:2001vk}
  K.~Higashijima and M.~Nitta,
  Prog.\ Theor.\ Phys.\  {\bf 103}, 635 (2000)
  [arXiv:hep-th/9911139];
  K.~Higashijima, T.~Kimura and M.~Nitta,
  Nucl.\ Phys.\  B {\bf 623}, 133 (2002)
  [arXiv:hep-th/0108084].

\bibitem{Berkovits:2004bw}
  N.~Berkovits and S.~A.~Cherkis,
  JHEP {\bf 0412}, 049 (2004)
  [arXiv:hep-th/0409243].

\bibitem{GNOWnine}  M.~Eto, T.~Fujimori, S.~B.~Gudnason, K.~Konishi, T.~Nagashima, M.~Nitta, K.~Ohashi and W.~Vinci,  in preparation.

\bibitem{quasi-NG}
  M.~Bando, T.~Kuramoto, T.~Maskawa and S.~Uehara,
  Phys.\ Lett.\  B {\bf 138}, 94 (1984); 
  Prog.\ Theor.\ Phys.\  {\bf 72}, 313 (1984);
  Prog.\ Theor.\ Phys.\  {\bf 72}, 1207 (1984);
  A.~C.~W.~Kotcheff and G.~M.~Shore,
  Int.\ J.\ Mod.\ Phys.\  A {\bf 4}, 4391 (1989); 
  K.~Higashijima, M.~Nitta, K.~Ohta and N.~Ohta,
  Prog.\ Theor.\ Phys.\  {\bf 98}, 1165 (1997)
  [arXiv:hep-th/9706219];
  M.~Nitta,
  Int.\ J.\ Mod.\ Phys.\  A {\bf 14}, 2397 (1999)
  [arXiv:hep-th/9805038].


\bibitem{Fractional}  M.~Eto, T.~Fujimori, S.~B.~Gudnason, K.~Konishi, T.~Nagashima, M.~Nitta, K.~Ohashi and W.~Vinci,  in preparation.
\bibitem{WVLump}
W.~Vinci,
 arXiv:0810.2449 [hep-th].

\bibitem{Benson:1994dp}
  K.~M.~Benson, A.~V.~Manohar and M.~Saadi,
  Phys.\ Rev.\ Lett.\  {\bf 74}, 1932 (1995)
  [arXiv:hep-th/9409042];
  K.~M.~Benson and M.~Saadi,
  Phys.\ Rev.\  D {\bf 51}, 3096 (1995)
  [arXiv:hep-th/9409109].

\bibitem{Callias:1977kg}
  C.~Callias,
  Commun.\ Math.\ Phys.\  {\bf 62}, 213 (1978).

\bibitem{Weinberg:1979er}
  E.~J.~Weinberg,
  Phys.\ Rev.\  D {\bf 19}, 3008 (1979).

\bibitem{Manton:1981mp}
 N.~S.~Manton,
 Phys.\ Lett.\  B {\bf 110}, 54 (1982).


\end{thebibliography}
\end{document}